\documentclass[11pt,letterpaper]{article}
\pdfoutput=1
\usepackage{soul}
\usepackage{jheppub}
\usepackage[utf8]{inputenc}
\usepackage[T1]{fontenc}
\usepackage{amsfonts}
\usepackage{amsmath}
\usepackage{tensor} 
\usepackage{mathtools} 
\allowdisplaybreaks[4]         
\usepackage{amssymb}   
\usepackage{euscript}       
\usepackage{color}          
\newcommand{\labell}[1]{\label{#1}} 

\newcommand{\req}[1]{(\ref{#1})} 

\newcommand{\bea}{\begin{eqnarray}}
\newcommand{\eea}{\end{eqnarray}}
\newcommand{\ba}{\begin{eqnarray}}
\usepackage{braket}
\newcommand{\ea}{\end{eqnarray}}

\newcommand{\beq}{\begin{equation}}
\newcommand{\eeq}{\end{equation} }
\newcommand{\beqa}{\begin{eqnarray}}

\newcommand{\eeqa}{\end{eqnarray}}
\newcommand{\beqar}{\begin{eqnarray*}}
\newcommand{\eeqar}{\end{eqnarray*}}
\newcommand{\e}{\epsilon}

\newcommand{\be}{\begin{equation}}
\newcommand{\ee}{\end{equation}}

\newcommand{\X}{\mathcal{X}}
\newcommand{\W}{\mathcal{W}}
\newcommand{\R}{\mathcal{R}}
\newcommand{\Z}{\mathcal{Z}}
\newcommand{\A}{\mathcal{A}}

\newcommand{\C}{\mathcal{C}}
\newcommand{\D}{\mathcal{D}}

\newcommand{\G}{\mathcal{G}}

\renewcommand{\c}{$c$}

\title{ \boldmath Quasinormal modes of NUT-charged black branes in the AdS/CFT correspondence}

\author[a]{Pablo A. Cano,}
\author[b]{and David Pere\~niguez}

\affiliation[a]{Instituut voor Theoretische Fysica, KU Leuven.\\
Celestijnenlaan 200D, B-3001 Leuven, Belgium \vspace{0.1cm}}
\affiliation[b]{Instituto de F\'isica Te\'orica UAM/CSIC.\\ C/Nicol\'as Cabrera, 13-15, C.U. Cantoblanco, 28049 Madrid, Spain}

\emailAdd{pabloantonio.cano@kuleuven.be}
\emailAdd{david.perenniguez@uam.es}

\date{\today}
\abstract{We study the scalar, electromagnetic and gravitational perturbations of planar AdS$_4$ black holes with NUT charge. In the context of the AdS/CFT correspondence, these solutions describe a thermal quantum field theory embedded in a G\"odel-type universe with closed time-like curves. For a given temperature and NUT charge, two different planar Taub-NUT solutions exist, but we show that only the one with a positive specific heat contributes to the Euclidean saddle point in the path integral.  By using the Newman-Penrose formalism, we then derive the master equations satisfied by scalar, electromagnetic and gravitational perturbations in this background, and show that the corresponding equations are separable. Interestingly, the solutions pile up in the form of Landau levels, and hence are characterized by a single quantum number $q$.
We determine the appropriate boundary conditions satisfied by the master variables and using these we compute the quasinormal modes of scalar and gravitational perturbations. On the other hand, electromagnetic perturbations depend on a free parameter whose determination is problematic. 
We find that all the scalar and gravitational QNM frequencies lie in the lower half of the complex plane, indicating that these Taub-NUT spacetimes are stable. We discuss the implications of these results in the light of the AdS/CFT correspondence.

 }

\begin{document} 
\maketitle

\flushbottom

\section{Introduction}\labell{Introduction}
Motivated by the AdS/CFT correspondence \cite{Maldacena,Witten,Gubser}, the study of asymptotically anti-de Sitter (AdS) black holes has been a major field of research in the last two decades. According to this correspondence, black hole solutions in the bulk of AdS are dual to a thermal quantum field theory living in the boundary of the spacetime and whose temperature is given by the Hawking's temperature of the black hole. 
In this context, the holographic dictionary can be applied to gain a great deal of information about the hydrodynamics of strongly-coupled plasmas by studying the properties of the black hole solutions \cite{Son:2007vk,Gubser:2009md,Hartnoll:2009sz,Herzog:2009xv}. In particular, perturbations of different fields in the background of a black hole geometry can be used to compute transport coefficients and correlators in the dual theory, and thus providing us with valuable results that can be difficult to obtain by first principles in the quantum theory. 

In a black hole, the late-time behaviour of perturbations is ruled by the quasinormal modes (QNMs), which are solutions satisfying an outgoing boundary condition at the horizon (\textit{i.e.}, absence of waves coming from the horizon) plus --- in the context of AdS/CFT --- Dirichlet boundary conditions at infinity--- see the reviews \cite{Berti:2009kk,Konoplya:2011qq}. Quasinormal modes only exist for a discrete set of complex frequencies, called the QNM frequencies, and whose imaginary part determines the damping time. 
The QNMs of black holes defined in this way correspond to the poles of the retarded Green functions of the dual theory and therefore they characterize the response of the dual plasma under perturbations \cite{Birmingham:2001pj,Son:2002sd,Starinets:2002br,Policastro:2002se,Herzog:2002fn,Nunez:2003eq,Kovtun:2005ev,Baier:2007ix}.

A large part of the literature on this topic has focused on AdS$_{5}$ solutions --- see the previous references --- and especially on black holes with a planar horizon, since these are dual to a 4-dimensional CFT in flat space. In this paper, nonetheless, we are interested in AdS$_4$ geometries. As a matter of fact, the AdS$_{4}$/CFT$_{3}$ correspondence is well-motivated \cite{Aharony:2008ug} and it can indeed be relevant for certain condensed-matter systems that behave effectively as $2+1$ dimensional \cite{Hartnoll:2009sz,Herzog:2009xv}. 
The quasinormal modes of 4-dimensional Schwarzschild-AdS black holes were studied in Refs.~\cite{Horowitz:1999jd,Cardoso:2001bb,Cardoso:2003cj,Musiri:2003rs}, while those of black holes with planar, toroidal and cylindrical topologies were first computed in Refs.~\cite{Cardoso:2001vs,Miranda:2005qx}.  The results on the latter were later revised and extended in Ref.~\cite{Miranda:2008vb} by implementing the boundary conditions required by holography.  
On the other hand, the quasinormal modes of large Kerr-AdS black holes were analyzed in \cite{Giammatteo:2005vu}. 

In addition to these cases, there is a family of gravitational solutions that has not been yet fully exploited in holography: black holes with NUT charge  \cite{Taub:1950ez,Newman:1963yy,Hawking:1998ct,Chamblin:1998pz}. 
Taub-NUT\footnote{We use the term ``Taub-NUT'' to refer indistinctly to both NUT-type and bolt-type solutions.}  solutions have the distinct property of being only locally asymptotically AdS, which translates into the fact that the boundary is no longer (locally) conformally flat. Thus, NUT charge breaks conformal invariance of the dual theory, and this may allow us to probe non-trivial aspects of the CFT. For instance, Euclidean AdS-Taub-NUT solutions describe CFTs placed on squashed spheres  \cite{Hawking:1998ct,Chamblin:1998pz}, and studying how the free energy depends on the NUT charge has led to interesting results both in supersymmetric \cite{Imamura:2011wg,Martelli:2012sz} and non-supersymmetric \cite{Bobev:2016sap,Bobev:2017asb,Bueno:2018yzo,Bueno:2020odt} setups. 

Lorentzian Taub-NUT solutions, on the other hand, have been less studied in the context of holography due to their seemingly pathological properties. Indeed, these solutions contain Misner strings and closed time-like curves \cite{Misner:1963fr,Manko:2005nm}, and they give rise to an apparent failure of the first law of thermodynamics \cite{Astefanesei:2004ji}. 
However, there is a renewed interest in ``rehabilitating'' these spacetimes. On the one hand, Ref.~\cite{Clement:2015cxa} has shown that freely falling observers do not experience any of these pathologies, since there are no closed time-like geodesics and Misner strings are invisible to those observers --- see also \cite{Clement:2015aka}. On the other hand, the thermodynamic description of Taub-NUT solutions has been finally understood on the basis that Misner strings are acceptable and that, accordingly, the NUT charge should be regarded as an independent thermodynamic variable \cite{Kubiznak:2019yiu,Bordo:2019tyh,Bordo:2019rhu} --- see also \cite{Ciambelli:2020qny}. 

Lorentzian AdS-Taub-NUT solutions give indeed rise to interesting boundary theories. In Refs.~\cite{Leigh:2011au,Leigh:2012jv} it was noted that, unlike the Kerr-AdS solution, NUT-charged solutions describe fluids with vorticity, and hence explore a qualitatively different aspect of the dual theory.  More recently, Ref.~\cite{Kalamakis:2020aaj} initiated the study of scalar perturbations of spherical Taub-NUTs in connection to holography, finding that the result is dramatically dependent on whether the Misner string is regarded as physical or not. 
In this work, we will consider instead the case of planar Taub-NUT black holes \cite{Astefanesei:2004kn} ---  we recall that, just like in the case of AdS black holes, NUT-charged solutions can have either spherical, planar or hyperbolic transverse sections. We consider this case to be particularly interesting for two main reasons. 
First, the planar NUT solutions are free of Misner strings, so that one gets rid of all the difficulties and ambiguities introduced by these objects. Second, these solutions are a generalization of the planar black holes, and hence the boundary metric can be considered as a continuous deformation of flat space.  More precisely, the boundary of these geometries is similar to a G\"odel universe \cite{Astefanesei:2004kn}, where the NUT charge controls the rotation. 
In this sense, it is interesting to see how the properties of the dual strongly-coupled plasma change as we increase the NUT charge.

In this paper, we explore this question by computing the quasinormal mode spectrum of planar Taub-NUT black holes.  We shall perform an analysis of (massless) scalar, electromagnetic and gravitational perturbations, providing --- to the best of our knowledge --- the first complete calculation of quasinormal modes of black holes with NUT charge. 

The paper is organized as follows
\begin{itemize}
\item In Section~\ref{sec:taubnut} we review the planar Taub-NUT geometries, establishing their basic properties, their thermodynamics description and introducing the Newman-Penrose formalism that we use in the next section. 
\item In Section~\ref{sec:pert} we perform perturbation theory on these geometries. The case of a scalar field is considered first and we note an interesting analogy between the angular separation of the QNMs and Landau quantization. We then use the Newman-Penrose formalism to derive separable equations for the master electromagnetic and gravitational variables. 
\item In Section~\ref{sec:bdry} we study the boundary conditions for QNMs. Imposing Dirichlet boundary conditions on the electromagnetic and gravitational perturbations, we derive the form of the boundary conditions on the master Newmann-Penrose variables. We find that, besides the QNM frequency, the QNMs depend on another parameter related to the polarization, and which has to be  determined by solving simultaneously the equations for both NP variables. In the gravitational case we determine analytically this polarization parameter by using the Teukolsy-Starobinsky identities, and hence we reduce the problem to solving only one equation with fixed boundary conditions. On the other hand, we find that the electromagnetic NP variables satisfy degenerate equations, and therefore the  polarization parameter cannot be determined. 
\item We compute the QNM frequencies of scalar and gravitational perturbations in Section~\ref{sec:QNMs}. Despite the breaking of parity, the spectra of both types of perturbations is symmetric under the change of sign of the NUT charge. We obtain an analytic approximation for a special family of gravitational QNMs, that we call pseudo-hydrodynamic modes, whose frequency vanishes in the zero NUT charge limit. In addition, we provide strong evidence that no unstable mode exists. 

\item We present our conclusions in Section \ref{sec:conc}.
\end{itemize}

\section{Planar Taub-NUT black holes and their holographic dual}\label{sec:taubnut}
We consider Einstein gravity with a negative cosmological constant,

\begin{equation}
S=\frac{1}{16\pi G}\int d^4x\sqrt{|g|}\left[R+\frac{6}{L^2}\right]
\end{equation}

\noindent
In this paper, we are interested in the following solution of Einstein's theory, corresponding to a Taub-NUT black hole with planar topology \cite{Chamblin:1998pz},
\begin{equation}\label{eq:NUTBB}
ds^2=-V(r)\left(dt+\frac{2n}{L^2}xdy\right)^2+\frac{dr^2}{V(r)}+\frac{r^2+n^2}{L^2}\left(dx^2+dy^2\right)
\end{equation}
where $n$ is the NUT charge, the function $V(r)$ is given by

\begin{equation}
V(r)=\frac{(r-r_{+}) \left(3 n^4+6 n^2 r r_{+}+r r_{+} \left(r^2+r r_{+}+ r_{+}^2\right)\right)}{L^2  r_{+} \left(n^2+r^2\right)}\, ,
\end{equation}
and the coordinates $(x,y)$ span $\mathbb{R}^2$. 
For $n=0$ this solution reduces to the AdS black brane, but nevertheless it has some remarkable properties that we review next.
 First of all, this solution conserves all the symmetries of the black brane, corresponding to time translations and the symmetries of $\mathbb{R}^2$, with the difference that the latter now act non-trivially in the time variable. The corresponding four Killing vectors read
\begin{equation}\label{eq:Killing}
\begin{aligned}
\xi_{(t)}&=\partial_{t}\, ,\\
\xi_{(1)}&=-\frac{2n}{L^2}y\partial_{t}+\partial_{x}\, ,\\
\xi_{(2)}&=\partial_{y}\, ,\\
\xi_{(3)}&=\frac{n}{L^2}(x^2-y^2)\partial_{t}+y\partial_{x}-x\partial_{y}\, .
\end{aligned}
\end{equation}
Note that these symmetries allow one to consider quotients of this solution by discrete groups. For instance one may take $y$ to be periodic, in which case the black hole would have cylindrical topology. We will restrict to the case of $(x,y)$ spanning the plane. 

The event horizon of the black hole is located at $r=r_{+}>0$, which is a Killing horizon for $\xi_{(t)}$. The corresponding surface gravity  reads
\begin{equation}
\kappa=\frac{1}{2}V'(r_+)=\frac{3(n^2+r_{+}^2)}{2L^2r_{+}}\, .
\end{equation}
One can see that the function $V(r)$ is strictly positive for $r_{+}<r<\infty$ and hence there are no other horizons for $\partial_{t}$. There are, however, horizons for the other Killing vectors, which indicate the presence of closed timelike curves (CTCs). For instance, the norm of $\xi_{(2)}$ reads
\begin{equation}
\xi_{(2)}^2=\frac{r^2+n^2}{L^2}-V(r)\left(\frac{2nx}{L^2}\right)^2\, ,
\end{equation}
and hence it becomes timelike if $x$ is large. The symmetries of this spacetime imply that there are CTCs around any point (in the region $r>r_{+}$), but, however,  there are no closed timelike geodesics \cite{Brecher:2003rv,Clement:2015cxa}, so that the solution is possibly less pathological than one would expect.  On the other hand, unlike the spherical Taub-NUT solutions, these NUT black branes do not possess Misner singularities.

At infinity, the metric function $V(r)$ behaves as $V(r)=r^2/L^2+\mathcal{O}(1)$, and hence the boundary metric at $r\rightarrow\infty$ is conformally equivalent to
\begin{equation}\label{bdrymetric}
d\hat{s}^2=-\left(dt+\frac{2n}{L^2}xdy\right)^2+dx^2+dy^2\, .
\end{equation}
This metric is not conformally flat, and therefore the solution is only asymptotically locally AdS. In the boundary theory, this means that conformal invariance is broken. However, the boundary still has many symmetries --- given by \req{eq:Killing} --- and one can see that it is a homogeneous space corresponding to a Lorentzian continuation of Nil space --- the group manifold of Heisenberg's group. Indeed, note that the translational Killing vectors satisfy the Heisenberg's algebra
\begin{equation}
\left[\xi_{(t)},\xi_{(1)}\right]=\left[\xi_{(t)},\xi_{(2)}\right]=0\, ,\quad \left[\xi_{(1)},\xi_{(2)}\right]=\frac{2n}{L^2}\xi_{(t)}\, .
\end{equation}
On a more physical perspective, the metric \req{bdrymetric} can be interpreted as a rotating universe, very similar to the non-trivial $(2+1)$-dimensional section of the famous G\"odel solution \cite{Godel:1949ga}, the paradigmatic example of a universe with closed timelike curves.\footnote{More precisely, the metric \req{bdrymetric} is the equatorial section of the Som-Raychaudhuri solution \cite{10.2307/2415999}, as originally noted in \cite{Astefanesei:2004kn}. Both metrics have qualitatively similar properties.}
Hence, when one applies the holographic dictionary to these solutions, one is probing the dynamics of a quantum theory placed in this exotic spacetime. Although the existence of a globally defined timelike Killing vector allows one to define a Hamiltonian, performing quantum field theory in this background is challenging due to its unusual causal structure \cite{Leahy:1982dj,Novello:1992hp,Radu:2001jq,Brecher:2003rv}. In this sense, holography can be used to gain some insight about the behaviour of a quantum theory in such spacetime. 
Besides, the dual CFT would be in a thermal state whose properties are determined by the thermodynamic quantities of the black hole, that we review next.

\subsection{Thermodynamics}
The temperature of the NUT-charged black branes is given by Hawking's result $T=\kappa/(2\pi)$, so that
\begin{equation}\label{eq:Temp}
T=\frac{3(n^2+r_{+}^2)}{4\pi L^2r_{+}}\, .
\end{equation}
One can see that, for a given $n$, the temperature reaches its minimum value for $r_{+}=|n|$, in whose case we have $T=T_*$, where
\begin{equation}
T_*=\frac{3|n|}{2\pi L^2}\, .
\end{equation}
On the other hand, the temperature diverges both for $r_{+}\rightarrow 0$ and $r_{+}\rightarrow\infty$. Hence, when $T>T_*$, there are two different black hole solutions with the same $T$ and $n$.  This allows us to distinguish three different families of solutions, corresponding to $n<-r_{+}$, $-r_{+}<n<r_{+}$ or $n>r_{+}$. 
We can also identify the mass of the solution by analyzing the behaviour near infinity. In fact, one can just apply the usual the ADM result which tells us that the total energy $E$ can be identified by looking at the $1/r$ term in the asymptotic expansion of $V$. In particular, the coefficient of that term should be equal to $-8 \pi G L^2 E/V_2$, where $V_2$ is the volume of the transverse space, $V_2=\int dx dy$. Note that in this case $V_2$ is infinite, and hence it is more appropriate to talk about energy density $\rho=E/V_2$, rather than total energy. This quantity, in fact, can be interpreted as the energy density in the boundary CFT.  The expansion of $V(r)$ reads
\begin{equation}
V(r)=\frac{r^2}{L^2}+\frac{5 n^2}{L^2}-\frac{r_{+}^4+6 n^2 r_{+}^2-3 n^4}{L^2 rr_{+}}+\mathcal{O}\left(\frac{1}{r^2}\right)
\end{equation}
and therefore, we get

\begin{equation}
\rho=\frac{r_{+}^4+6 n^2 r_{+}^2-3 n^4}{8\pi G L^4 r_{+}}\, .
\end{equation}
On the other hand, the entropy of the black hole is given by $S=A/(4G)$, but since this area of the horizon is divergent, it is again convenient to work in terms if the entropy density, $s=S/V_{2}$, which reads

\begin{equation}
s=\frac{r_{+}^2+n^2}{4GL^2}
\end{equation}
Now, an apparent puzzle in the case of these solutions is that the first law of thermodynamics does not seem to hold, \textit{i.e.}, we get $d\rho\neq T ds$ when varying the previous expressions with respect to $r_{+}$. However, the reason is that the NUT charge should also be interpreted as a thermodynamical variable which will modify the first law.  For a long time this was a source of confusion in the case of spherical Taub-NUT black holes, since regularity of the Euclidean geometry imposes a restriction between NUT charge and temperature \cite{Astefanesei:2004ji}. Only recently it was realized that one can achieve a full-cohomegeneity first law for spherical NUTs by allowing the NUT charge to vary independently. 
In the case of planar NUT black holes, however, there is no restriction between $n$ and $T$, and it is natural to treat the NUT charge as an additional thermodynamic variable. To the best of our knowledge, the existence of a first law in the case of planar Taub-NUT solutions was first reported in  \cite{Bueno:2018uoy}.

In order to complete the thermodynamic characterization of these planar NUT black holes, we must compute the free energy from the Euclidean on-shell action. The Euclidean solution is obtained, not only by Wick-rotating the time coordinate, $t=i\tau$, but also the NUT charge, $\hat n=i n$. In that case the metric reads

\begin{equation}\label{eq:NUTBBE}
ds^2_{E}=V(r)\left(d\tau+\frac{2\hat{n}}{L^2}xdy\right)^2+\frac{dr^2}{V(r)}+\frac{r^2-\hat{n}^2}{L^2}\left(dx^2+dy^2\right)\, .
\end{equation}
It is important to note that, in Euclidean signature, only the solutions with $r_{+}^2\ge \hat n^2$ are regular, which means that the Lorentzian solutions with $n^2>r_{+}^2$ do not have an Euclidean description. This suggests that for a given $T>T_{*}$ only the solution with $r_{+}^2\ge n^2$ should be taken into account in the path integral, and hence that it is the dominant saddle. 
Let us also mention that, in the literature, the Euclidean solutions with $r_{+}^2=\hat{n}^2$ are called Taub-NUT, while the rest are Taub-bolt. However, we shall make no distinctions since the former can be considered as a limit of the latter.  

The free energy can be computed from the following well-posed and regularized Euclidean action
\begin{equation}
I_{E}=-\frac{1}{16\pi G}\int d^4x\sqrt{|g|}\left[R+\frac{6}{L^2}\right]-\frac{1}{8\pi G}\int d^3x\sqrt{h}\left[K-\frac{2}{L}-\frac{L}{2}\mathcal{R}\right]\, ,
\end{equation}
where $K$ is the extrinsic curvature of the boundary and $\mathcal{R}$ is the Ricci scalar of the boundary's intrinsic. The free energy $F=T I_{E}$ reads
\begin{equation}
F=-\frac{V_2(r_{+}^4+3\hat{n}^4)}{16\pi G L^4 r_{+}}\, .
\end{equation}
Let us then define the free-energy density $\varepsilon=F/V_2$ and express this result in terms of the Lorentzian NUT charge $n$,

\begin{equation}
\varepsilon=-\frac{(r_{+}^4+3n^4)}{16\pi G L^4 r_{+}}\, .
\end{equation}
Now, it turns out that, instead of $n$, the thermodynamic relations are most naturally written in terms of the variable $\theta=\frac{1}{n}$.
Then, using the chain rule one can compute the derivatives of the free energy at constant $\theta$ and $T$, which read

\begin{align}
s=&-\left(\frac{\partial \varepsilon}{\partial T}\right)_{\theta}=\frac{r_{+}^2+n^2}{4GL^2}\, ,\\
\psi=&-\left(\frac{\partial \varepsilon}{\partial \theta}\right)_{T}=\frac{3n^3(r_{+}^2-n^2)}{8\pi G L^4 r_{+}}\, .
\end{align}
We check that $s$ indeed coincides with the Bekenstein-Hawking entropy density. On the other hand, $\psi$ is a new thermodynamic potential conjugate to $\theta$. Making use of these results, one observes that the energy $\rho$ computed according to the ADM prescription, coincides with the double Legendre transform of the free energy with respect to $T$ and $\theta$. 

\begin{equation}
\rho=\varepsilon+T s+\theta \psi\, .
\end{equation}
This is is not the standard definition of internal energy, which suggests that, in the presence of NUT charge, the potentials $\varepsilon$ and $\rho$ probably have a different thermodynamic interpretation. 
In any case, this result implies that $\rho$ satisfies the following first law,

\begin{equation}
d\rho=Tds+\theta d\psi\, .
\end{equation}

Finally, we can study the thermodynamic stability of these solutions. One can first can compute the specific heat at constant $\theta$, 
\begin{equation}
C_{\theta}=T\left(\frac{\partial s}{\partial T}\right)_{\theta}=\frac{r_{+}^2(n^2+r_{+}^2)}{2GL^2(r_{+}^2-n^2)}\, ,
\end{equation}
and one can see that $C_{\theta}>0$ as long as $r_{+}^2>n^2$, implying thus stability when $n$ is held fixed. More generally, one can study the concavity of the free energy, for which one may compute the second variation of $\varepsilon$, which reads
\begin{equation}
\delta^2\varepsilon=-\frac{2 \pi  r_+^3}{3G \left(r_+^2-n^2\right)}\delta T^2-\frac{2n^3 \left(n^2+r_+^2\right)}{2 GL^2 \left(r_+^2-n^2\right)}\delta T\delta\theta+\frac{3 n^4\left(3 r_+^4-10 n^2 r_+^2+3 n^4\right)}{8\pi G L^4 r_{+}\left(r_+^2-n^2\right)}\delta\theta^2\, .
\end{equation}
The solution will be thermally stable if this is a negative-definite quadratic from, but we can see that this never happens because the term with $\delta\theta^2$ is positive for $r_{+}^2>n^2$, while the one of $\delta T^2$ is only negative in that region. Therefore, these planar Taub-NUT black holes are only thermodynamically stable under changes of the temperature but not under changes of $n$.

\subsection{Newman--Penrose formalism}
The description of perturbations on a black hole spacetime is a task of extraordinary complexity. The linearized equations governing first order field components on local coordinates are considerably involved already in the simplest backgrounds such as Schwarzschild's black hole, and almost intractable in more realistic cases like Kerr's spacetime. In addition, it is far from obvious how the large amount of gauge symmetry should be fixed. Teukolsky's seminal work \cite{Teukolsky:1973ha} constituted a major breakthrough in the clarification of these issues. Considering an algebraically special background space, of Petrov type $D$ (e.g. Schwarzschild and Kerr spacetimes), he derived decoupled equations for perturbations of several kinds and, furthermore, these admit solutions in separable form. One of the elements underlying such a remarkable success is the Newman--Penrose (NP) formalism  \cite{Newman:1961qr}. In particular, it provides a very natural formulation of Petrov's classification, as well as the Goldberg--Sachs theorem, and this translates into the vanishing of several NP variables of the background. It is in this situation that the equations decouple and, in addition, become gauge invariant. 

The study of perturbations on the background \eqref{eq:NUTBB} can be conveniently performed in the NP formalism. A Newman--Penrose frame on a pseudo--Riemannian space\footnote{We will be following the conventions in \cite{Stephani:2003tm}.} is a complex tetrad $\bold{e}_{a}$,
\begin{equation}
\bold{e}_{1}=\bold{m},\ \ \ \bold{e}_{2}=\overline{\bold{m}},\ \ \  \bold{e}_{3}=\bold{l},\ \ \ \bold{e}_{4}=\bold{k}, 
\end{equation}
composed of two real, null vectors $\bold{k}$ and $\bold{l}$, and one complex, null vector $\bold{m}$ together with its conjugate $\overline{\bold{m}}$, so that
\begin{equation}
\bold{k}\cdot\bold{k}=\bold{l}\cdot\bold{l}=\bold{m}\cdot\bold{m}=0,
\end{equation}
and these are further subject to the normalization conditions
\begin{equation}
\bold{k}\cdot\bold{l}=-\bold{m}\cdot\overline{\bold{m}}=-1, \ \ \ \bold{k}\cdot\bold{m}=\bold{l}\cdot\bold{m}=0.
\end{equation}
When acting as operators on functions $\varphi$, it is customary to give particular names to the vectors of the NP basis,
\begin{equation}
D\varphi:=k^{\mu}\partial_{\mu}\varphi,\ \ \ \ \Delta\varphi:=l^{\mu}\partial_{\mu}\varphi,\ \ \ \ \delta\varphi:=m^{\mu}\partial_{\mu}\varphi,\ \ \ \ \delta^{*}\varphi:=\overline{m}^{\mu}\partial_{\mu}\varphi.
\end{equation}
A convenient choice for the space \eqref{eq:NUTBB} is
\begin{align}\label{eq:NPframe}
\bold{k}=k^{\mu}\partial_{\mu}=\frac{1}{V}\left(\partial_{t}+V\partial_{r}\right), \ \ \bold{l}=l^{\mu}\partial_{\mu}=\frac{1}{2}\left(\partial_{t}-V\partial_{r}\right), \\
\bold{m}=m^{\mu}\partial_{\mu}=iL\frac{e^{-i\arctan{(r/n)}}}{\sqrt{2(n^{2}+r^{2})}}\left(\partial_{x}+i\partial_{y}-i\frac{2nx}{L^{2}}\partial_{t}\right).
\end{align}
The vectors $\bold{k}$ and $\bold{l}$ are geodesic and shear-free so that the following spin coefficients vanish
\begin{equation}
\kappa=\sigma=\nu=\lambda=0\, .
\end{equation}
By the Goldberg--Sachs theorem it follows that the space must be of Petrov type $D$, so four out of the five Weyl scalars vanish
\begin{equation}
\Psi_{0}=\Psi_{1}=\Psi_{3}=\Psi_{4}=0.
\end{equation}
In addition, the frame has been chosen to be parallelly  propagated along $\bold{k}$, \textit{i.e.}
\begin{equation}
\nabla_{\bold{k}}\bold{k}=\nabla_{\bold{k}}\bold{l}=\nabla_{\bold{k}}\bold{m}=0,
\end{equation}
a property that implies the vanishing of additional spin coefficients.

As shown by Teukolsky \cite{Teukolsky:1973ha}, the vanishing of these quantities makes perturbation theory tractable on such a background  --- we will make use of those results in next section. For that, we need the spin connection, whose non-vanishing components read

\begin{equation}
\rho=-\Gamma_{142}=\frac{-1}{r+in}, \ \ \mu= \Gamma_{231}=-\frac{V}{2(r+in)},  \ \  \gamma=  \frac{1}{2}\left(\Gamma_{433}-\Gamma_{123}\right)=\frac{1}{4}\left(V'+\frac{2ni}{n^{2}+r^{2}}V\right).
\end{equation}
On the other hand, the Ricci tensor has the same components as the metric $R_{12}=-R_{34}=-6/L^2$, 
and the only non-vanishing Weyl scalar, $\Psi_2$, reads

\begin{equation}
\Psi_{2}=-C_{\mu\nu\rho\sigma}k^{\mu}m^{\nu}l^{\rho}\overline{m}^{\sigma}=-C_{4132}=-\frac{1-3\epsilon i}{2L^{2}}\left(\frac{1+i\epsilon}{(r/r_{+})+i\epsilon}\right)^{3}\, ,
\end{equation}
where $\epsilon=n/r_{+}$. 
This completes the enumeration of non-vanishing NP variables of the space \eqref{eq:NUTBBE}, in the frame \eqref{eq:NPframe}. 
 

\section{Perturbation theory}\label{sec:pert}
In this section we study scalar, electromagnetic and gravitational perturbations around the planar NUT black holes introduced in the previous section. By using the Newman-Penrose formalism, we will show that in all cases the perturbations can be analyzed through a few master variables that satisfy decoupled equations. 
Once the problem is reduced to a decoupled equation for a scalar variable $\Psi$, one can try to separate variables. Now, an important difference with respect to the NUT-neutral case is that the translational Killing vectors $\xi_{(1)}$, $\xi_{(2)}$ do not commute, $[\xi_{(1)},\xi_{(2)}]\neq 0$, and as usual these do not commute with the rotational vector $\xi_{(3)}$. Hence, one cannot fully separate the equations a priori, and at best one can choose the variable $\Psi$ to be an eigenfunction for one of the sets of commuting Killing vectors
\begin{equation}
\left\{\xi_{(t)},\,\xi_{(1)}\right\}\, ,\quad \left\{\xi_{(t)},\,\xi_{(2)}\right\}\, ,\quad  \left\{\xi_{(t)},\,\xi_{(3)}\right\}\, .
\end{equation}
In the coordinates in which \req{eq:NUTBB} is expressed, the vector $\xi_{(2)}=\partial_{y}$ is a coordinate vector and hence it is appropriate to choose the set $\left\{\xi_{(t)},\,\xi_{(2)}\right\}$. Due to the symmetries of the metric, this is completely equivalent to choosing the set $\left\{\xi_{(t)},\,\xi_{(1)}\right\}$.To see this, notice that the transformation $t'=t+\frac{2n}{L^2}xy$, $x'=-y$, $y'=x$ leaves the metric invariant while setting $\xi_{(1)}=\partial_{y}'$. 
On the other hand, the analysis of quasinormal modes using the set $\left\{\xi_{(t)},\,\xi_{(3)}\right\}$ is more obscure, but one expects again that the results would be equivalent. 
From now on we assume that our perturbations are eigenfunctions of $\xi_{(t)}$ and $\xi_{(2)}$, and hence we have

\begin{equation}\label{eq:separation}
 \left . 
      \begin{matrix} 
\xi_{(t)}\Psi&=&-i\omega\Psi\\
\xi_{(2)}\Psi&=&ik\Psi
      \end{matrix} 
   \right \}\quad
   \Rightarrow\quad \Psi=e^{-i\left(\omega t-k y\right)}h(r,x)\, .
\end{equation}
In addition, the dependence on the $x$ and $r$ coordinates can be further separated as we show below. 

\subsection{Scalar perturbations}
Let us consider first the case of a massless scalar field $\phi$ in the background of \eqref{eq:NUTBB} satisfying the wave equation,
\begin{equation}
\nabla^{\mu}\nabla_{\mu}\phi=0\, .
\end{equation}
As we just discussed above, we separate the $t$ and $y$ coordinates according to 

\begin{equation}
\phi=e^{-i\left(\omega t-k y\right)}\frac{\psi(r,x)}{\sqrt{r^2+n^2}}\, ,
\end{equation}
where the factor of $1/\sqrt{r^2+n^2}$ is conventional. Then,  we find that $\psi$ satisfies the following equation:

\begin{align}
-V\frac{\partial}{\partial r}\left(V\frac{\partial \psi}{\partial r}\right)+\psi\left[- \omega ^2+ \frac{V}{r^2+n^2}\left(V'r+\frac{n^2 V}{ \left(r^2+n^2\right)}\right)\right]& \\
+\frac{V L^2}{\left(n^2+r^2\right)} \left[-\frac{\partial^2\psi}{\partial x^2}+\left(k+\frac{2 n x \omega}{L^2} \right)^2\psi \right]&=0\, .
\end{align}
We then note that this equation admits for separable solutions 

\begin{equation}
\psi(r,x)=Y(r)\mathcal{H}(x)\, ,
\end{equation}
where $Y(r)$ and $\mathcal{H}(x)$ satisfy respectively the following equations

\begin{align}\label{eq:radials}
-V\frac{d}{d r}\left(V\frac{dY}{dr}\right)+Y\left[- \omega ^2+ \frac{V}{r^2+n^2}\left(2L^2\mathcal{E}+V'r+\frac{n^2 V}{ \left(r^2+n^2\right)}\right)\right]=&0\, ,\\
-\frac{1}{2}\frac{d^2\mathcal{H}}{dx^2}+\frac{1}{2}\left(k+\frac{2 n x \omega}{L^2} \right)^2\mathcal{H}=&\mathcal{E}\mathcal{H}\, ,
\label{eq:angulars}
\end{align}
where $\mathcal{E}$ is a constant. In the case of vanishing NUT charge, this constant can take any value, as it is related to the wavenumber in the $x$ direction, which can be chosen freely. This situation changes dramatically in the presence of NUT charge. Indeed, we observe a quite remarkable fact: the equation \req{eq:angulars} is identical to that of a quantum harmonic oscillator, where the point of equilibrium is located at $x_0=-kL^2/(2n\omega)$, and where the corresponding mass and frequency are $m=1$, $\omega_{os}^2=(2n\omega/L^2)^2$. There is an even more accurate analogy with Landau quantization that we explore below. 
Now we search for regular solutions such that $\mathcal{H}(x)\rightarrow 0$ at $x\rightarrow\pm\infty$, and this leads to the familiar results for the eigenfunctions and eigenvalues of the harmonic oscillator. There is a catch, though, since we have to take into account that $\omega$ is complex and that $n$ can have either sign. Thus, we must distinguish between the cases $\text{Re}(n\omega)>0$ and $\text{Re}(n\omega)<0$. Introducing 
\begin{equation}
s=\operatorname{sign}\left[\text{Re}(n\omega)\right]\, ,
\end{equation}
we have that the physically relevant solution of \req{eq:angulars} reads

\begin{equation}\label{eq:solang}
\mathcal{H}_{q}(x)=e^{-s\frac{n\omega}{L^2}\left(x+\frac{kL^2}{2n\omega}\right)^2}H_q\left(\sqrt{\frac{sn\omega}{L^2}}\left(x+\frac{kL^2}{2n\omega}\right)\right)\, ,\quad q=0,1,\ldots,
\end{equation}
where $H_q(z)$ are the Hermite's polynomials. The eigenvalues $\mathcal{E}_{q}$ read in turn

\begin{equation}\label{eq:quant}
\mathcal{E}_{q}=\frac{sn\omega}{L^2}(1+2q)\, .
\end{equation}
Thus, unlike the NUT-neutral case, we obtain a quantization condition on the angular part of the perturbations, and hence the spectrum of quasinormal modes will be discrete. Also note that the eigenvalues $\mathcal{E}_{q}$ are independent from the wavenumber in the $y$ direction, $k$, and hence the quasinormal modes will be degenerate. 

Now we can bring this result to the radial equation \req{eq:radials}, and it also proves useful to perform the following redefinitions
\begin{equation}\label{eq:dimensionless}
z=\frac{r_{+}}{r}\, ,\quad \epsilon=\frac{n}{r_{+}}\, ,\quad \hat \omega=\frac{L^2\omega}{r_{+}}\, ,\quad \hat V=\frac{L^2V}{r_{+}^2}\, .
\end{equation}
Then, the radial equation reads

\begin{align}\label{eq:radials}
\hat{V}z^2\frac{d}{dz}\left(\hat{V}z^2\frac{dY}{dz}\right)+Y \left[\hat\omega ^2- \frac{z^2\hat V}{1+z^2\epsilon^2}\left(2s \epsilon\hat\omega(1+2q)-\partial_{z}\hat{V}z+\frac{\epsilon^2z^2 \hat{V}}{ \left(1+\epsilon^2\right)}\right)\right]=&0\, 
\end{align}
Notice that the only free parameters in this equation are $\epsilon$, the dimensionless frequency $\hat\omega$ and the index $q$. 

\subsubsection*{Relation to Landau quantization}
Interestingly, the perturbations in the Taub-NUT backgrounds organize in an analogous way to the Landau levels of a charged particle moving in a uniform magnetic field.  In order to establish this analogy, let us first note that we can write the metric \req{eq:NUTBB} in a gauge-invariant form as 
\begin{equation}\label{eq:NUTBB2}
ds^2=-V(r)\left(dt+A\right)^2+\frac{dr^2}{V(r)}+\frac{r^2+n^2}{L^2}\left(dx^2+dy^2\right)\, ,
\end{equation}
where $A$ is a 1-form satisfying $dA=\frac{2n}{L^2}dx\wedge dy$. Thus, coordinate transformations of the form $t\rightarrow t+f(x,y)$ can be reabsorbed as gauge transformations $A\rightarrow A-df$. Then, one can in fact interpret this $A$ as a uniform magnetic field with magnitude $B=2n/L^2$.  Let us then consider a particle of charge $e$ moving in the $(x,y)$ plane (in flat space) in the background of this field. The Hamiltonian is given by 

\begin{equation}
\mathbf{H}=\frac{1}{2}\pi_{x}^2+\frac{1}{2}\pi_{y}^2\, ,
\end{equation}
where, in the gauge $A=B x dy$, the momenta read

\begin{equation}
\pi_{x}=-i\partial_{x}\, ,\quad \pi_{y}=-i\partial_{y}-e x B\, .
\end{equation}
Then, the Sch\"ordinger equation $\mathbf{H}\psi=E\psi$ yields

\begin{equation}
\left[-\frac{1}{2}\partial_{x}^{2}+\frac{1}{2}\left(-i\partial_{y}-e x B\right)^2\right]\psi=E\psi\, ,
\end{equation}
and by using $-i\partial_{y}\psi=k\psi$  we get the same equation \req{eq:angulars} we got for the angular part of the perturbations in the Taub-NUT geometry, provided one identifies the charge of the particle with the frequency of the perturbation as $e=-\omega$.
Thus, the transverse $(x,y)$ part of the quasinormal modes of the Taub-NUT background are eigenfunctions of this Hamiltonian and have the same quantization, which is given by the Landau levels $q=0,1,\ldots$. As we show next, electromagnetic and gravitational perturbations organize in a similiar fashion. 
Clearly, this analogy can be traced back to the fact that NUT charge is the gravitational equivalent of magnetic charge. 

\subsection{Electromagnetic and Gravitational perturbations}

Let us now address the study of perturbations electromagnetic and gravitational perturbations. Thus, we consider a vector field $A_{\mu}$ satisfying Maxwell equations in the background of \req{eq:NUTBB}
\begin{equation}
\nabla_{\mu}F^{\mu\nu}=0\, ,\qquad F_{\mu\nu}=2\partial_{[\mu}A_{\nu]}\, ,
\end{equation}
and a metric perturbation $\tilde g_{\mu\nu}=g_{\mu\nu}+h_{\mu\nu}$ satisfying the linearized Einstein's equations

\begin{equation}
G_{\mu\nu}^{L}\left[h_{\alpha\beta}\right]-\frac{3}{L^2}h_{\mu\nu}=0\, .
\end{equation}

\noindent
While the symmetries of the \req{eq:NUTBB} may still allow one to perform a complete decomposition of $A_{\mu}$ and $h_{\mu\nu}$ --- see 
\cite{Young:1983dn,Warnick:2006ih,Holzegel:2006gn} for the case of $\mathrm{SU(2)}$ symmetry and \cite{Krtous:2018bvk} for electromagnetic perturbations in the Kerr-NUT-(A)dS spacetime --- we find that the Newman-Penrose formalism offers a possibly clearer way to compute perturbations. 

In the NP frame, the field strength of the Maxwell field is described by three independent (complex) components that are customarily denoted as,
\begin{equation}
\phi_{0}=F_{\mu\nu}k^{\mu}m^{\nu}=F_{41},\ \ \ \phi_{1}=\frac{1}{2}F_{\mu\nu}\left(k^{\mu}l^{\nu}+\overline{m}^{\mu}m^{\nu}\right)=\frac{1}{2}(F_{43}+F_{21}),\ \ \ \phi_{2}=F_{\mu\nu}\overline{m}^{\mu}l^{\nu}=F_{23}.
\end{equation}
Since the background considered here is neutral, the $\phi_i$ correspond to linear electromagnetic perturbations. 
In addition, since the metric \req{eq:NUTBB} is of Petrov type D and our NP frame \req{eq:NPframe} has $\bold{k}$ and $\bold{l}$ aligned with the repeated principal null directions, we can apply directly the results from Teukolsky \cite{Teukolsky:1973ha}. These imply that $\phi_0$ and $\phi_2$ satisfy two decoupled equations that read
\begin{align}\label{eq:Maxwell}
\begin{aligned}
\left[\left(D-2\rho-\rho^{*}\right)\left(\Delta +\mu-2\gamma\right)-\delta\delta^{*}\right]\phi_{0}=&0\, ,\\
\left[\left(\Delta+\gamma-\gamma^{*}+2\mu+\mu^{*}\right)\left(D-\rho\right)-\delta^{*}\delta\right]\phi_{2}=&0
\end{aligned}
\end{align}

On the other hand, gravitational perturbations are described by small changes in the NP frame, $\bold{e}_{a}\rightarrow \bold{e}_{a}+\bold{e}_{a}^{(1)}+...$, which, in turn, induce changes in the NP variables introduced above, \textit{e.g.} $\Psi_{a}\rightarrow\Psi_{a}+\Psi_{a}^{(1)}+...$ for the Weyl scalars $\Psi_{a}$. Since $\Psi_{0}$, $\Psi_{1}$, $\Psi_{3}$ and $\Psi_{4}$ all have a vanishing background value, it follows that they are already linearized. In particular the scalars $\Psi_{0}$ and $\Psi_{4}$, which are defined as 

\begin{align}\label{def:psi0}
\Psi_{0}=&C_{\mu\nu\rho\sigma}k^{\mu}m^{\nu}k^{\rho}m^{\sigma}\, ,\\\label{def:psi4}
\Psi_{4}=&C_{\mu\nu\rho\sigma}l^{\mu}\overline{m}^{\nu}l^{\rho}\overline{m}^{\sigma}\, ,
\end{align}
satisfy two decoupled equations, 

\begin{equation}\label{eq:Grav}
\begin{aligned}
\left[(D-4\rho-\rho^{*})(\Delta -4\gamma+\mu)-\delta\delta^{*}-3\Psi_{2}\right]\Psi_{0}=&0\, ,\\
\left[\left(\Delta+3\gamma-\gamma^{*}+4\mu+\mu^{*}\right)\left(D-\rho \right)-\delta^{*}\delta -3\Psi_{2}\right]\Psi_{4}=&0
\end{aligned}
\end{equation}

We now search for separable solutions of the variables $(\phi_0, \phi_2)$ and $(\Psi_0, \Psi_4)$. For any of these --- call it $\psi$ --- we can separate the dependence on $t$ and $y$ as in \req{eq:separation}, so that we write $\psi=e^{-i(\omega t-k y)}\mathcal{H}(x)R(r)$. On the other hand, the dependence on the coordinate $x$ in \req{eq:Maxwell} and \req{eq:Grav} only appears in the operator $\delta\delta^{*}$, which reads

\begin{equation}
\delta\delta^{*}=\frac{L^2}{2(n^2+r^2)}\left[\partial_{x}^2+\left(\partial_{y}-\frac{2nx}{L^2}\partial_{t}\right)^2+i\frac{2n}{L^2}\partial_{t}\right]\, .
\end{equation}

\noindent
Thus, demanding that $\delta\delta^{*}\psi=\lambda(r)\psi$ leads to the same equation for $\mathcal{H}$ as in the scalar case, given by \req{eq:angulars}. Likewise, by imposing regularity of $\mathcal{H}$ at infinity we obtain the Hermite functions \req{eq:solang} and therefore we get
\begin{equation}
\delta\delta^{*}\psi=-\frac{L^2}{n^2+r^2}\left(\mathcal{E}_{q}-\frac{n\omega}{L^2}\right)\psi\, ,\quad \delta^{*}\delta\psi=-\frac{L^2}{n^2+r^2}\left(\mathcal{E}_{q}+\frac{n\omega}{L^2}\right)\psi\, ,
\end{equation}
where the eigenvalues $\mathcal{E}_{q}$ are those in \req{eq:quant}.  Finally, it is possible to write \eqref{eq:Maxwell} and \eqref{eq:Grav} in a symmetric form \cite{Chandrasekhar:1985kt} by introducing the radial functions $Y_{\pm 1}$ and $Y_{\pm 2}$ as, 

\begin{equation}\label{Ypm1def}
\phi_{0}=\frac{e^{-2i\arctan{(r/n)}}}{V\sqrt{n^2+r^2}}e^{-i\omega t+i k y}\mathcal{H}_{q}(x)Y_{+1}(r),\ \ \ \phi_{2}=\frac{1}{\sqrt{n^{2}+r^{2}}}e^{-i\omega t+i k y}\mathcal{H}_{q}(x)Y_{-1}(r)\, ,
\end{equation}
and
\begin{equation}\label{Ypm2def}
\Psi_{0}=\frac{e^{-4i\arctan{(r/n)}}}{V^{2}\sqrt{n^2+r^2}}e^{-i\omega t+i k y}\mathcal{H}_{q}(x)Y_{+2}(r),\ \ \ \Psi_{4}=\frac{1}{\sqrt{n^{2}+r^{2}}}e^{-i\omega t+i k y}\mathcal{H}_{q}(x)Y_{-2}(r)\, .
\end{equation}

\noindent
Then Eqs.~\eqref{eq:Maxwell} and \eqref{eq:Grav} yield the following master equations for the radial variables
\begin{equation}\label{eq:mastereqs}
\Lambda^{2} Y_{\pm S}(r)+S P(r)\Lambda_{\pm}Y_{\pm S}(r)-\left(\frac{2L^2\mathcal{E}_{q}}{r^2+n^2}+Q_{S}(r)\right)V(r)Y_{\pm S}(r)=0\, ,
\end{equation}
where $S=1$ for electromagnetic perturbations and $S=2$ for gravitational ones. Here we have introduced the differential operators

\begin{equation}
 \Lambda_{\pm}=\frac{d}{dr_{*}}\pm i\omega,\ \ \ \ \ \Lambda^{2}=\frac{d^{2}}{dr_{*}^{2}}+\omega^{2}\, ,\quad \text{where}\quad \frac{d}{dr^{*}}=V\frac{d}{dr}\, ,
\end{equation}
and $P(r)$ and $Q_{S}(r)$ are functions given by

\begin{equation}
P=-V'+\frac{2 (r-2 i n) V}{n^2+r^2}\, ,
\end{equation}
and 
\begin{equation}
Q_{S}=\begin{cases}
\displaystyle\frac{3 n^2 V}{\left(n^2+r^2\right)^2}\ \ \ &(S=1)\\
&\\
\displaystyle\frac{\left(12 n^2+8 i n r-r^2\right) V}{\left(n^2+r^2\right)^2}-\frac{(4 i n-r) V'}{n^2+r^2}-\frac{V''}{2} \ \ \ &(S=2)
\end{cases}
\end{equation}

\noindent
These equations can be written in a dimensionless way by introducing $z$,  $\epsilon$ and $\hat\omega$ as in \req{eq:dimensionless}, which implies that the dimensionless QNM frequencies $\hat\omega$ will only depend on $\epsilon$ and the level $q$. 
In order to obtain these frequencies, the radial equations \eqref{eq:mastereqs} must be supplemented with suitable boundary conditions, which we determine in the following section.

\section{Boundary conditions}\label{sec:bdry}
Once we have determined the master equations governing the perturbations of scalar, electromagnetic and gravitational fields, we are interested in studying the corresponding quasinormal modes, which are determined by a specific choice of boundary conditions. 
At the horizon of the black holes, these modes satisfy the condition of behaving as outgoing waves, while the conditions at the boundary of AdS can be chosen in different ways. For instance, one might impose the master variables to vanish at infinity \cite{Cardoso:2001vs,Miranda:2005qx}.
 However, we are interested in making contact with the AdS/CFT correspondence, and in that case the boundary conditions are uniquely determined \cite{Dias:2009ex,Dias:2013sdc,Cardoso:2013pza}. According to AdS/CFT, the different operators of the boundary theory couple to the normalizable mode of the perturbations in the bulk, and therefore we must make sure that only those modes are excited if we want to interpret the corresponding quasinormal modes as perturbations of a plasma in the dual theory. 

\noindent
In order to study the boundary conditions it is useful to introduce first the coordinate
\begin{equation}
z=\frac{r_{+}}{r}\, ,
\end{equation}
so that the metric can be written as

\begin{equation}
ds^2=\frac{1}{z^2}\left[-\frac{r_{+}^2 f(z)}{L^2}\left(dt+\frac{2n}{L^2}xdy\right)^2+\frac{dz^2 L^2}{f(z)}+\frac{r_{+}^2+z^2n^2}{L^2}\left(dx^2+dy^2\right)\right]
\end{equation}

\begin{equation}
f(z)=\frac{L^2}{r_{+}^2}z^2V(r_{+}/z)\, .
\end{equation}
In this way, infinity corresponds to $z=0$, while the horizon is placed at $z=1$. 
On the other hand, the tortoise coordinate $r_*$ is defined by 
\begin{equation}
r_*=-\int \frac{dz L^2}{r_{+}f(z)}\, ,
\end{equation}
and we note that near the horizon $z=1$ it reads

\begin{equation}\label{eq:tortnh}
r_*\approx \frac{1}{4\pi T} \log(1-z)\, ,
\end{equation}
where $T$ is the Hawking temperature \req{eq:Temp}.

\subsection{Scalar field}
In the near-horizon region $z=1$, the solution to the radial scalar equation \req{eq:radials} can be expanded in a Frobenius series
\begin{equation}
Y(z)=(1-z)^{\alpha}\left[c_0++c_1(1-z)+c_2(1-z)^2+\ldots\right]\, .
\end{equation}
The indicial equation has the following two solutions for $\alpha$,

\begin{equation}
\alpha_{\pm}=\pm\frac{i\hat\omega}{3(1+\epsilon^2)}\, ,
\end{equation}
and taking into account \req{eq:tortnh} and that $\hat\omega=L^2\omega/r_{+}$, we get that the solution behaves as $Y\sim e^{4\pi T\alpha_{\pm} r_{*}}=e^{\pm i\omega r_{*}}$. Since the solution must behave as an outgoing wave at the horizon, we must choose the root $\alpha_{-}$. 

On the other hand, near the AdS boundary $z=0$ we find that there are two independent modes:
\begin{equation}
Y(z)=a z^2+b z^{-1}\left(1+\mathcal{O}(z)\right)\quad \text{when} \quad z\rightarrow 0\, .
\end{equation}
We keep  the normalizable mode, which is the one that couples to a scalar field in the dual theory, and hence we have to impose that $Y(0)=0$. The conditions at infinity and at the horizon can only be satisfied simultaneously by a discrete set of complex frequencies $\omega$: the quasinormal mode frequencies. 

\subsection{Electromagnetic field}\label{subsec:bdryA}
In the case of the electromagnetic field, the analysis of the boundary conditions in the near-horizon are analogous to the scalar case. Again one finds that the NP variables $\phi_0$, $\phi_2$ can be expanded in a Frobenius series near $z=1$, and imposing the condition of outgoing waves one finds the following solutions for the radial functions $Y_{\pm1}$:
\begin{equation}\label{eq:Y1nh}
Y_{\pm1}\sim(1-z)^{\alpha_{\pm1}} \quad\text{when}\quad z\rightarrow 1, 
\end{equation}
where 
\begin{equation}
\alpha_{+1}=-\frac{i\hat\omega}{3(1+\epsilon^2)}\, , \quad \alpha_{-1}=1-\frac{i\hat\omega}{3(1+\epsilon^2)}\, .
\end{equation}

\noindent
The analysis of boundary conditions at infinity, on the other hand, is much involved than in the case of a scalar field. By analyzing the solutions of the radial equations \req{eq:mastereqs} for $Y_{\pm1}$, we see that the two independent solutions behave near $z=0$ as 
\begin{equation}\label{eq:Y1as}
Y_{\pm 1}(z)=a_{\pm1}+b_{\pm1} z \quad \text{when} \quad z\rightarrow 0\, ,
\end{equation}
where $a_{\pm1}$ and $b_{\pm1}$ are constants. Now, the boundary conditions are not imposed directly on the NP variables but on the perturbation of the Maxwell field $A_{\mu}$, so we must study how these relate. 
Let us for into account that we can always choose a gauge in which the $z$-component the vector vanishes $A_{z}=0$.  Then, the solutions to Maxwell equations near $z=0$ behave as $A_a \sim A^{(1)}_{a}+zA^{(2)}_{a}$, where $a$ denotes the boundary indices $a=t,x,y$. Therefore, Dirichlet boundary conditions imply that $A^{(1)}_{a}=0$, and we only keep the mode that decays at infinity. Separating variables, this means that we can write the vector asymptotically as

\begin{equation}\label{eq:vecpert}
A_{a}=ze^{-i(\omega t-k y)}\gamma_{a}(x)+\mathcal{O}(z^3)\, ,
\end{equation}
where $\gamma_{a}$ are certain functions and one can check that the following term in the $z$-expansion is indeed $\mathcal{O}(z^3)$. Now, the functions $\gamma_{a}$ are not arbitrary, but we find that Maxwell equations impose the following constraint, 

\begin{equation}\label{eq:gammaMax}
 \left(\omega -\frac{2 n x}{L^2} \left(k +\frac{2 n x \omega}{L^2} \right)\right)\gamma_{t}-i \gamma_{x}'+\left(k +\frac{2 n x \omega}{L^2} \right)\gamma_{y}=0\, .
\end{equation}
On the other hand, we are searching for solutions such that the NP variables $\phi_0$ and $\phi_2$ are separated, and this will impose, too, conditions on the $\gamma_a$. Computing $\phi_0$ and $\phi_2$ from the vector perturbation \req{eq:vecpert} we find that

\begin{align}
\hat\phi_0&=e^{-i(\omega t-k y)}\left[A_{+1}+B_{+1}z+\mathcal{O}(z^2)\right]\, ,\\
\hat\phi_2&=e^{-i(\omega t-k y)}\left[A_{-1}+B_{-1} z+\mathcal{O}(z^2)\right]\, ,
\end{align}
where $\hat\phi_{0,2}$ are defined as

\begin{equation}
\hat{\phi}_0=V\sqrt{n^2+r^2}e^{2i\arctan{(r/n)}}\phi_{0}\, ,\qquad  \hat{\phi}_2=\sqrt{n^{2}+r^{2}}\phi_2\, , 
\end{equation}
and the coefficients $A_{\pm1}$, $B_{\pm1}$ read

\begin{align}
A_{\pm 1}=&\frac{2^{\pm1/2}}{4L}\left(-\frac{2 i n x \gamma _t}{L^2}\pm \gamma _x+i\gamma _y\right)\, ,\\
B_{\pm 1}=&\frac{2^{\pm1/2}}{4  L r_+}\left[L^2 \gamma
   _t'+\left(\mp k -\frac{2 n^2 x}{L^4}\right) L^2\gamma _t+i \left(\mp n+L^2 \omega \right) \gamma _x+\left(n\mp L^2 \omega \right) \gamma _y\right]\, .
\end{align}
Now, on the other hand, if both $\phi_0$ and $\phi_2$ can be separated, then the result should read

\begin{align}
\hat\phi_0&=e^{-i(\omega t-k y)}\mathcal{H}_{q_{+1}}(x)\left[a_{+1}+b_{+1}z+\mathcal{O}(z^2)\right]\, ,\\
\hat\phi_2&=e^{-i(\omega t-k y)}\mathcal{H}_{q_{-1}}(x)\left[a_{-1}+b_{-1} z+\mathcal{O}(z^2)\right]\, ,
\end{align}
where we have taken into account \req{eq:Y1as} and where $\mathcal{H}_{q_{\pm 1}}(x)$ are the eigenfunctions in \req{eq:solang}, with two possibly different levels $q_{+1}$ and $q_{-1}$ for each of the variables. Thus, we obtain a system of four equations for the variables $\gamma_{a}$ and the four constants $a_{\pm1}$, $b_{\pm1}$, 

\begin{equation}\label{eq:ABvec}
A_{\pm1}=a_{\pm 1}\mathcal{H}_{q_{\pm 1}}\, ,\quad B_{\pm1}=b_{\pm 1}\mathcal{H}_{q_{\pm 1}}\, .
\end{equation}
Together with \req{eq:gammaMax}, we have to solve a system of five equations which is not guaranteed to have solutions. In order to simplify the computations, at this point it is interesting to note that we can set $k=0$ without loss of generality.  In fact, the change of variables 
\begin{align}\label{eq:isom}
\hat x=&x-\sigma\, ,\qquad
\hat t=t+\frac{2n}{L^2}\sigma y
\end{align}
leaves invariant the background metric and therefore is a symmetry of the linearized equations.  On the other hand it transforms the perturbation $A_a$ as follows

\begin{equation}
\hat{A}_{a}=z e^{-i\left(\omega \hat{t}- \hat{k} y\right)}\hat{\gamma}_{a}(\hat x)\, ,\quad\text{where}\quad \hat k=k+\frac{2n\omega\sigma}{L^2}\, ,
\end{equation}
and 

\begin{align}
\hat{\gamma}_{\hat{t}}&=\gamma_{t}\, ,\quad \hat{\gamma}_{\hat{x}}=\gamma_{x}\, ,\quad 
\hat{\gamma}_{y}=\gamma_{y}-\frac{2n\sigma}{L^2}\gamma_{t}\, .
\end{align}
Therefore, by choosing $\sigma=-k L^2/(2n\omega)$ we get $\hat k=0$. Equivalently, we can always work with the solution with $k=0$ and generate another solution with $k\neq 0$ by applying the isometric transformation \req{eq:isom}. Thus, from now on we set $k=0$. 

One can see that from the five equations in \req{eq:gammaMax} and \req{eq:ABvec} it is possible to obtain explicitly the values of $\gamma_{t}$, $\gamma_{t}'$, $\gamma_{x}$, $\gamma_{x}'$ and $\gamma_{y}$, but of course, in order for this to be an actual solution, $\gamma_{t}'$ and $\gamma_{x}'$ should in fact be the derivatives of $\gamma_{t}$ and $\gamma_{x}$. As it turns out, this only happens when the following constraints meet. First, the two levels $q_{+1}$ and $q_{-1}$ must be related according to 

\begin{equation}\label{eq:levelvector}
q_{+1}=q_{-1}+2s\, ,
\end{equation}
where we recall that $s=\operatorname{sign}\left[\text{Re}(n\omega)\right]$. Thus we have $q_{-1}=0,1,2,...$ for $s=1$ and $q_{-1}=2,3,4,...$ for $s=-1$. On the other hand, the ratios of the constants $a_{\pm}$, $b_{\pm}$,

\begin{equation}
\lambda_{\pm1}=\frac{b_{\pm1}}{a_{\pm1}}
\end{equation}
must be related according to 

\begin{equation}\label{eq:lambda1}
\lambda_{-1}=\frac{\lambda_{+1}\left(2 q \epsilon -\hat{\omega }+\epsilon \right)-i \left(2 (2 q+3) \hat{\omega }
   \epsilon -\hat{\omega }^2+3 \epsilon ^2\right)}{-i \lambda_{+1} +(2 q+5) \epsilon -\hat{\omega }}\, .
\end{equation}
where $q$ is 
\begin{equation}
q=\begin{cases}
q_{-1} \quad&\text{if}\quad s=1\, ,\\
-1-q_{-1} \quad&\text{if}\quad s=-1\,
\end{cases}
\end{equation}

\noindent
Note that this is all we need in order to characterize the boundary conditions, since the overall normalization of $Y_{\pm1}$ is not relevant when searching for quasinormal modes. 
Now, consistency of the system of equations requires an additional constraint that involves such overall normalization, 

\begin{equation}\label{eq:a1a-1}
2\epsilon\frac{a_{-1}}{a_{+1}}=
\begin{cases}
i \lambda_{+1} -(2 q_{-1}+5) \epsilon +\hat{\omega }\quad&\text{if}\quad s=1\, ,\\
-\frac{\left(i \lambda_{+1}+(2 q_{-1}-3) \epsilon +\hat{\omega }\right)}{4 (q_{-1}-1) q_{-1}}\quad&\text{if}\quad s=-1\, .
\end{cases}
\end{equation}
In that case, the explicit solution for the $\gamma_{a}$ reads

\begin{align}
\gamma_{t}&=-\frac{i a_{+1} L^3 \left(-i \lambda_{+1} -\hat{\omega }+\epsilon \right)}{\sqrt{2} r_{+} x \hat{\omega } \epsilon }\left[2(1+q_{-1})\mathcal{H}_{q_-}+\mathcal{H}_{2+q_-}\right]\, ,\\
\gamma_{x}&=\frac{\sqrt{2}a_{+1}L}{\epsilon}\left[\left(-i \lambda_{+1} +\epsilon(2 q_{-1}+5)  -\hat{\omega }\right)\mathcal{H}_{q_-}+\epsilon \mathcal{H}_{2+q_-}\right]\, ,\\\notag
\gamma_{y}&=-\frac{i \sqrt{2} a_{+1} L}{\hat{\omega } \epsilon }\Big[\left(2 (q_{-1}+1) \epsilon ^2+\lambda_{+1}  \left(i \hat{\omega }-2 i (q_{-1}+1) \epsilon \right)-(4 q_{-1}+7) \hat{\omega } \epsilon +\hat{\omega }^2\right)\mathcal{H}_{q_-}\\
&+\epsilon  (\epsilon -i \lambda_{+1} )\mathcal{H}_{2+q_-}\Big]\, ,
\end{align}
for $s=1$, and there is a similar solution for $s=-1$. 

Then, in order to find the electromagnetic quasinormal modes, the idea would be to simultaneously solve the radial equations \req{eq:mastereqs} for $Y_{+1}$ and $Y_{-1}$ with the levels $q_{\pm1}$ related according to \req{eq:levelvector} and with the boundary conditions given by \req{eq:Y1nh}, \req{eq:Y1as} and \req{eq:lambda1}. Note that, once $\epsilon$ and $q_{-1}$ are specified, the problem only contains two parameters, $\hat\omega$ and $\lambda_{+}$, and the hope is a solution exists only for discrete values of these quantities. Unfortunately, this is not the case, since the equations for  $Y_{+1}$ and $Y_{-1}$ are degenerate. In order to see this, we first note the following Maxwell equations in the NP formalism
\begin{equation}
(D-2\rho)\phi_1=\delta^{*}\phi_0\, ,\quad (D-\rho)\phi_2=\delta^{*}\phi_1\, .
\end{equation}
Combining these it is possible to derive the following relation between $\phi_0$ and $\phi_2$,

\begin{equation}
\hat{\delta}^{*}\hat{\delta}^{*}\phi_{0}=R(D-\rho)R(D-2\rho)\phi_2\, ,
\end{equation}
where 

\begin{equation}
R=\frac{i\sqrt{2(r^2+n^2)}}{L}e^{-i \arctan(r/n)}\, ,\quad \hat\delta^{*}=R\delta^{*}=\partial_{x}-i\partial_{y}+i\frac{2nx}{L^2}\partial_{t}\, .
\end{equation}
Then, by using the decomposition \req{Ypm1def} one first derives the relation between the levels $q_{\pm1}$ given in \req{eq:levelvector}\footnote{Interestingly, the operators $\hat{\delta}^{*}$ and $\hat{\delta}$ act as the ladder operators of the harmonic oscillator, so they raise and lower the Landau level $q$.}, and one also obtains a relation between $Y_{+1}$ and $Y_{-1}$, 

\begin{align}\notag
Y_{+1}=&-\frac{Y_{-1}\left((2 q_{-1}+1) \epsilon  \left(3 z^4 \epsilon ^4-6 z^2 \epsilon ^2+z^3 \left(-3 \epsilon ^4+6 \epsilon ^2+1\right)-1\right)+\hat{\omega} \left(z^2 \epsilon ^2+1\right)^2\right)}{2 (q_{-1}+1) (q_{-1}+2) (z-1) \epsilon  \left(3 z^3 \epsilon ^4+z^2 \left(6 \epsilon ^2+1\right)+z+1\right)}\\
&-\frac{i \left(z^2 \epsilon ^2+1\right) Y_{-1}'}{2 (q_{-1}+1) (q_{-1}+2) \epsilon }\, ,\\\notag
Y_{-1}=&-\frac{Y_{+1}\left((2 q_{-1}+5) \epsilon  \left(3 z^4 \epsilon ^4-6 z^2 \epsilon ^2+z^3 \left(-3 \epsilon ^4+6 \epsilon ^2+1\right)-1\right)+\hat\omega \left(z^2 \epsilon ^2+1\right)^2\right)}{2 (z-1) \epsilon  \left(3 z^3 \epsilon ^4+z^2 \left(6 \epsilon ^2+1\right)+z+1\right)}\\
&+\frac{i \left(z^2 \epsilon ^2+1\right)Y_{+1}'}{2 \epsilon }\, ,
\end{align}
where we have used the master equations \req{eq:mastereqs}. One can see that these relations map the solutions of $Y_{\pm1}$ with the boundary conditions \req{eq:Y1nh} into each other and they imply that the asymptotic behaviour of these functions is always related according to \req{eq:lambda1} --- independenly of the boundary conditions imposed on the vector $A_{\mu}$. Therefore, both equations are degenerate and the value of $\lambda_{+1}$ (or $\lambda_{-1}$) cannot be found in this way. In the case of vanishing NUT charge, one can decouple the electromagnetic perturbations in modes of definite parity, which are achieved only for two specific values of $\lambda_{+1}$ ($\lambda_{-1}$). However, NUT charge breaks all reflection symmetries of the background, and therefore we do not have a similar decomposition of the perturbations. Hence, we seem to be unable to  determine the polarization parameter $\lambda_{\pm1}$, which would suggest that the spectrum of QNMs depends continuously on this parameter. Clearly, more research in this direction is needed in order to understand the puzzling properties of electromagnetic perturbations in these geometries. By now, we will focus on the gravitational case, for which a similar method does work.

\subsection{Gravitational field}
Let us finally turn to the case of the boundary conditions for gravitational perturbations. In the near-horizon region we find that the outgoing-wave condition leads to the following form of the radial functions $Y_{\pm2}$, 
\begin{equation}\label{eq:Y2nh}
Y_{\pm2}\sim(1-z)^{\alpha_{\pm2}} \quad\text{when}\quad z\rightarrow 1, 
\end{equation}
where 
\begin{equation}
\alpha_{+2}=-\frac{i\hat\omega}{3(1+\epsilon^2)}\, , \quad \alpha_{-2}=2-\frac{i\hat\omega}{3(1+\epsilon^2)}\, .
\end{equation}

On the other hand, the discussion on boundary conditions at infinity proceeds analogously to the electromagnetic case. First, by analyzing the solutions of the Newman-Penrose variables $Y_{\pm 2}$, one can see that near the boundary they behave as
\begin{equation}\label{eq:Y2as}
Y_{\pm 2}(z)=a_{\pm2}+b_{\pm2} z \quad \text{when} \quad z\rightarrow 0\, .
\end{equation}
The integration constants  $a_{\pm2}$ and $b_{\pm2}$ will be then ultimately related to the boundary conditions imposed on the metric perturbation. Let us consider a metric perturbation $g_{\mu\nu}\rightarrow g_{\mu\nu}+h_{\mu\nu}$ in the geometry of these NUT black branes. Due to gauge freedom, we can always choose a gauge in which $h_{\mu z}=0$, so that the non-vanishing components are those transverse to the $z$ direction, $h_{ab}$. Then, near $z=0$, the metric perturbation $h_{ab}$ has two modes,

\begin{equation}
h_{ab}=zh^{(1)}_{ab}+z^{-2}\left(h^{(2)}_{ab}+\mathcal{O}(z)\right)\quad \text{when} \quad z\rightarrow 0\, .
\end{equation}
The holographic dictionary tells us that the renormalizable mode is the one coupled to the dual stress-energy tensor, $T^{ab}$, and therefore we set $h^{(2)}_{ab}=0$. Now we can use the fact that $\partial_{t}$ and $\partial_{y}$ are Killing vectors in order to separate variables, so that we have

\begin{equation}\label{eq:hab}
h_{ab}=z e^{-i\left(\omega t- k y\right)}\gamma_{ab}(x)+\mathcal{O}(z^3) \quad \text{when} \quad z\rightarrow 0\, .
\end{equation}
However, just like in the case of electromagnetic perturbations, we can always set $k=0$ by performing the isometric transformation \req{eq:isom}. For the sake of completeness let us point out that the transformed metric perturbation reads
\begin{equation}
\hat{h}_{ab}=z e^{-i\left(\omega \hat{t}- \hat{k} y\right)}\hat{\gamma}_{ab}(\hat x)\, ,\quad \hat k=k+\frac{2n\omega\sigma}{L^2}
\end{equation}
where

\begin{align}
\hat{\gamma}_{\hat{t}\hat{t}}&=\gamma_{tt}\, ,\quad \hat{\gamma}_{\hat{t}\hat{x}}=\gamma_{tx}\, ,\quad \hat{\gamma}_{\hat{x}\hat{x}}=\gamma_{xx}\\
\hat{\gamma}_{\hat{t}y}&=\gamma_{ty}-\frac{2n\sigma}{L^2}\gamma_{tt}\, ,\quad \hat{\gamma}_{\hat{x}y}=\gamma_{xy}-\frac{2n\sigma}{L^2}\gamma_{tx}\, , 
\quad \hat{\gamma}_{yy}=\gamma_{yy}-\frac{4n\sigma}{L^2}\gamma_{ty}+\left(\frac{2n\sigma}{L^2}\right)^{2}\gamma_{tt}\, .
\end{align}
so that, by choosing $\sigma=-k L^2/(2n\omega)$ we get $\hat k=0$. Thus, let us set $k=0$ from now on. 

Next, we have to determine the equations satisfied by the ``polarization matrix'' $\gamma_{ab}$. By expanding the linearized Einstein equations around $z=0$, we find that the components of this matrix satisfy four equations, corresponding to to $G_{\mu z}+3/L^2 g_{\mu z}=0$. These yield
\begin{equation}\label{eq:gammaabeq}
\begin{aligned}
\frac{2n\omega}{L^2} x \gamma_{tx}+i \gamma_{tx}'-\omega (\gamma_{xx}+\gamma_{yy})&=0\, ,\\
\left(1-\frac{4n^2x^2}{L^4}\right)\gamma_{tt}+\frac{4nx}{L^2}\gamma_{ty}-\gamma_{xx}-\gamma_{yy}&=0\, ,\\
\frac{4n^2x}{L^4}\gamma_{tt}-\left(1-\frac{4n^2x^2}{L^4}\right)\gamma_{tt}'-i\omega\left(1-\frac{4n^2x^2}{L^4}\right)\gamma_{tx}&\\
-\frac{2n}{L^2}\gamma_{ty}-\frac{4nx}{L^2}\gamma_{ty}'-\frac{2in\omega x}{L^2}\gamma_{xy}+\gamma_{yy}'&=0\, ,\\
\omega \left(1-\frac{4n^2x^2}{L^4}\right)\gamma_{ty}-i \gamma{xy}'+\frac{2nx\omega}{L^2}\gamma_{yy}&=0\, ,
\end{aligned}
\end{equation}
where a prime denotes a derivative with respect to $x$. Let us now leave these equations for a moment to consider the NP variables $\Psi_{0}$ and $\Psi_{4}$. These can be computed from the metric perturbation $h_{\mu\nu}$ according to their definition in \req{def:psi0} and \req{def:psi4}. In doing this, one has to be careful to take into account not only the variation of the Weyl tensor, but also the variation in the frame, \textit{i.e},
\begin{equation}
\Psi_{0}=\delta C_{\mu\nu\rho\sigma}k^{\mu}m^{\nu}k^{\rho}m^{\sigma}+C_{\mu\nu\rho\sigma}\delta(k^{\mu}m^{\nu}k^{\rho}m^{\sigma})\, .
\end{equation}
However, since the only non-vanishing Weyl scalar in the background is $\Psi_{2}=-C_{\mu\nu\rho\sigma}k^{\mu}m^{\nu}l^{\rho}\overline{m}^{\sigma}$, which is obtained from the contraction with the four different frame vectors, it is clear that $C_{\mu\nu\rho\sigma}\delta(k^{\mu}m^{\nu}k^{\rho}m^{\sigma})=0$, because in this expression the Weyl tensor is always contracted twice either with $k$ or with $m$, and therefore no combination involving $\Psi_2$ appears. A similar argument holds for $\Psi_{4}$, and hence it is enough to keep only the variation of Weyl curvature when computing these scalars in the perturbed metric.
Using \req{eq:hab} and expanding near $z=0$ we find that 

\begin{align}
\hat{\Psi}_0&=e^{-i\omega t}\left[A_{+2}+B_{+2}z+\mathcal{O}(z^2)\right]\, ,\\
\hat{\Psi}_4&=e^{-i\omega t}\left[A_{-2}+B_{-2} z+\mathcal{O}(z^2)\right]\, ,
\end{align}
where $\hat{\Psi}_0$ and $\hat{\Psi}_4$ are the rescaled variables  
\begin{equation}
\hat{\Psi}_0=V^{2}\sqrt{n^2+r^2}e^{+4i\arctan{(r/n)}}\Psi_{0}\, ,\qquad  \hat{\Psi}_4=\sqrt{n^{2}+r^{2}}\Psi_4\, , 
\end{equation}
and the coefficients $A_{\pm2}$, $B_{\pm2}$ read

\begin{align}
A_{\pm2}&=-\frac{3\cdot2^{\pm1}}{L^2}\left[\frac{n^2x^2}{L^4}\gamma_{tt}\pm \frac{inx}{L^2}\gamma_{tx}-\frac{nx}{L^2}\gamma_{ty}-\frac{1}{4}(\gamma_{xx}-\gamma_{yy})\mp \frac{i}{2}  \gamma_{xy}\right]\\\notag
B_{\pm2}&=\mp\frac{3\cdot 2^{\pm1}}{L^2r_{+}}\Bigg[\pm\frac{i n}{2}  \left(1-\frac{2 n^2 x^2}{L^4}\right) \gamma _{{tt}}+\frac{nx}{2}\left(\frac{2n}{L^2}\mp\omega\right)(\gamma_{tx}\pm i\gamma_{ty})\\
   &+\frac{1}{4}  \left(n\mp L^2 \omega \right)\left(\pm i\gamma _{{xx}}-2 \gamma
   _{{xy}}\mp i \gamma _{{yy}}\right)+\frac{1}{4}
   \left(\pm 2 i n x \gamma _{{tt}}'-L^2 \left(\gamma _{{tx}}'+\pm i \gamma
   _{{ty}}'\right)\right)\Bigg]\, .
\end{align}

Now, when searching for quasinormal modes, we demand that the variables $\Psi_{0,4}$ be separable, and this gives us additional equations for the metric perturbation. If these are separable, then we have seen that they have the form

\begin{align}
\hat{\Psi}_0&=e^{-i\omega t}\mathcal{H}_{q_{+2}}(x)\left[a_{+2}+b_{+2}z+\mathcal{O}(z^2)\right]\, ,\\
\hat{\Psi}_4&=e^{-i\omega t}\mathcal{H}_{q_{-2}}(x)\left[a_{-2}+b_{-2} z+\mathcal{O}(z^2)\right]\, ,
\end{align}
where we are using \req{eq:Y2as}, and the levels $q_{\pm2}$ are allowed to be different. Thus, consistency with separability demands the following constraints 

\begin{equation}\label{eq:A2B2}
A_{\pm2}=a_{\pm2}\mathcal{H}_{q_{\pm2}}\, ,\quad B_{\pm2}=b_{\pm2}\mathcal{H}_{q_{\pm2}}\, .
\end{equation}
In total, \req{eq:A2B2} and \req{eq:gammaabeq} form a system of eight equations for the six variables $\gamma_{ab}$, and therefore it is an overdetermined system; in order for a solution to exist, the parameters $a_{\pm2}$, $b_{\pm2}$ and the levels $q_{\pm2}$ cannot be arbitrary.  By analyzing those equations, one can see that a solution exists only if the levels $q_{\pm2}$ are related according to 
\begin{equation}\label{eq:qpmrel}
q_{+2}=q_{-2}+4s\, ,
\end{equation}
so that $q_{-2}$ takes the values $q_{-2}=0,1,2,\ldots$ for $s=1$ and $q_{-2}=4,5,6,\ldots$ for $s=-1$. 
In addition, the ratios
\begin{equation}\label{rat}
\lambda_{\pm2}=\frac{b_{\pm2}}{a_{\pm2}}\, ,
\end{equation}
must be related according to

\begin{equation}\label{eq_lambda2rel}
\lambda_{-2}=\frac{M_q+P_q\lambda_{+2}}{Q_q+S_q\lambda_{+2}}\, ,
\end{equation}
where 
\begin{align}
M_{q}=&-i \left(\left(8 q^2+40 q+41\right) \hat\omega^2 \epsilon ^2-3 (2 q+5) \hat\omega^3 \epsilon +7 (2 q+5) \hat\omega \epsilon ^3+\hat\omega^4+2 \epsilon ^4\right)\, ,\\
P_{q}=&\left(2 q^2+2 q-5\right) \hat\omega \epsilon ^2-2 (2 q+3) \hat\omega^2 \epsilon +\hat\omega^3-2 \epsilon ^3\, ,\\
Q_{q}=&\left(2 q^2+18 q+35\right) \hat\omega \epsilon ^2-2 (2 q+7) \hat\omega^2 \epsilon +\hat\omega^3+2 \epsilon ^3\, ,\\
S_{q}=&i \left(-(2 q+5) \hat\omega \epsilon +\hat\omega^2-2 \epsilon ^2\right)\, ,
\end{align}
and where 

\begin{equation}
q=\begin{cases}
q_{-2} \quad&\text{if}\quad s=1\, ,\\
-1-q_{-2} \quad&\text{if}\quad s=-1\,
\end{cases}
\end{equation}
There is also a relation between the normalizations of $Y_{+2}$ and $Y_{-2}$, which reads

\begin{equation}
2\epsilon^2\hat\omega\frac{a_{-2}}{a_{+2}}=\left(2 q_{-2}^2+18 q_{-2}+35\right) \hat{\omega } \epsilon ^2-2 (2 q_{-2}+7) \hat{\omega }^2 \epsilon +\hat{\omega }^3+2 \epsilon ^3-i \lambda_{+2}  \left((2 q_{-2}+5) \hat{\omega } \epsilon -\hat{\omega }^2+2 \epsilon ^2\right)
\end{equation}
for $s=1$ and

\begin{equation}
2\epsilon^2\hat\omega\frac{a_{-2}}{a_{+2}}=\frac{\left(2 q_{-2}^2-14 q_{-2}+19\right) \hat{\omega } \epsilon ^2+2 (2 q_{-2}-5) \hat{\omega }^2 \epsilon +\hat{\omega }^3+2 \epsilon ^3+i \lambda_{+2}  \left((2 q_{-2}-3) \hat{\omega } \epsilon +\hat{\omega }^2-2 \epsilon ^2\right)}{16 (q_{-2}-3) (q_{-2}-2) (q_{-2}-1) q_{-2}}
\end{equation}
for $s=-1$, but this is irrelevant for the computation of quasinormal modes. Finally, one can obtain an explicit solution for the $\gamma_{ab}$ in terms of Hermite functions $\mathcal{H}_{p}$, which we show in Appendix \ref{app:metric}.

These results fix the boundary conditions up to the choice of the complex constant $\lambda_{+2}$ (and up to trivial rescalings of $Y_{\pm 2}$).  In the case of vanishing NUT charge, there are two admissible values of $\lambda_{+2}$ that give rise to quasinormal modes, and these correspond to choosing either parity odd or parity even polarizations. However, in the case at hands the background breaks parity, and hence one cannot determine a priori the value of $\lambda_{+2}$. Then, in order to find the quasinormal modes, one has to solve simultaneously the equations \req{eq:mastereqs} for $Y_{+2}$ and $Y_{-2}$ with the boundary conditions discussed above.
Unlike in the electromagnetic case, these equations are not degenerate, and the problem will only have solutions for a discrete set of values of $\omega$ (the quasinormal frequencies) and $\lambda_{+2}$ (which determine the polarization). 

Fortunately, it is possible to find an analytic result for the polarization parameter $\lambda_{+2}$ by using the so-called Teukolsky-Starobinsky identities (see e.g. \cite{Teukolsky:1974yv} and \cite{Chandrasekhar:1985kt} for a detailed analysis of those in Kerr's space, and \cite{Dias:2013sdc} for Kerr-(A)dS in the context of holography). These relate solutions of the $Y_{+2}$ variable with those of $Y_{-2}$, and vice-versa.  In order to find these identities, it is useful to introduce two new radial functions defined by

\begin{equation}\label{eq:radialvarR}
R_{(+2)}=\frac{(r^2+n^2)^{3/2}}{\Delta}e^{-4 i \arctan(r/n)}Y_{+2}\, ,\quad R_{(-2)}=\frac{(r+i n)^4}{(r^2+n^2)^{1/2}\Delta}Y_{-2}\, ,
\end{equation}
where $\Delta=(r^2+n^2)V(r)$. In terms of these, the radial equations for each level ($q_{+2}$ or $q_{-2}$ in each case) read

\begin{align}\label{eq:R+eq}
&\left[\mathcal{D}_{-1}\Delta\mathcal{D}^{\dagger}_{1}+6\left(\frac{r^{2}+n^{2}}{L^{2}}+i\omega r\right)-4\omega n\left(s(q_{+2}+1/2)-2\right)\right]R_{(+2)}^{q_{+2}}=0\\
&\left[\mathcal{D}_{-1}^{\dagger}\Delta\mathcal{D}_{1}+6\left(\frac{r^{2}+n^{2}}{L^{2}}-i\omega r\right)-4\omega n\left(s(q_{-2}+1/2)+2\right)\right]R_{(-2)}^{q_{-2}}=0\, ,
\label{eq:R+eq'}
\end{align}
where we have introduced the operators 

\begin{equation}
\mathcal{D}_{m}=\partial_{r}-i\omega\frac{r^{2}+n^{2}}{\Delta}+m\frac{\Delta'}{\Delta}, \ \ \ \ \mathcal{D}^{\dagger}_{m}=\partial_{r}+i\omega\frac{r^{2}+n^{2}}{\Delta}+m\frac{\Delta'}{\Delta}\, ,
\end{equation}
which satisfy the properties 
\begin{equation}
\mathcal{D}_{m}\Delta=\Delta \mathcal{D}_{m+1},\ \ \ \ \mathcal{D}_{m}^{\dagger}\Delta=\Delta \mathcal{D}_{m+1}^{\dagger}
\end{equation}

We see that the variables $R_{(+2)}^{q_{+2}}$ and $R_{(-2)}^{q_{-2}}$ satisfy conjugate equations when the levels $q_{+2}$ and $q_{-2}$ are related as in \req{eq:qpmrel}, and in that case it is possible to show the following relations (the TS identities): 

\begin{align}\label{eq:TSI1}
\mathcal{D}_{-1}^{\dagger}\Delta\mathcal{D}_{0}^{\dagger}\mathcal{D}_{0}^{\dagger}\Delta\mathcal{D}_{1}^{\dagger}R_{(+2)}^{\omega,q_{+2}}&=C_{(-2)}R_{(-2)}^{\omega,q_{+2}-4s}\, , \\
\mathcal{D}_{-1}\Delta\mathcal{D}_{0}\mathcal{D}_{0}\Delta\mathcal{D}_{1}R_{(-2)}^{\omega,q_{-2}}&=C_{(+2)}R_{(+2)}^{\omega,q_{-2}+4s} \, ,
\label{eq:TSI2}
\end{align}
where $C_{(\pm2)}$ are certain complex constants that can always be chosen as complex-conjugates of each other by an appropriate choice of normalization of the radial functions. These relations mean that given a solution $R_{(+2)}^{q_{+2}}$ of the $(+2)$ equation, then $\mathcal{D}_{-1}^{\dagger}\Delta\mathcal{D}_{0}^{\dagger}\mathcal{D}_{0}^{\dagger}\Delta\mathcal{D}_{1}^{\dagger}R_{(+2)}^{\omega,q_{+}}$ is a solution of the $(-2)$-equation with same frequency but Landau level $q_{-2}=q_{+2}-4s$, and similarly for the second identity.
We remark that these relations map the solutions of the radial equations into each other, but this does not necessarily mean that these relations are actually realized for generic perturbations --- proving that is much harder.  However, in the case of quasinormal modes, it is not difficult to see that the TS identities map the solutions with the correct boundary conditions at the horizon \req{eq:Y2nh} into each other. This means that, at least when searching for quasinormal modes, the TS identities do hold.  These identities allow us to obtain the value of $\lambda_{+2}$ and to reduce the problem of finding QNMs to solving one equation for one variable. 

To show this, consider the asymptotic behavior of $Y_{+2}$ with generic Robin boundary conditions,
\begin{equation}\label{eq:Y2lamb}
Y_{+2}=a_{+2}(1+\lambda_{+2}z+\mathcal{O}(z^2))\, .
\end{equation}
Using the relations \req{eq:radialvarR} and \req{eq:TSI1} one finds that the variable $Y_{-2}$ then satisfies\footnote{In performing this map one has to be careful to include the $\mathcal{O}(z^2)$ terms (not shown above) in $Y_{+2}$.} 

\begin{equation}\label{eq:Ym2lamb}
Y_{-2}=a_{-2}(1+\lambda_{-2}z+\mathcal{O}(z^2))\, ,
\end{equation}
for certain $a_{-2}$, and where $\lambda_{-2}$ reads

\begin{equation}\label{eq:lambda2rel2}
\lambda_{-2}=\frac{\hat{M}_{\hat q}+\hat{P}_{\hat q}\lambda_{+2}}{\hat{Q}_{\hat q}+\hat{S}_{\hat q}\lambda_{+2}}\, ,
\end{equation}

where

\begin{align}
\notag
\hat{M}_{\hat q}=&-4 i \Big(\left(8 \hat{q}^2-24 \hat{q}+9\right) \hat{\omega }^2 \epsilon ^2+7 (2 \hat{q}-3) \hat{\omega } \epsilon ^3+(9-6 \hat{q}) \hat{\omega }^3 \epsilon +\hat{\omega }^4\\
&+2 \epsilon  \left(9 i \epsilon ^4+25 \epsilon ^3-18 i \epsilon ^2-3 i\right)\Big)\, ,\\
\hat{P}_{\hat q}=&4 \left(2 \hat{q}^2-14 \hat{q}+19\right) \hat{\omega } \epsilon ^2-8 (2 \hat{q}-5) \hat{\omega }^2 \epsilon +4 \hat{\omega }^3-9 i \epsilon ^4-32 \epsilon ^3+18 i \epsilon ^2+3 i\, ,\\
\hat{Q}_{\hat q}=&4 \left(2 \hat{q}^2+2 \hat{q}-5\right) \hat{\omega } \epsilon ^2+8 (1-2 \hat{q}) \hat{\omega }^2 \epsilon +4 \hat{\omega }^3+9 i \epsilon ^4+32 \epsilon ^3-18 i \epsilon ^2-3 i\, ,\\
\hat{S}_{\hat q}=&4 i \left((3-2 \hat{q}) \hat{\omega } \epsilon +\hat{\omega }^2-2 \epsilon ^2\right)\, ,
\end{align}
and where in this case we are defining

\begin{equation}
\hat q=\begin{cases}
q_{+2} \quad&\text{if}\quad s=1\, ,\\
-1-q_{+2} \quad&\text{if}\quad s=-1\,
\end{cases}
\end{equation}
Comparing with \req{eq_lambda2rel} we have two relations between $\lambda_{+2}$ and $\lambda_{-2}$, so we can determine both parameters. We get two different solutions, which read

\begin{align}\notag
\lambda_{+2}^{(\pm)}&=\frac{i }{(3-2 \hat{q}) \hat{\omega } \epsilon +\hat{\omega }^2-2 \epsilon ^2}\bigg(2 \hat{q}^2 \hat{\omega } \epsilon ^2+2 \hat{q} \hat{\omega } \epsilon ^2-4 \hat{q} \hat{\omega }^2 \epsilon +\hat{\omega }^3-5 \hat{\omega } \epsilon ^2+8 \epsilon ^3+2 \hat{\omega }^2 \epsilon\\
\label{eq:lambda2sol}
& \mp 2 \epsilon ^2 \sqrt{(\hat{q}-3) (\hat{q}-2) (\hat{q}-1) \hat{q} \hat{\omega }^2+9 \epsilon ^2} \bigg)\, ,\\
\lambda_{-2}^{(\pm)}&=-\lambda_{+2}^{(\pm)}-8i\epsilon\, .
\label{eq:lambdam2sol}
\end{align}
Each of these solutions corresponds to one of the two possible polarization modes of gravitational waves.

 As a check of our computations we may consider the limit of vanishing NUT charge. In order to recover the perturbations for the planar black hole with momentum $\vec{k}$ one should take the limit $n\rightarrow 0$ and $q_{\pm 2}\rightarrow \infty$ in a way in which $4sn\omega q_{\pm2}/L^2\rightarrow \vec{k}^2$. By doing so, we get the following limiting values of $\lambda_{\pm2}$,
 
 \begin{align}
 \lambda_{+2}^{(-)}=- \lambda_{-2}^{(-)}=&-\frac{i \left(\hat k^2-2 \hat\omega^2\right)}{2 \hat \omega}\, ,\\
 \lambda_{+2}^{(+)}=- \lambda_{-2}^{(+)}=&\frac{2 i \hat\omega\left(\hat k^2 -\hat\omega^2\right)}{\hat k^2-2 \hat\omega^2}\, ,
 \end{align}
where $\hat k=L^2 \vec{k}/r_{+}$ is the dimensionless momentum. It is not difficult to check that these coefficients precisely correspond the appropriate boundary conditions for the variables $Y_{\pm2}$ for odd $(-)$ and even $(+)$ parity perturbations of the black brane, respectively --- see \cite{Miranda:2008vb}.  

Let us also mention that, instead of establishing a boundary condition for the radial Teukolsky variables by computing them from the metric perturbation, it is possible to go the other way around by using the so-called Hertz potentials. These are related to the Teukoslky variables and allow one to reconstruct the metric perturbation from them. In this way, one can determine what boundary conditions for the radial variables give rise to Dirichlet boundary conditions for the metric perturbation. We study this alternative method in Appendix~\ref{app:Hertz}, finding perfect agreement with our results above. 

In sum, we have found that, in order to find the gravitational QNMs, we have to solve the radial equation \req{eq:mastereqs} for $Y_{+2}$  with the boundary conditions given by \req{eq:Y2nh}, \req{eq:Y2lamb} and \req{eq:lambda2sol} (or equivalently, the equation for $Y_{-2}$ with the conditions \req{eq:Y2nh}, \req{eq:Ym2lamb} and \req{eq:lambdam2sol}).\footnote{The normalization constants $a_{\pm2}$ in \req{eq:Y2lamb} and \req{eq:Ym2lamb} are irrelevant for the definition of the QNMs.} This problem only has solutions for a discrete set of complex frequencies, which are the quasinormal-mode frequencies.

\section{Quasinormal modes}\label{sec:QNMs}
Having reduced the study of perturbations to a one-dimensional problem given by the radial equations \req{eq:radials} and \req{eq:mastereqs} and having determined the boundary conditions that the corresponding variables must satisfy, we are now ready to compute the quasinormal modes. 
Before showing the explicit results, we can first determine some general properties of the quasinormal mode frequencies $\omega$. First, note that the dimensionless frequencies $\hat\omega$ will only depend on $\epsilon$ and on the level $q$ (plus on the overtone number, which we omit). Therefore, the actual frequencies scale linearly with the size of the black brane for fixed $\epsilon$,

\begin{equation}\label{generalscaling1}
\omega=\frac{\hat{\omega}_{q}(\epsilon)}{L^{2}}r_{+}.
\end{equation}
In other words, since $\epsilon=n/r_{+}$, we conclude that the QNM frequencies are homogeneous functions of degree 1 of $r_+$ and $n$. 
From the point of view of the dual CFT, however, the quantities $r_{+}$ and $n$ do not have a direct interpretation, and instead the physically relevant quantities in the boundary theory are the ratio $n/L^2$ --- see  \req{bdrymetric} --- and the temperature $T$ given by \req{eq:Temp}. The QNM frequencies are then homogeneous functions of $T$ and $n/L^2$, and they can be conveniently expressed in terms of the dimensionless ratio

\begin{equation}
\xi=\frac{3n}{2\pi T L^2},
\end{equation}
which satisfies $-1\le\xi\le1$. Then, the QNM frequencies read

\begin{equation}\label{generalscaling2}
\omega=2\pi\frac{\xi^{2} \hat{\omega}_{q}\left(\epsilon(\xi)\right)}{3\left(1-\sqrt{1-\xi^{2}}\right)}T,
\end{equation}
where $\epsilon$ and $\xi$ are related by\footnote{Notice that given $\xi$, there are actually two compatible values of $\epsilon$, given by $\epsilon_{\pm}(\xi)=\frac{1}{\xi}\left(1\pm \sqrt{1-\xi^{2}}\right)$. However, only the $(-)$ branch contributes to the Euclidean saddle point and thus we will focus on this case.}
\begin{equation}
\epsilon(\xi)=\frac{1}{\xi}\left(1- \sqrt{1-\xi^{2}}\right).
\end{equation}
Thus, we shall study $\omega/T$ as a function of $\xi$. 
The frequencies feature in addition a symmetry under the exchange of sign of $n$ (or $\xi$, equivalently). Namely, we have
\begin{equation}\label{eq:QNM-n}
\omega_{q}(-n)=-\omega^{*}_{q}(n),
\end{equation}
meaning that given a QNM frequency $\omega_{q}(n)$ of the solution with NUT charge $n$, then $-\omega^{*}_{q}(n)$ is a frequency of the solution with charge $-n$. This result can be obtained by noticing that the complex-conjugate variables $Y_{\pm S}^*$ satisfy the same equations and boundary conditions as $Y_{\pm S}$ with $\omega\rightarrow -\omega^{*}$ and $n\rightarrow-n$. Thus, there is a correspondence between the QNMs with $\text{Re}(\omega)>0$ and NUT charge $n$ and those with  $\text{Re}(\omega)<0$ and NUT charge $-n$, and vice-versa. Hence, it is sufficient to focus on studying the QNMs with $\text{Re}(\omega)>0$ for both positive and negative $n$. In the case of scalar QNMs, one can also see that the frequencies are actually symmetric under the change of sign of $n$, $\omega_{\rm scalar}(n)=\omega_{\rm scalar}(-n)$, because the radial equation is invariant under the change of sign of $n$. For the gravitational perturbations, however, one can see that the replacement $n\rightarrow-n$ is not a symmetry of the master equations \req{eq:mastereqs}, nor of the boundary conditions \req{eq:lambda2sol}. Thus, in principle one should not expect the spectrum of quasinormal modes to be identical for positive and negative $n$.

In order to compute the QNM frequencies, we use the following method. Taking into account the boundary conditions we have determined, we first expand the corresponding variables $Y_{S}$ near the horizon using a Frobenius series and asymptotically using a Taylor expansion. This gives us two approximate solutions $Y_{S}^{\rm +}(z)$ and $Y_{S}^{\infty}(z)$ valid in the regions $z\sim 1$ and $z\sim 0$, respectively. One must then try to glue both solutions, but this only will be possible if $\hat\omega$ is a QNM frequency. One may use directly the asymptotic expansions $Y_{S}^{\rm +}(z)$ and $Y_{S}^{\infty}(z)$ to find the QNM frequencies by imposing the glueing condition $Y_{S}^{\rm \infty}\partial_{z}Y_{S}^{\rm +}-\partial_{z}Y_{S}^{\rm \infty}Y_{S}^{\rm +}\big|_{z_{\rm joint}}=0$,  for some intermediate $z_{\rm joint}$. This yields an algebraic equation for $\omega$, whose solutions should converge to the QNM frequencies when the number of terms in the asymptotic expansions tend to infinity, However, we have found that the convergence is not very good as we increase the NUT charge, and in order to improve the accuracy of our results we use a numerical integration. Thus, we use the near-horizon expansion $Y_{S}^{\rm +}(z)$ to initialize the numerical method at some $z_{\rm ini}$ close to $z=1$, and we numerically integrate the solution up to some $z_{\rm end}$ close to $z=0$. Then, we compute the Wronskian 
\begin{equation}
W_{S}=Y_{S}^{\rm \infty}\partial_{z}Y_{S}^{\rm num}-\partial_{z}Y_{S}^{\rm \infty}Y_{S}^{\rm num}\Big|_{z_{\rm end}}, 
\end{equation}
and we search for solutions of $W_{S}=0$. In the case of the scalar field, $S=0$, we have a single equation, $W_0=0$, that determines the QNM frequencies.  In the gravitational case, $S=\pm 2$, we can use any of the two equations $W_{2}=0$ or $W_{-2}=0$. As discussed in the previous section, the two variables $Y_{\pm2}$ are isospectral when provided with their own set of boundary conditions, \req{eq:lambda2sol} and \req{eq:lambdam2sol}, respectively, so it is enough to work with only one of them; for instance, $Y_{+2}$. In the electromagnetic case we have seen in section~\ref{subsec:bdryA} that both variables $Y_{\pm 1}$ are isospectral for \emph{any} choice of the free polarization parameter $\lambda_{+1}$, which seems to indicate that the QNMs depend continuously on this parameter. We will leave this case for future developments, and we will focus here on the scalar and gravitational QNMs. 


In order to understand the structure of these quasinormal modes, it is important to see how they relate to those of the black brane. One can see that, in the limit of vanishing NUT charge, we should recover the quasinormal modes of vanishing momentum ($\hat k=0$) of the black brane. Also, note that the spectrum becomes independent of the level $q$ in that limit, and hence an infinite number of modes $\omega_{q}$ of different $q$ collapse to the same mode. 
On the other hand, it is not clear that one can recover the QNMs of black branes with arbitrary momentum in the limit of $n\rightarrow0$. Note that this momentum can be identified as 
\begin{equation}\label{eq:momentum}
\hat k^2=\lim_{n\rightarrow0}2\mathcal{E}_{q}=\lim_{n\rightarrow0}2sn\omega(1+2q)/L^2\, ,
\end{equation}
thus, in order to get a non-vanishing value one must take simultaneously $n\rightarrow 0$ and $q\rightarrow\infty$ in a way that $qn$ remains finite in that limit. However, the resulting value of $\hat k^2$ would be in general complex unless one chooses $q$ to be complex as well, but in that case the connection with the QNMs of Taub-NUT black holes is broken. Hence, one should not expect to recover all the QNMs of the planar black holes in a continuous way. In any case, as a test for our method, we have checked that in this limit we reproduce the correct values for the axial and polar gravitational QNM frequencies, as shown in tables 3 and 2 of Refs~\cite{Cardoso:2001vs} and \cite{Miranda:2008vb}, respectively.
Let us now present our results.

\subsection{Scalar}

\begin{figure*}[t!]
	\begin{center}
		\includegraphics[width=0.49\textwidth]{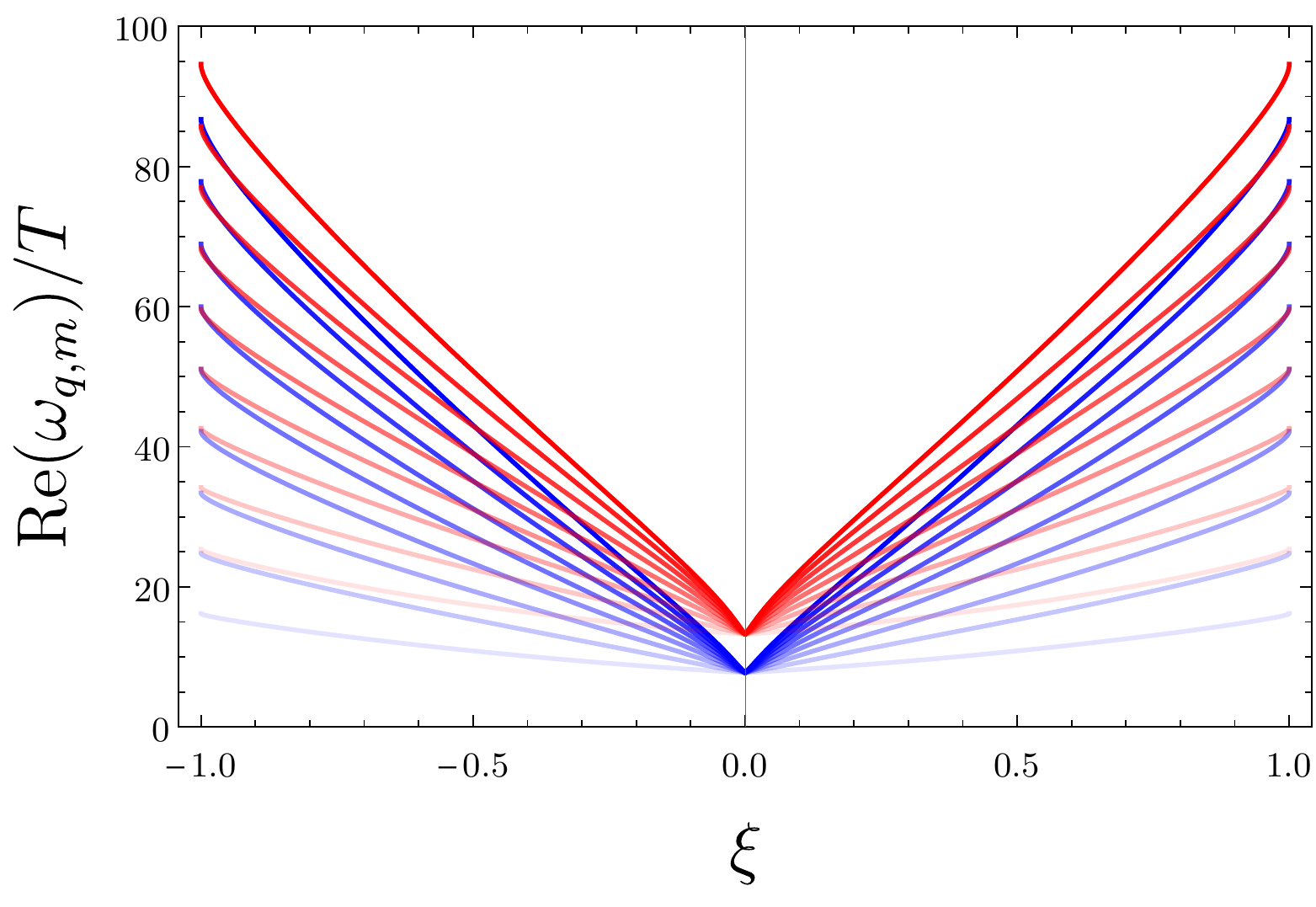}
		\includegraphics[width=0.49\textwidth]{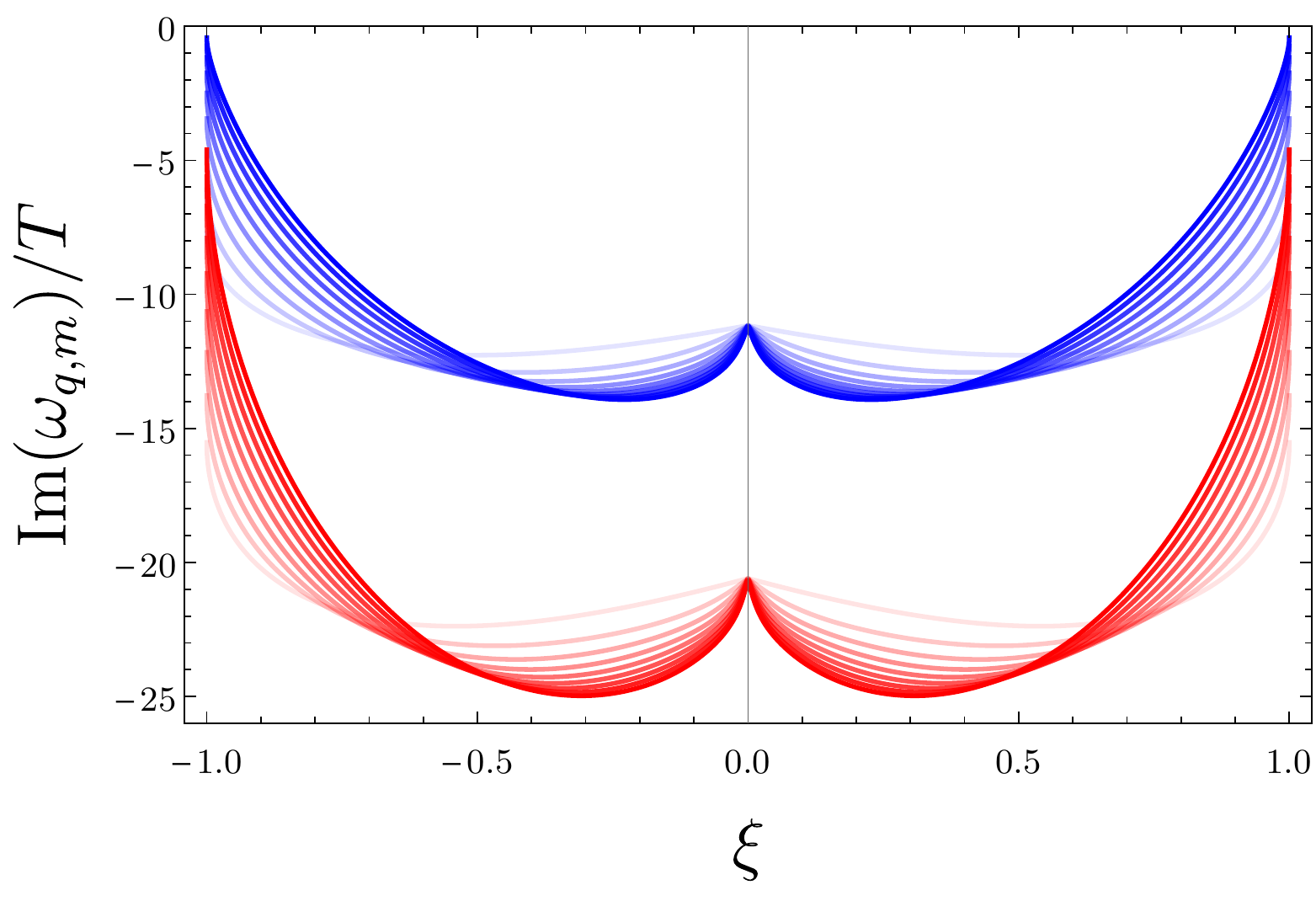}
		\caption{Real and imaginary parts of the scalar QMN frequencies $\omega_{q,m}/T$ as a function of $\xi=\frac{3n}{2\pi T L^2}$. In order of increasing opacity the curves correspond to the levels $q=0,1,...,8$. The fundamental mode $(m=0)$ is shown in blue and the first overtone $(m=1)$ in red. }
		\label{fig:scalar}
	\end{center}
\end{figure*}

We start with the simple case of a massless scalar field.  For every value of $\xi$ and the level $q$, there is an infinite family of QNMs $\omega_{q,m}$, where, for decreasing order of the imaginary part we label these modes by $m=0,1,\ldots$. The one with the largest imaginary part is the fundamental mode $(m=0)$ and the rest are overtones. 

In figure \ref{fig:scalar} we show the fundamental mode and the first overtone for the scalar QNM frequencies for the levels $q=0,1,\ldots 8$. As discussed above, we see that in the limit $\xi\rightarrow 0$ all the modes with different $q$ collapse to the same corresponding mode of the black brane. As a check, we get that 

\begin{align}
 \omega_{q,0}(0)&\approx (7.75-11.2i)T\approx (1.85-2.66 i)r_{+}/L^2\, ,
\end{align}
which agrees with the fundamental mode of the black brane when $r_{+}^2>>\vec{k}^2$ \cite{Cardoso:2001vs}.
As we can see in figure \ref{fig:scalar}, the real part of $\omega$ grows almost linearly with $\xi$ (or $n$), while the imaginary part has a non-monotonic dependence. Also, note that these frequencies are symmetric for $\xi\rightarrow-\xi$. 
For $\xi\sim\pm 1$ we see that $\text{Im}(\omega_{q,m})\sim0$ for large $q$, but our numeric results suggest that it never becomes positive, and therefore, scalar perturbations are stable for the whole range of $\xi$. We recall that the results in Fig.~\ref{fig:scalar} refer to the branch of black holes with positive specific heat, $r_{+}^2>n^2$. 
We have briefly looked to case of $r_{+}^2<n^2$, and for those black holes our results indicate that all the quasinormal modes have very small imaginary parts $\text{Im}(\omega_{q,m})\sim0$, but that still do not cross 0.

\subsection{Gravitational}

Let us now turn to the most interesting case of gravitational modes. We recall that these come in two different classes, $\omega^{(\pm)}_q$, corresponding to the two different polarization modes given in \req{eq:lambda2sol}.
In addition, in analogy with the case of the black brane, we may distinguish two families of modes according to the their behaviour in the limit $n\rightarrow 0$.

\subsubsection{Pseudo-hydrodynamic mode}

We find that for every level $q$ there is a special mode such that $\omega_{q}\rightarrow 0$ in the limit $n\rightarrow 0$. We recall that, in the case of the black brane, both axial and polar perturbations contain a hydrodynamic mode, \textit{i.e}, one whose frequency vanishes when $\vec{k}\rightarrow 0$ \cite{Miranda:2008vb}. 
In the presence of NUT charge, one cannot talk about hydrodynamic modes because the spectrum of quasinormal modes is discrete, and thus we refer to the modes $\omega_{q}$ that vanish for $n\rightarrow 0$ as ``pseudo-hydrodynamic''. These must be in fact related to the hydrodynamic modes of the black brane. 

\begin{figure*}[t!]
	\begin{center}
		\includegraphics[width=0.49\textwidth]{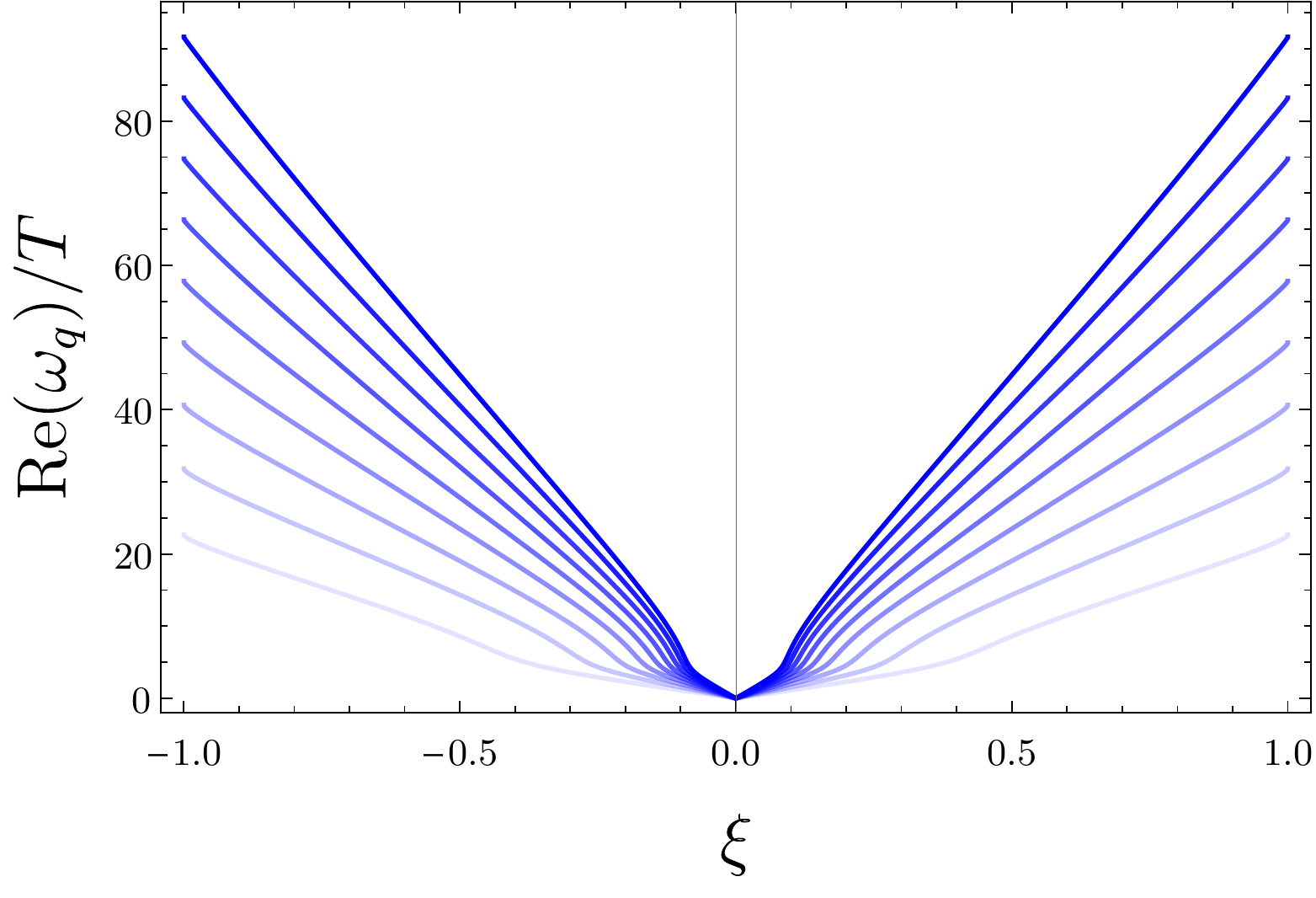}
		\includegraphics[width=0.49\textwidth]{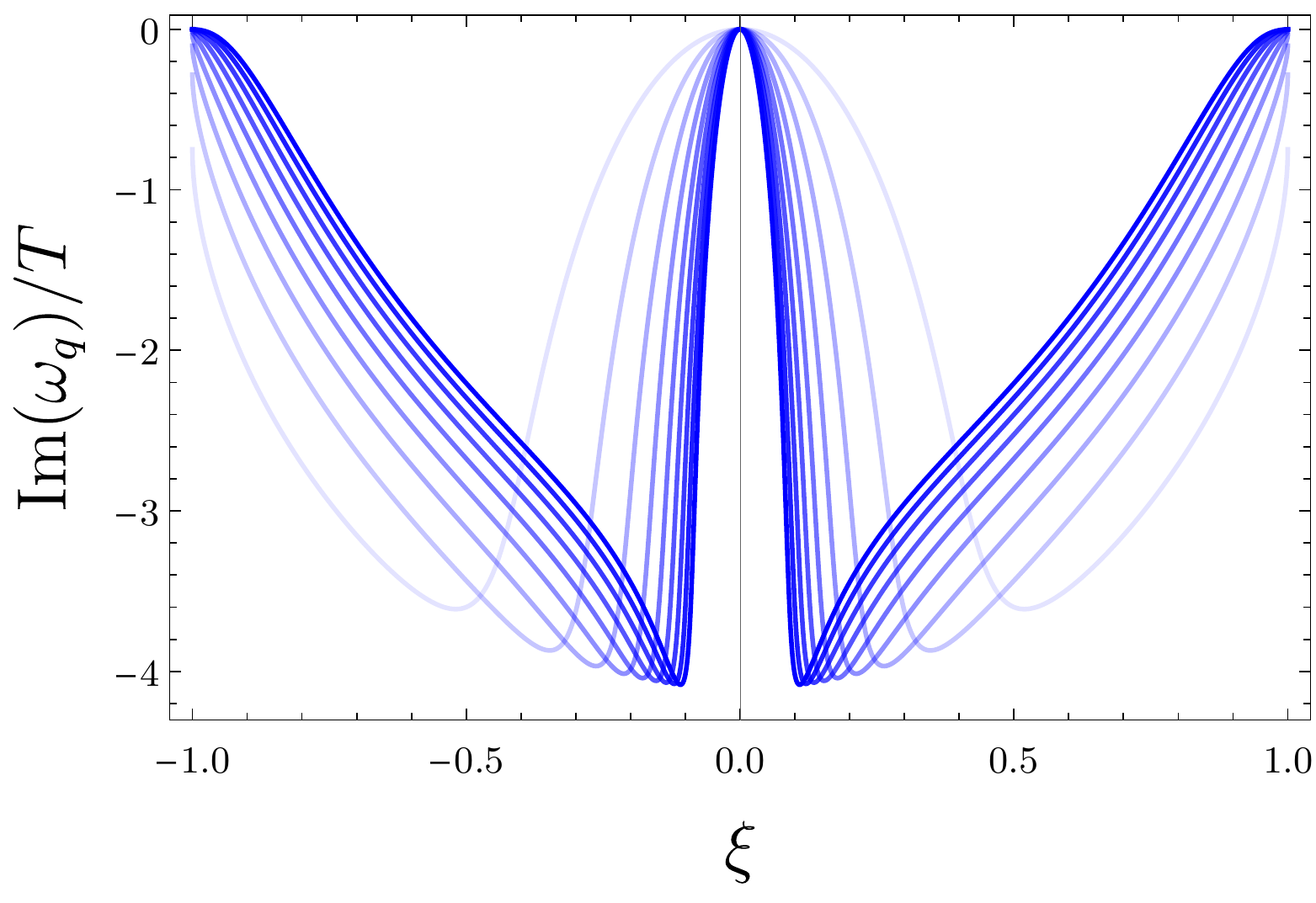}
		\includegraphics[width=0.49\textwidth]{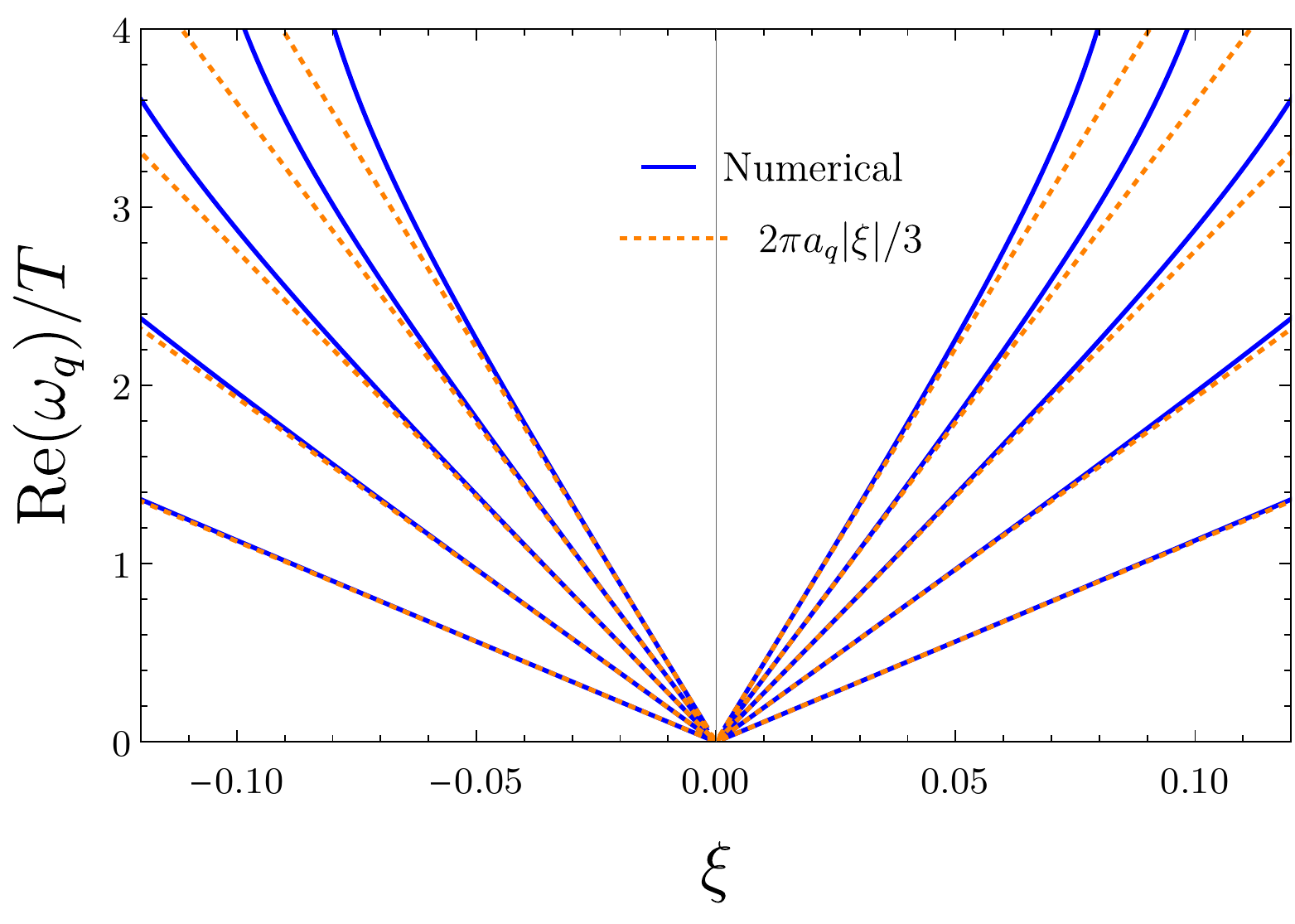}
		\includegraphics[width=0.49\textwidth]{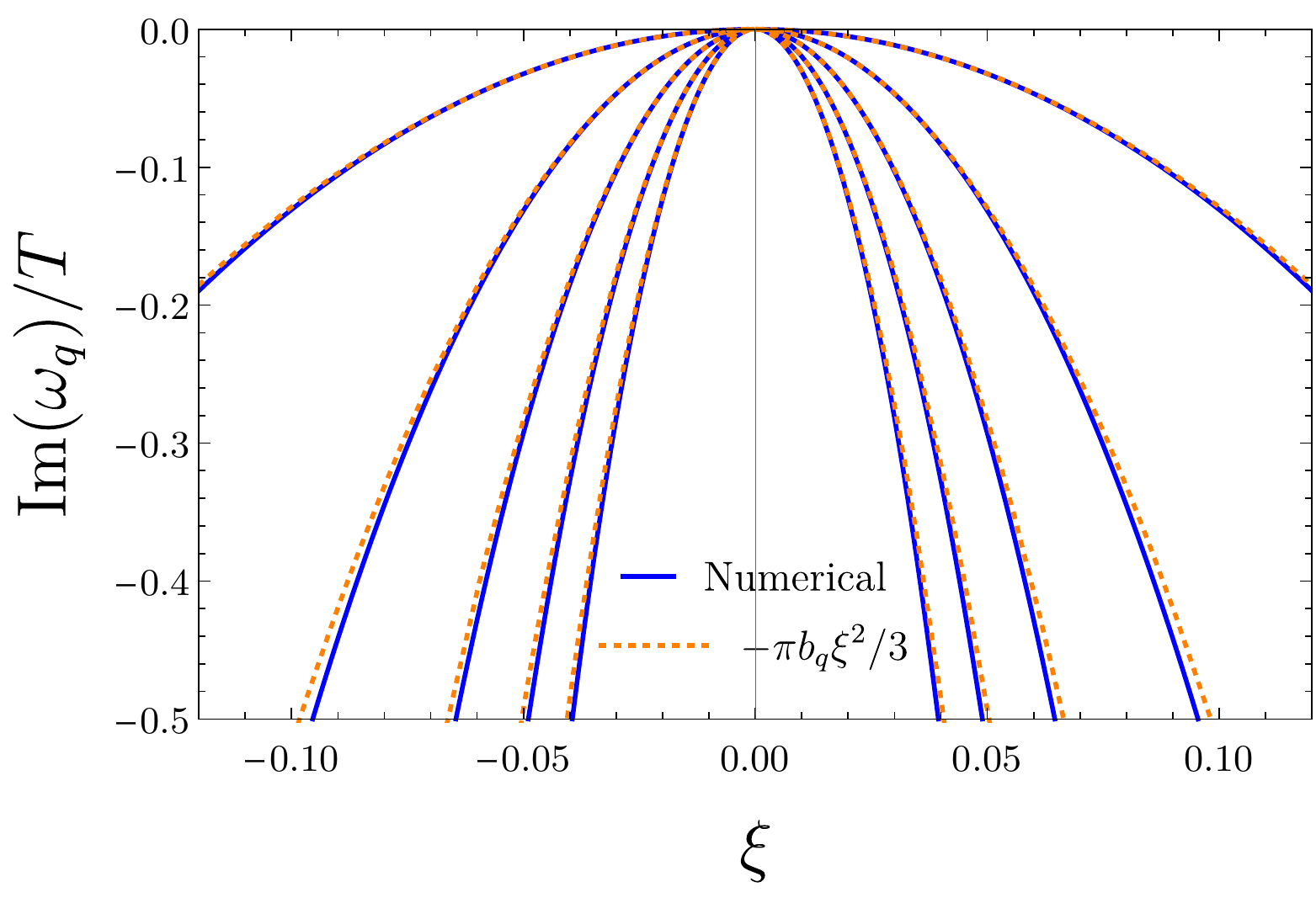}
		\caption{Pseudo-hydrodynamic QNM frequencies of gravitational perturbations as a function of $\xi=\frac{3n}{2\pi T L^2}$. Top row: in order of increasing opacity we show the levels $q=0,1,...,8$, where $q=q_{+2}-4$ if $\text{Re}(n\omega)>0$ and $q=q_{+2}$ if $\text{Re}(n\omega)<0$. Bottom row: behaviour near $\xi=0$ and comparison with the analytic result \req{omegasmall2}. In order to facilitate the visualization in that case we only show the modes with $q=0,2,4,6,8$.}
		\label{fig:grav}
	\end{center}
\end{figure*}

We find that these pseudo-hydrodynamic modes only exist for the $(+)$ polarization --- we comment on the absence of these modes for $(-)$ polarization below --- and we show their corresponding quasinormal frequencies  in Fig.~\ref{fig:grav} for the levels $q=0,\ldots 8$ (where $q=q_{+2}-4$ if $\text{Re}(n\omega)>0$ and $q=q_{+2}$ if $\text{Re}(n\omega)<0$).  
As we can see, the real part behaves linearly with $\xi$ near $\xi=0$, while the imaginary part is quadratic in that region. For larger values of $\xi$ the real part of $\omega_{q}$ transitions to a different linear dependence, while the imaginary part has a non-monotonic behaviour. Indeed, after reaching a minimum value, the imaginary part grows and becomes close to 0 for $\xi=\pm1$. We observe that for larger $q$, the imaginary part becomes even smaller near $\xi=\pm1$, but interestingly it does not become positive, which indicates  that there are no unstable modes --- we study the stability of these solutions below. 
Another property of these QNM frequencies that is worth remarking is that they are symmetric under the exchange $\xi\rightarrow-\xi$. 
This indicates that the exchange of sign of the NUT charge must be indeed a hidden symmetry of the equations \req{eq:mastereqs} with the boundary conditions \req{eq:lambda2sol} and \req{eq:lambdam2sol}.

Let us now focus on the region $\xi<<1$. We can actually obtain analytic approximations for the pseudo-hydrodynamic QNMs in this limit. Recalling first the boundary conditions \req{eq:Y2nh}, we can expand the function $Y_{+2}$ near $z=1$ as
\begin{align}\label{eq:HHmethod0}
Y_{+2}(z)=(1-z)^{-\frac{i\hat\omega}{3(1+\epsilon^2)}}\sum_{i=0}^{\infty}c_{i}(1-z)^i\, .
\end{align}
Using the master equation \req{eq:mastereqs} one can then find explicitly the values of all the coefficients $c_{i}$  in terms of the first one up to a given order $i=i_{\rm max}$. We can then implement a method similar the one of Horowitz and Hubeny  \cite{Horowitz:1999jd} and glue this expansion with the one in \req{eq:Ym2lamb} at $z=0$, which yields the equation
\begin{equation}\label{HHmethod}
\lambda^{(-)}_{+2}Y_{+2}(0)-Y'_{+2}(0)=0\, ,
\end{equation}
where we recall that $\lambda^{(-)}_{+2}$ is given by \req{eq:lambda2sol}. In general, this method can be used to obtain an approximate solution to the QNM frequencies, but in the limit $|\xi|<<1$ we can obtain an analytic result.  Taking into account the input from the numerical result, we will have
\begin{equation}\label{omegasmall1}
\hat\omega_{q}=\epsilon a_{q}-i b_{q}\epsilon^2+\mathcal{O}(\epsilon^3)\, ,
\end{equation}
for some coefficients $a_{q}$ and $b_{q}$ (near $\xi=0$ the relation between this variable and $\epsilon$ is simply $\xi\approx 2\epsilon$). Without loss of generality, let us consider $\epsilon>0$ and $\text{Re}(\omega)>0$. Then, in the limit $\epsilon\rightarrow 0$, one can see that the expansion \req{eq:HHmethod0} collapses to a polynomial, and we have $\lim_{\epsilon}Y'_{+2}(0)=\tfrac{3}{2}\lim_{\epsilon}Y_{+2}(0)$. On the other hand, for generic values of $a_q$, the coefficient  $\lambda^{(+)}_{+2}$ is $\mathcal{O}(\epsilon)$ in that limit, and hence the equation \req{HHmethod} is not satisfied. The only way in which $\lambda^{(+)}_{+2}$  does not vanish at $\epsilon=0$ is when the denominator in \req{eq:lambda2sol} is of order $\epsilon^3$, and it is easy to see that this happens when 
\begin{equation}
a_{q}^2-(5+2q)a_{q}-2=0\, ,
\end{equation}
where we recall that for $\text{Re}(n\omega)>0$ we are defining $q=q_{+2}-4$, which takes the values $q=0,1,2,\ldots$. The positive root of this equation yields the following value for $a_q$,
\begin{align}
a_{q}&=\frac{1}{2} \left(2 q+5+\sqrt{4 q^2+20 q+33}\right)\, .
\end{align}
With this choice, one can solve the equation \req{HHmethod} order by order in the $\epsilon$ expansion.
For, say, $i_{\rm max}=3$, one finds 
\begin{equation}
b_{q}=\frac{4}{3}\left(q^2+5 q+4+\frac{(q+2) (q+3) (2 q+5)}{\sqrt{4 q^2+20 q+33}}\right)\, .
\end{equation}
Then it is easy to check that this result does not change for larger values of $i_{\rm max}$, and thus this value of $b_q$ is exact.
This leads to the following expression for the physical frequency $\omega$

\begin{equation}\label{omegasmall2}
\omega_{q}=\frac{2\pi T}{3}\left[a_{q}\xi-\frac{ib_{q}}{2}\xi^2\right]=\frac{n a_{q}}{L^2}-i\frac{3b_{q}n^2}{4\pi TL^4}\, ,
\end{equation}
which is valid when $q|\xi|<<1$. As we show in the second row of Fig.~\ref{fig:grav}, these expressions match the numeric results with great accuracy. Finally, it is interesting to analyze what happens in the limit $n\rightarrow 0$ and $q\rightarrow\infty$ such that $qn$ remains finite. 
In that case we have
\begin{equation}\label{omegaaprox}
\omega\approx \frac{2nq}{L^2}-i\frac{2(nq)^2}{\pi TL^4}\, .
\end{equation}
On the other hand, we recall that in this limit we can identify a momentum for the perturbations $\hat{k}$ according to  \req{eq:momentum}, which yields

\begin{equation}
\hat{k}^2=\frac{4nq\omega}{L^2}\, .
\end{equation}
Moreover, this is a real momentum when $|n|q<<1$.  Combining this expression with \req{omegaaprox} we obtain the following effective dispersion relation for small $\hat{k}$

\begin{equation}\label{dispersion}
\omega\approx \frac{\hat{k}}{\sqrt{2}}-i\frac{\hat{k}^2}{8 \pi T}\, .
\end{equation}
This is precisely the dispersion relation for the hydrodynamic mode of polar perturbations in the absence of NUT charge \cite{Miranda:2008vb}. 
Hence, the pseudo-hydrodynamic modes of the NUT-charged black holes are analogous to that mode of the black brane. 
One may wonder why we do not obtain other modes similar to the hydrodynamic mode of axial perturbations (which correspond to the $(-)$ family in \req{eq:lambda2sol}). The reason is that such mode is purely damped, and according to the identification \req{eq:momentum} we would need to choose $q$ to be imaginary in order to recover a solution of the black brane with real momentum.  Thus, that mode is simply not present in the Taub-NUT planar black holes. 

When $\xi$ becomes larger, we cannot obtain an analytic result for the frequencies, but we can obtain a reasonable good approximation for the real part. In fact, we observe that the real part of the dimensionless frequencies $\hat\omega_{q}$ is a linear function of $\epsilon$, and a fit to the numerical data shows that the slope is proportional to $q$. Namely, we get
\begin{equation}
\text{Re}(\hat\omega_{q})\approx 4.04(q+3)\epsilon\, ,
\end{equation}
plus a constant term that is much smaller. Interestingly, this seems to work not only for $\epsilon\le1$, but for arbitrarily large $\epsilon$. Now, when we take into account \req{generalscaling2}, we deduce that the dimensionful frequencies $\omega_q$ are also a linear function of $\xi$

\begin{equation}
\text{Re}(\omega_{q})\approx \frac{2\pi T\xi}{3}4.04(q+3)\approx \frac{4.04n}{L^2}(q+3)\, .
\end{equation}

\subsubsection{Ordinary quasinormal modes}

\begin{figure}[t!]
	\begin{center}
		\includegraphics[width=0.49\textwidth]{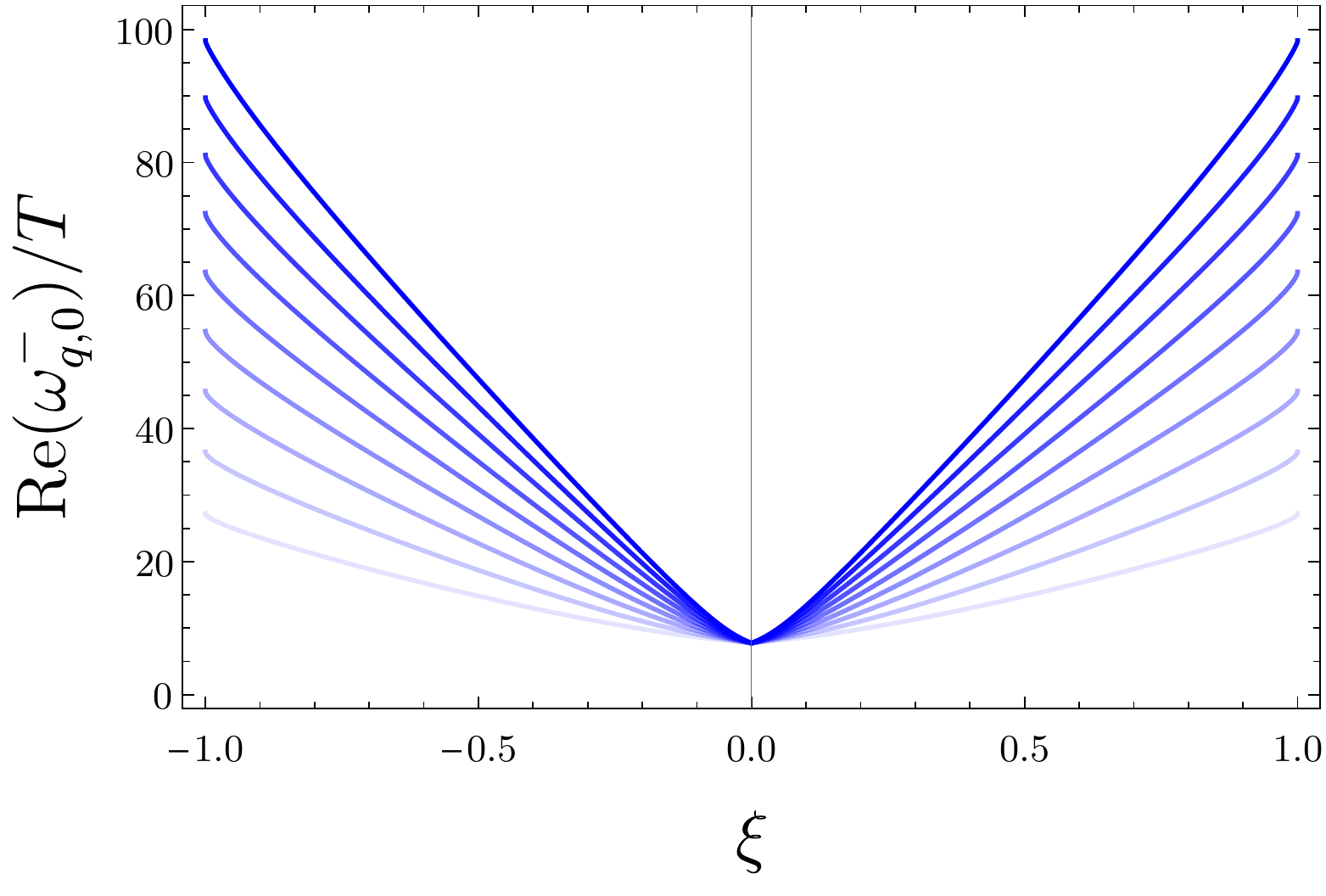}
		\includegraphics[width=0.49\textwidth]{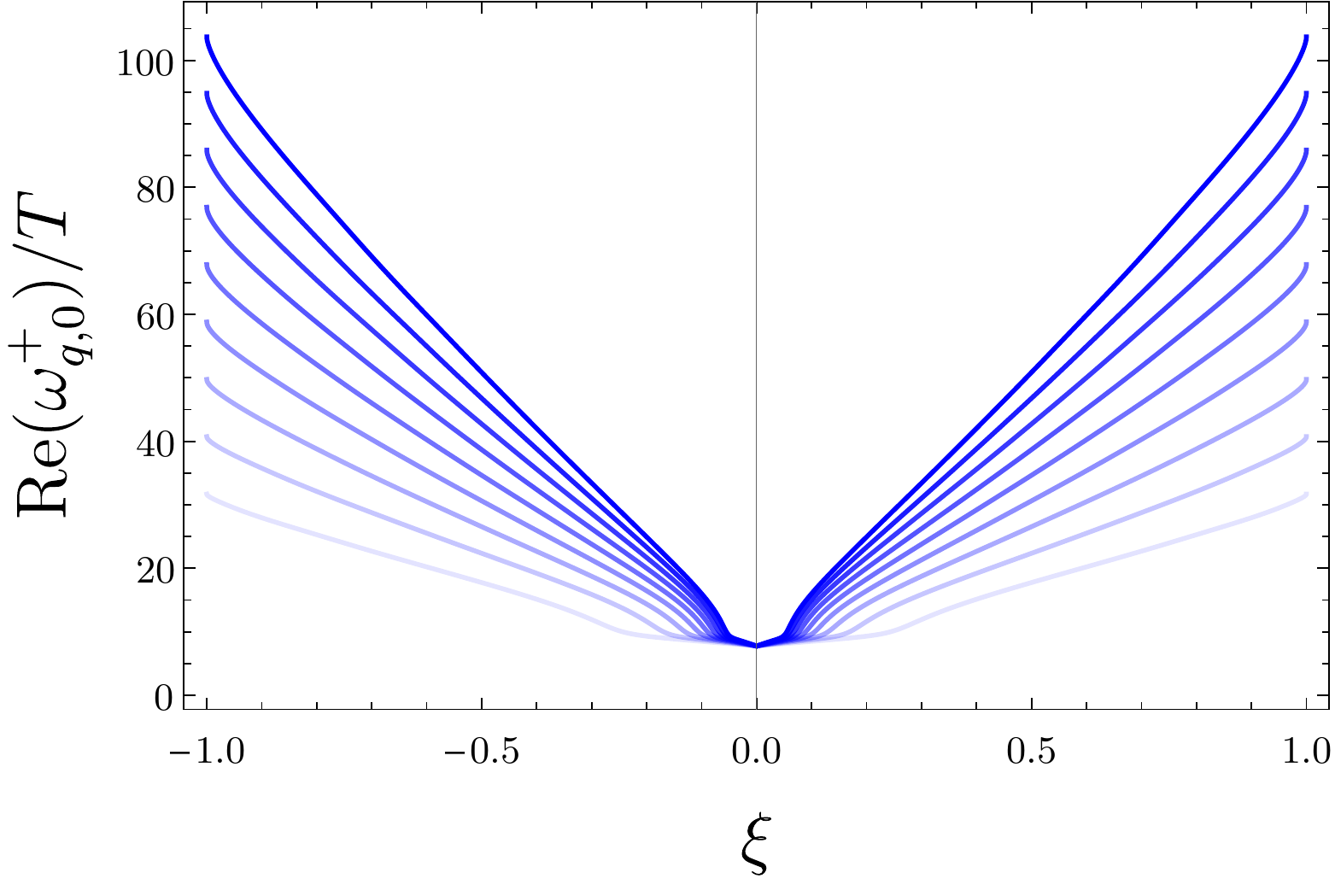}
		\includegraphics[width=0.49\textwidth]{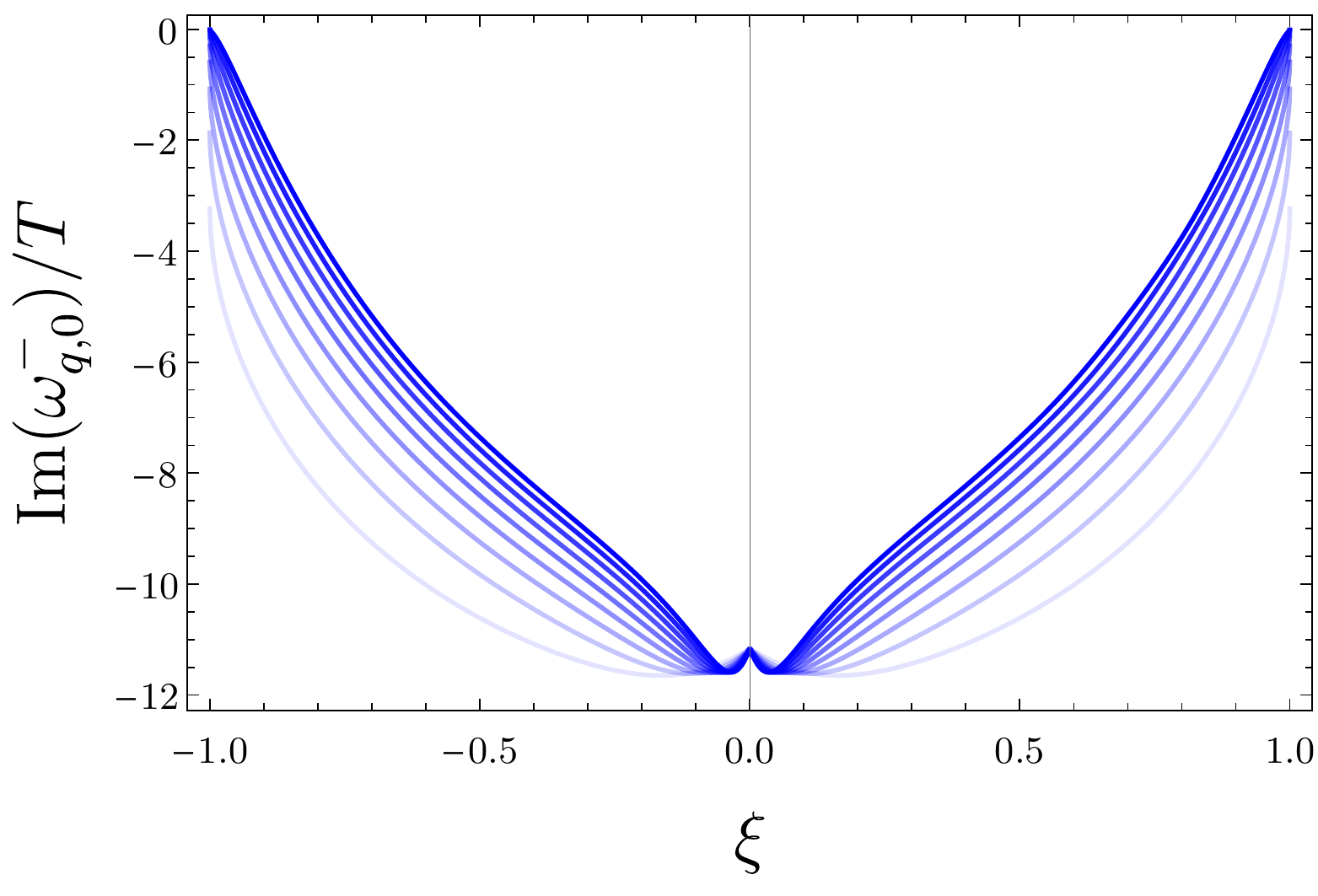}
		\includegraphics[width=0.49\textwidth]{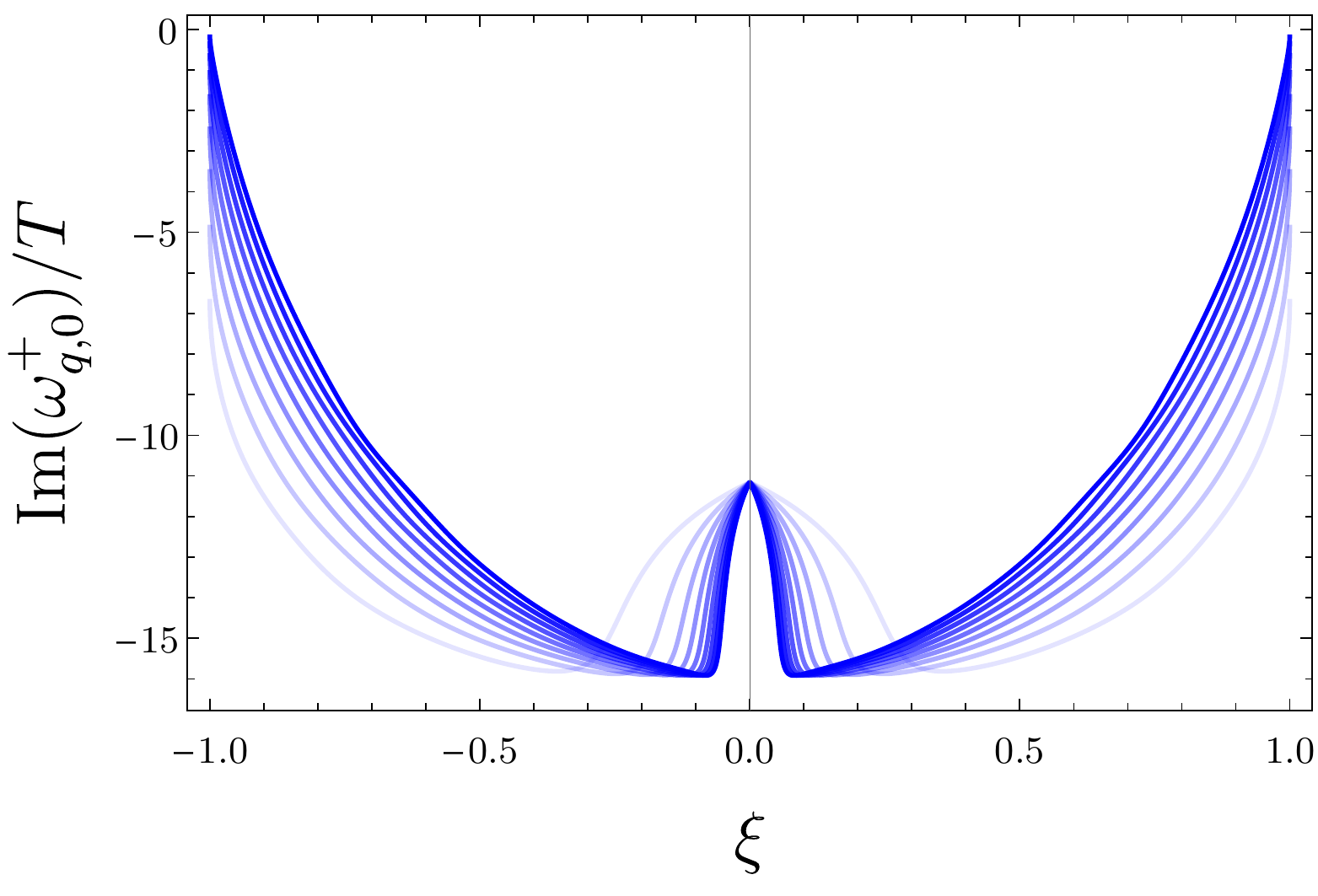}
		\caption{Ordinary gravitational quasinormal modes: we show the lowest overtones $m=0$ of both families $\omega^{-}_{q,0}$ (left) and $\omega^{+}_{q,0}$ (right) as a function of $\xi=\frac{3n}{2\pi T L^2}$.  In order of increasing opacity the curves correspond to the levels $q=0,1,\ldots, 8$.  }
		\label{fig:grav2}
	\end{center}
\end{figure}

The rest of the gravitational quasinormal modes have frequencies that tend to a constant, non-vanishing value in the limit $n\rightarrow 0$. These are labeled by the polarization type $\pm$ defined in \req{eq:lambda2sol}, the Landau level $q$ and the overtone number $m=0,1,2,...$, and we denote them $\omega^{\pm}_{q,m}$. As already remarked before, the values of these frequencies for $n\rightarrow 0$ will correspond to the black brane's QNM frequencies at vanishing momentum. It is known that the polar and axial QNMs of the black brane become degenerate when the momentum tends to zero \cite{Miranda:2008vb}, which means that, in our case,  both classes of modes  $\omega^{+}_{q,m}$ and  $\omega^{-}_{q,m}$ also become degenerate. For the same $m$, the frequencies of all the modes in the two families collapse to the same value,
\begin{equation}
\lim_{n\rightarrow 0}\omega^{+}_{q,m}(n)=\lim_{n\rightarrow 0}\omega^{-}_{q',m}(n)\equiv \omega_{m}(0)\quad \forall \,\,q,q'
\end{equation}

In Fig.~\ref{fig:grav2} we show the lowest ($m=0$) QNMs for a few levels $q$, where the first thing we notice is that  the spectrum is again symmetric for $\xi>0$ and $\xi<0$. The structure of the QNM frequencies as a function of $\xi$ is somewhat similar to the one of the pseudo-hydrodynamic modes, with the real part scaling almost linearly with $q$ for most of the range of $\xi$. In particular, we have the following fits to the real parts of the dimensionless frequencies

\begin{equation}
\text{Re}(\hat\omega_{q}^{+})\approx (16.6+4.37q)\epsilon+c^{+}_{q}\, ,\quad \text{Re}(\hat\omega_{q}^{-})\approx (14.9+4.27q)\epsilon\, +c^{-}_{q}\, ,
\end{equation}
where the constant terms are small. When we use \req{generalscaling2}, this produces an almost linear relation between $\omega_{q}$ and $\xi$, although the non-vanishing constant terms introduce non-linearities near $\xi=\pm1$. 
On the other hand,  the imaginary part becomes very small as $\xi\rightarrow\pm1$, but as before, we do not observe any mode becoming unstable. In addition, for every value of $\xi$ and $q$, the imaginary parts of these modes are larger (in absolute value) than those of the pseudo-hydrodynamic modes, and hence there is no level crossing. 
In the opposite limit, at $\xi=0$, all the modes collapse to $\omega^{\pm}_{q,0}(0)\approx (1.849-2.664 i)r_{+}/L^2$, which agrees with the first ordinary mode of the black brane in the limit of vanishing momentum \cite{Cardoso:2001vs,Miranda:2005qx,Miranda:2008vb}.

\subsubsection{Stability}
So far, all the modes we have found are stable, meaning that their associated frequencies lie in the lower half of the complex plane. In order to show that the Taub-NUT solution is (linearly) stable one must prove that this property holds for every quasinormal modes. 
Here we provide evidence that this is indeed the case, but for future analyses it would be important to provide a solid proof of this fact. 

\begin{figure}[t!]
	\begin{center}
		\includegraphics[width=0.65\textwidth]{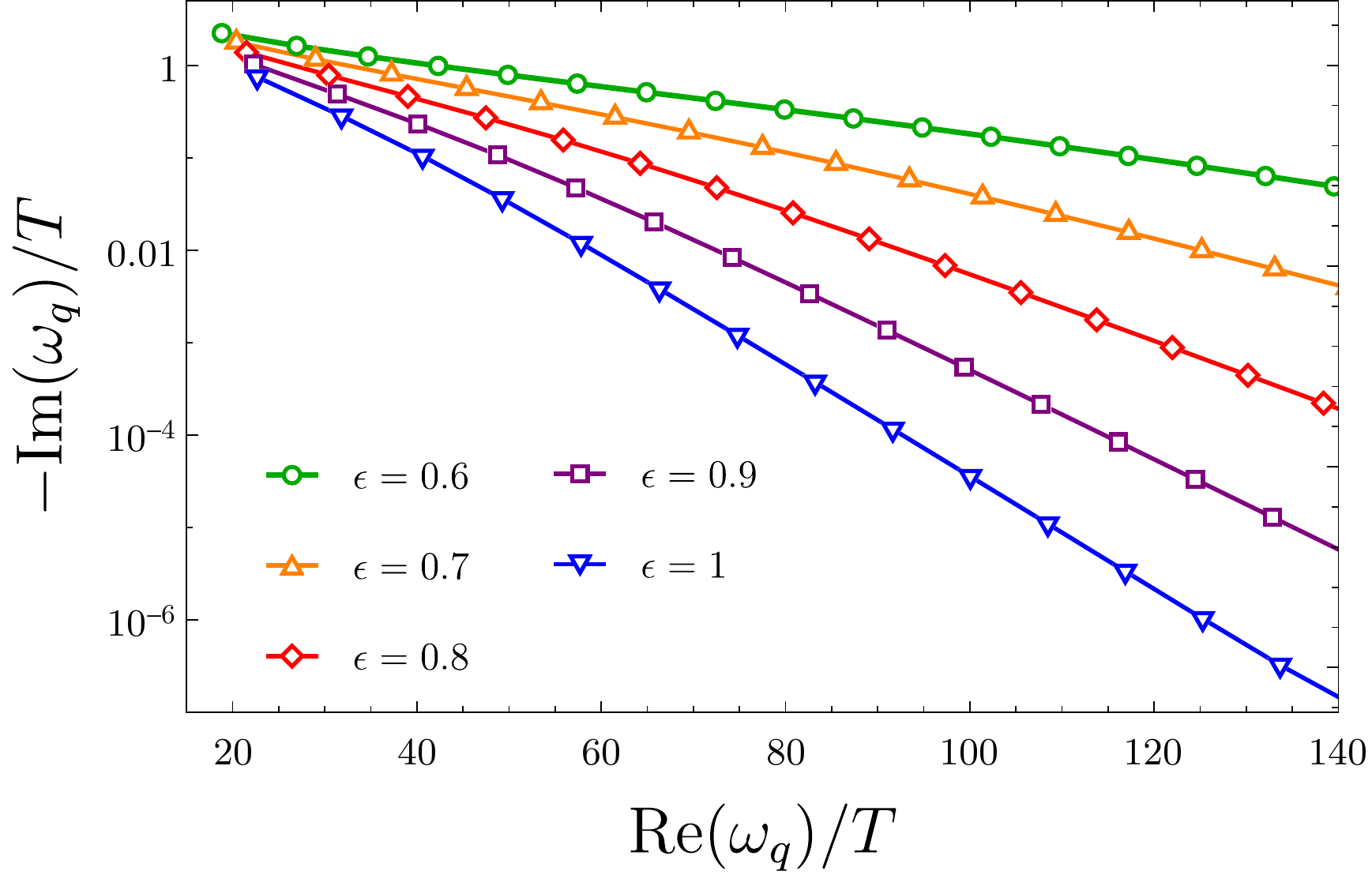}
		\caption{Trajectories in the complex plane of the QNM frequencies $\omega_{q}$ of lowest imaginary part (corresponding to the pseudo-hydrodynamic modes) for a few values of $\epsilon$. For large $q$ the imaginary part tends to zero exponentially, but it never becomes positive.}
		\label{fig:largeq}
	\end{center}
\end{figure}

As we have seen, the quasinormal modes with the lowest imaginary part are the pseudo-hydrodynamic ones, and the imaginary part becomes smaller as we increase $q$. Therefore, we should analyze the behaviour of these modes when $q\rightarrow\infty$. In Fig.~\ref{fig:largeq} we have plotted the trajectories in the complex plane of these modes for many values of $q$ and a some selected values of $\epsilon=n/r_{+}$. Thanks to the logarithmic scale in the vertical axis, we can see clearly that the imaginary part tends to zero exponentially with $q$ and that it also decreases when $\epsilon$ grows. Indeed, a fit to the numerical data reveals that the imaginary part of the QNM frequencies $\omega_{q}$ for large $q$ is well approximated by 
\begin{equation}
\text{Im}(\omega_{q}(\epsilon))\approx -T A(\epsilon) e^{-(2.2\epsilon-1.05)q}\, ,
\end{equation}
which is valid as long as $\epsilon$ is not far from $1$. For smaller values of $\epsilon$, the imaginary part is larger (in absolute value) and therefore, the negativity of $\text{Im}(\omega_{q}(\epsilon))$ for $\epsilon=1$ implies the stability of all the modes with $\epsilon\le 1$. However, the asymptotic behaviour for $q\rightarrow\infty$ is difficult to access for small $\epsilon$, since it requires going to larger and larger $q$, in which case our numeric method becomes less accurate. 
In any case, our data suggests that the imaginary part of $\omega_{q}$ ultimately decays exponentially with $q$ for any value of $\epsilon$. 
Thus, the conclusion is that the lowest-lying modes for every $q$ are stable for every $|\epsilon|\le 1$, and by extension all the modes are.  This signals that, despite the apparent pathological properties of the NUT-charged spacetimes, they actually give rise to stable and well-defined dynamics. 

\begin{figure}[t!]
	\begin{center}
		\includegraphics[width=0.49\textwidth]{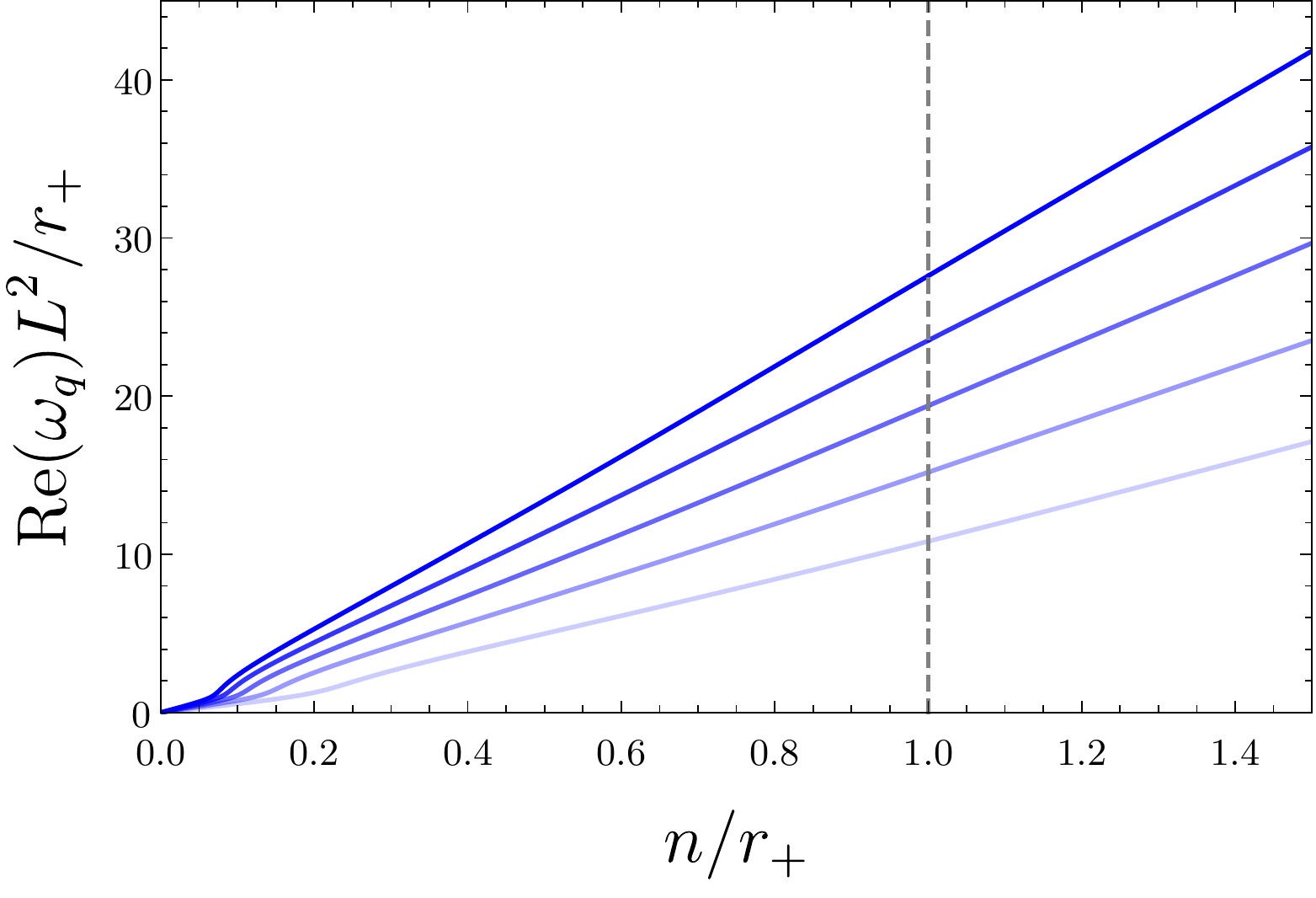}
		\includegraphics[width=0.49\textwidth]{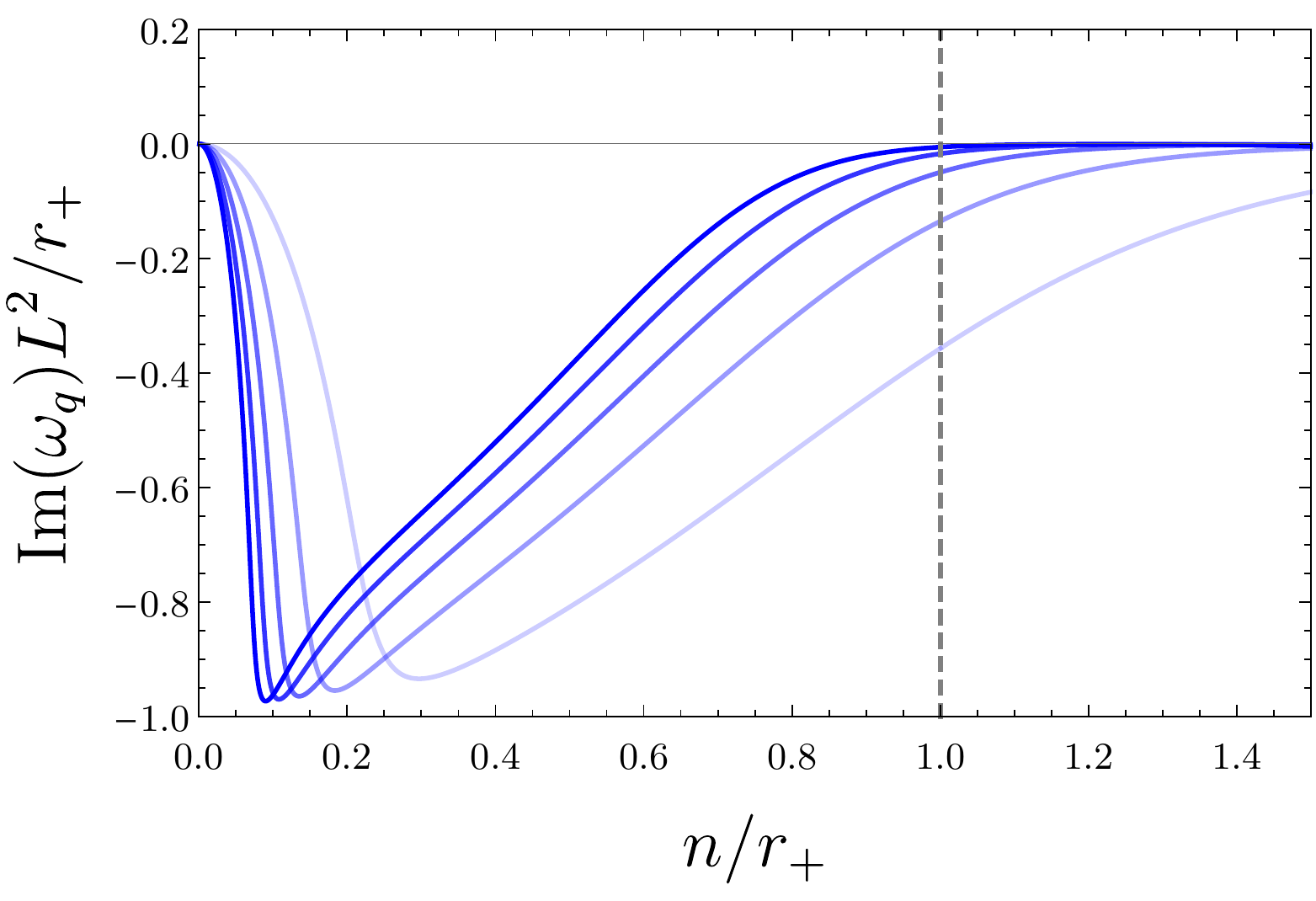}
		\caption{Pseudo-hydrodynamic quasinormal modes extended for $n/r_{+}>1$ for the levels $q=0,...,4$. There is nothing special at the point $n=r_{+}$ and the modes keep on being stable beyond it. However, their imaginary parts become exponentially small as we increase $n$.}
		\label{fig:grav3}
	\end{center}
\end{figure}

Finally, although we have focused on the case $|\epsilon|\le 1$ because it is the relevant one for holography, one may wonder what happens if we take even larger values of the NUT charge $|\epsilon|\ge 1$. In fact, since those solutions do not posses an Euclidean continuation, one may think that they could be unstable. In Fig.~\ref{fig:grav3}, we show the lowest gravitational QNM for a few values of $q$ as a function of $\epsilon=n/r_{+}$, extended  beyond $\epsilon=1$. We observe nothing special going on at that point, and in fact, the modes keep on being stable as we increase $\epsilon$. 
Nevertheless, Fig.~\ref{fig:grav3} shows that Taub-NUT solutions with increasingly large NUT charge have more quasinormal modes with extremely small imaginary parts, and it would be interesting to study if this could eventually give rise to a non-trivial instability when nonlinearities are taken into account.

\section{Conclusions}\label{sec:conc} 
We have performed a thorough analysis of the quasinormal modes of the planar Taub-NUT spacetimes given by \req{eq:NUTBB}. As we discussed, these describe the linear response to perturbations of a strongly-coupled plasma placed in the geometry \req{bdrymetric}, corresponding to a G\"odel-type universe with closed time-like curves. 

Our analysis revealed that QNMs in this background organize analogously to the Landau levels of a charged particle in a uniform magnetic field. 
Thus, unlike in the case of planar black holes, the spectrum of QNM frequencies is discrete and labeled by a unique quantum number $q$ (the Landau level). On the other hand, the QNMs are infinitely degenerate in the momentum $k$ along the isometric direction, which we chose to be $y$. 
Another novel aspect introduced by the NUT charge is that all the reflection symmetries of the spacetime are broken, which implies that one cannot decompose the perturbations of fields with spin into modes of definite parity. This leads to the appearance of an additional ``polarization parameter'' $\lambda_{+2}$ characterizing the gravitational QNMs. This parameter has to be determined together with the corresponding QNM frequency $\omega$ by solving simultaneously the equations for the two NP variables  $\Psi_0$ and $\Psi_4$. By using the Teukolsky-Starobinsky identities, we have been able to determine this parameter, which has two admissible values $\lambda_{+2}^{(\pm)}$ -- see Eq.~\req{eq:lambda2sol}. In the limit of vanishing NUT charge, these values give rise to modes with odd and even parity in the background of the black brane. Then, the boundary conditions for each of the NP variables are fully determined and it is enough to solve the radial equation for one of them to find the QNMs.
Finally, despite parity violation, we found that the spectrum of gravitational QNM frequencies is symmetric under the change of sign of the NUT charge. In addition, there is a conjugation symmetry that relates the positive-frequency modes of the solution with charge $n$ to the negative-frequency ones of the solution with charge $-n$, and vice-versa --- see Eq.~\req{eq:QNM-n}.

In the case of electromagnetic perturbations we have shown that a similar method does not work, since the equations for the NP variables $\phi_0$ and $\phi_2$ are degenerate. Thus the corresponding polarization parameter $\lambda_{+1}$ cannot be determined in this way or by parity arguments. This may lead to the conclusion that the spectrum of QNMs depends continuously on this parameter, but this issue certainly deserves further research. Perhaps analyzing the perturbations in terms of the vector field rather than in terms of the Newman-Penrose variables could shed light on this problem. 

Our numerical results on the scalar and gravitational QNM frequencies show that all of them lie in the lower half of the complex plane, and hence no instabilities are found despite the exotic causal structure of these spacetimes. Thus, this constitutes yet another step into the rehabilitation of Lorentzian spacetimes with NUT charge, in line with Refs.~\cite{Clement:2015cxa,Clement:2015aka,Kubiznak:2019yiu,Bordo:2019tyh,Bordo:2019rhu}. 
If we now apply the AdS/CFT correspondence, this result tells us not only that one should be able to perform quantum field theory in the background of the causality-violating metric \req{bdrymetric}, but that it should be possible to obtain sensible answers. 
Hence, it would now be interesting to perform a direct QFT computation in \req{bdrymetric} to try to reproduce the results obtained from holography. In particular, we managed to obtain an analytic expression for the pseudo-hydrodynamic modes in the limit of small NUT charge --- see Eq.~\req{omegasmall2}. As we have shown, that result generalizes the standard dispersion relation for the sound mode in  flat space to the case of the background \req{bdrymetric} when $n/L^2<<T$. It would be extremely interesting to attempt a derivation of that relation by studying the perturbations of a fluid in such background. 

Let us close our paper by commenting on other directions that should be considered. As we already mentioned, one should try to understand better the properties of electromagnetic QNMs. On the other hand, we have focused mainly on the scalar and gravitational modes with lowest imaginary part, but it would be interesting to complete the classification of QNMs by analyzing the overtone structure and the highly damped modes. In addition, even though we have provided compelling numerical evidence that no unstable QNMs exist, it would be important to offer a mathematical proof of this fact. 
Finally, it would also be worth extending these results to the case of Taub-NUT solutions of different topologies --- the spherical case is particularly interesting due to the interplay with the Misner string \cite{Kalamakis:2020aaj} --- or to higher dimensions. Hopefully these will offer further insight on the role of NUT charge in the AdS/CFT correspondence.

\section*{Acknowledgements}
We are glad to thank Vitor Cardoso and Karl Landsteiner for insightful discussions and comments.  The work of PAC is supported by a postdoctoral fellowship from the Research Foundation - Flanders (FWO grant 12ZH121N). DP is funded by a “Centro de Excelencia Internacional UAM/CSIC” FPI pre-doctoral grant

\appendix

\section{Asymptotic form of the metric perturbation}\label{app:metric}
As we have seen, the metric perturbation satisfying Dirichlet boundary conditions can be written near the boundary as

\begin{equation}
h_{ab}=ze^{-i\omega t}\gamma_{ab}(x)+\mathcal{O}(z^3)\, ,
\end{equation}
where we are already setting $k=0$ without loss of generality. The equations of motion allow one to express the component $\gamma_{xx}$ in terms of the rest as 
\begin{equation}
\gamma_{xx}=\left(1-\frac{4n^2x^2}{L^4}\right)\gamma_{tt}+\frac{4nx}{L^2}\gamma_{ty}-\gamma_{yy}. 
\end{equation}
Then, it is convenient to introduce a new matrix $\sigma_{ab}$ as follows $\gamma_{ab}=e^{-sn\omega x^2/L^2} \sigma_{ab}$. One finds that the equations of motion together with the separability conditions on the NP variables imply that $\sigma_{ab}$ is given by a finite sum of Hermite polynomials. In the case $s=1$ it reads

\begin{align}
\sigma_{tt}&=-\frac{10 a_{+2} L^2}{3r_{+} \hat\omega \epsilon}
 H_{q+2}(\hat x) \left(-(2 q+7) \hat\omega \epsilon +\hat\omega^2+i \lambda_{+2}  \hat\omega-2 \epsilon  (\epsilon -i \lambda_{+2} )\right)\, ,\\\notag
 \sigma_{tx}&=-\frac{5 i a_{+2} L^2 }{3 \sqrt{2}r_{+}  \left(\hat{\omega } \epsilon^3 \right)^{1/2}} \Big[\epsilon 
   H_{q_{-2}+3}\left(\hat{x}\right) \left(-i \lambda _{+2}-\hat{\omega }+\epsilon
   \right)\\
   &+2 H_{q_{-2}+1}\left(\hat{x}\right) \left(i \hat{\omega } \lambda
   _{+2}+\left(q_{-2}+1\right) \epsilon  \left(\epsilon -i \lambda
   _{+2}\right)-\left(3 q_{-2}+10\right) \hat{\omega } \epsilon +\hat{\omega
   }^2\right)\Big]\, ,\\\notag
   \sigma_{ty}&=\frac{5 a_{+2} L^2}{3 \sqrt{2}r_{+} 
   \left(\hat{\omega } \epsilon \right)^{3/2}}
 \Big[2 H_{q_{-2}+1}\left(\hat{x}\right) \left(4
   \left(q_{-2}+2\right) \epsilon ^2 \left(\epsilon -i \lambda
   _{+2}\right)+\hat{\omega }^2 \left(\left(-5 q_{-2}-14\right) \epsilon +i \lambda
   _{+2}\right)\right.\\\notag
   &\left.+\hat{\omega } \epsilon  \left(\left(4 q_{-2}^2+23 q_{-2}+29\right)
   \epsilon -i \left(3 q_{-2}+5\right) \lambda _{+2}\right)+\hat{\omega
   }^3\right)\\
   &+\epsilon  H_{q_{-2}+3}\left(\hat{x}\right) \left(-i \hat{\omega }
   \lambda _{+2}+4 \epsilon  \left(\epsilon -i \lambda _{+2}\right)+\left(4
   q_{-2}+13\right) \hat{\omega } \epsilon -\hat{\omega }^2\right)\Big]\, ,\\\notag
   \sigma_{xy}&=\frac{5 i a_{+2} L^2}{6 r_{+} \hat{\omega } \epsilon ^2} \Big[2 H_{q_{-2}}\left(\hat{x}\right) \left(-2 q_{-2}
   \left(q_{-2}+2\right) \epsilon ^2 \left(\epsilon -i \lambda
   _{+2}\right)+\hat{\omega }^2 \left(-2 \left(3 q_{-2}+8\right) \epsilon +i \lambda
   _{+2}\right)\right.\\\notag
   &\left.+\hat{\omega } \epsilon  \left(\left(8 q_{-2}^2+44 q_{-2}+55\right)
   \epsilon -i \left(4 q_{-2}+7\right) \lambda _{+2}\right)+\hat{\omega
   }^3\right)-\epsilon^2  \left(\epsilon -i \lambda _{+2}\right)
   H_{q_{-2}+4}\left(\hat{x}\right)\\
   & -2\epsilon  H_{q_{-2}+2}\left(\hat{x}\right) \left(i \hat{\omega
   } \lambda _{+2}+2 \left(q_{-2}+2\right) \epsilon  \left(\epsilon -i \lambda
   _{+2}\right)-\left(4 q_{-2}+13\right) \hat{\omega } \epsilon +\hat{\omega
   }^2\right)\Big]\, ,\\
   \sigma_{yy}&=-\frac{5 a_{+2} L^2}{6 r_{+}  \hat{\omega }^2 \epsilon ^2} \Big[2 H_{q_{-2}}(x) \left(-2 \hat{\omega } \epsilon ^2
   \left(\left(4 q_{-2}^3+28 q_{-2}^2+54 q_{-2}+29\right) \epsilon -i \left(4
   q_{-2}^2+10 q_{-2}+5\right) \lambda _{+2}\right)\right.\\\notag
   &\left.-8 \left(q_{-2}^2+3 q_{-2}+2\right)
   \epsilon ^3 \left(\epsilon -i \lambda _{+2}\right)+\hat{\omega }^3 \left(-2 \left(4
   q_{-2}+9\right) \epsilon +i \lambda _{+2}\right)\right.\\\notag
   &\left.   +\hat{\omega }^2 \epsilon 
   \left(\left(18 q_{-2}^2+82 q_{-2}+83\right) \epsilon -3 i \left(2 q_{-2}+3\right)
   \lambda _{+2}\right)+\hat{\omega }^4\right)\\\notag
   &+  \epsilon^2 
   H_{q_{-2}+4}(x) \left(-4 \epsilon  \left(\epsilon -i \lambda _{+2}\right)-4
   \left(q_{-2}+3\right) \hat{\omega } \epsilon +\hat{\omega }^2\right)\\
   &-2\epsilon
   H_{q_{-2}+2}(x) \left(\hat{\omega }-2 \left(2 q_{-2}+5\right) \epsilon \right)
   \left(i \hat{\omega } \lambda _{+2}-2 \epsilon  \left(\epsilon -i \lambda
   _{+2}\right)-\left(2 q_{-2}+7\right) \hat{\omega } \epsilon +\hat{\omega
   }^2\right)\Big]\, ,
    \end{align}
while for $s=-1$ the solution is 

\begin{align}
\sigma_{tt}&=\frac{17 a_{+2} L^2 H_{q_{+2}+2}\left(\hat{x}\right) \left(i \hat{\omega } \lambda
   _{+2}+\hat{\omega } \epsilon  \left(2 q_{+2}+3\right)-2 \epsilon  \left(\epsilon -i
   \lambda _{+2}\right)+\hat{\omega }^2\right)}{6 r_{+} \hat{\omega } \epsilon 
   \left(q_{+2}+1\right) \left(q_{+2}+2\right)}\, ,\\\notag
\sigma_{tx}&=\frac{17 i a_{+2} L^2 \hat{\omega }}{12 \sqrt{2} r_{+} \left(q_{+2}+1\right) \left(q_{+2}+2\right)
   \left(q_{+2}+3\right) \left(-\hat{\omega }\epsilon\right)^{3/2}} \Big[2 \epsilon  \left(q_{+2}^2+5
   q_{+2}+6\right) \left(i \lambda _{+2}+\hat{\omega
   }-\epsilon \right)\\
   &\times H_{q_{+2}+1}\left(\hat{x}\right)+H_{q_{+2}+3}\left(\hat{x}\right) \left(i \hat{\omega } \lambda
   _{+2}-\epsilon  \left(q_{+2}+4\right) \left(\epsilon -i \lambda
   _{+2}\right)+\hat{\omega } \epsilon  \left(3 q_{+2}+5\right)+\hat{\omega
   }^2\right)\Big]\, ,\\\notag
\sigma_{ty}&=-\frac{17 a_{+2} L^2}{12 \sqrt{2} r_{+}
   \left(q_{+2}+1\right) \left(q_{+2}+2\right) \left(q_{+2}+3\right) \left(-\hat{\omega
   } \epsilon \right)^{3/2}} \Big[H_{q_{+2}+3}\left(\hat{x}\right) \left(-4 \epsilon ^2
   \left(q_{+2}+3\right) \left(\epsilon -i \lambda _{+2}\right)\right.\\\notag
   &\left.+\hat{\omega }^2
   \left(\epsilon  \left(5 q_{+2}+11\right)+i \lambda _{+2}\right)+\hat{\omega }
   \epsilon  \left(\epsilon  \left(4 q_{+2}^2+17 q_{+2}+14\right)+i \lambda _{+2}
   \left(3 q_{+2}+10\right)\right)+\hat{\omega }^3\right)\\
   &+2 \epsilon  \left(q_{+2}^2+5
   q_{+2}+6\right) H_{q_{+2}+1}\left(\hat{x}\right) \left(i \hat{\omega } \lambda
   _{+2}+\hat{\omega } \epsilon  \left(4 q_{+2}+7\right)-4 \epsilon  \left(\epsilon -i
   \lambda _{+2}\right)+\hat{\omega }^2\right)\Big]\, ,\\\notag
\sigma_{xy}&=\frac{17 i a_{+2} L^2}{48 r_{+} \hat{\omega } \epsilon ^2 \left(q_{+2}+1\right)
   \left(q_{+2}+2\right) \left(q_{+2}+3\right) \left(q_{+2}+4\right)}    \Big[H_{q_{+2}+4}\left(\hat{x}\right) \left(-2 \epsilon ^2
   \left(q_{+2}^2+8 q_{+2}+15\right)\right.\\\notag
   &\left.\times \left(\epsilon -i \lambda
   _{+2}\right)+\hat{\omega }^2 \left(2 \epsilon  \left(3 q_{+2}+7\right)+i \lambda
   _{+2}\right)+\hat{\omega } \epsilon  \left(\epsilon  \left(8 q_{+2}^2+36
   q_{+2}+35\right)+i \lambda _{+2} \left(4 q_{+2}+13\right)\right)\right.\\\notag
   &\left.+\hat{\omega
   }^3\right)-8 \epsilon ^2 \left(q_{+2}^4+10 q_{+2}^3+35 q_{+2}^2+50 q_{+2}+24\right)
   \left(\epsilon -i \lambda _{+2}\right) H_{q_{+2}}\left(\hat{x}\right)\\
   &+4 \epsilon 
   \left(q_{+2}^2+7 q_{+2}+12\right) H_{q_{+2}+2}\left(\hat{x}\right) \left(i
   \hat{\omega } \lambda _{+2}-2 \epsilon  \left(q_{+2}+3\right) \left(\epsilon -i
   \lambda _{+2}\right)+\hat{\omega } \epsilon  \left(4 q_{+2}+7\right)+\hat{\omega
   }^2\right)\Big]\, ,\\\notag
\sigma_{yy}&=\frac{17a_{+2} L^2}{48 r_{+} \hat{\omega }^2} \Bigg[-\frac{H_{q_{+2}+4}\left(\hat{x}\right)}{\epsilon ^2 \left(q_{+2}+1\right) \left(q_{+2}+2\right)
   \left(q_{+2}+3\right) \left(q_{+2}+4\right)} \Big(\\\notag
   &2
   \hat{\omega } \epsilon ^2 \left(\epsilon  \left(4 q_{+2}^3+32 q_{+2}^2+74
   q_{+2}+41\right)+i \lambda _{+2} \left(4 q_{+2}^2+30 q_{+2}+55\right)\right)\\\notag
   &-8
   \epsilon ^3 \left(q_{+2}^2+7 q_{+2}+12\right) \left(\epsilon -i \lambda
   _{+2}\right)+\hat{\omega }^3 \left(\epsilon  \left(8 q_{+2}+22\right)+i \lambda
   _{+2}\right)\\\notag
   &+\hat{\omega }^2 \epsilon  \left(\epsilon  \left(18 q_{+2}^2+98
   q_{+2}+123\right)+3 i \lambda _{+2} \left(2 q_{+2}+7\right)\right)+\hat{\omega
   }^4\Big)\\\notag
   &-\frac{4
   H_{q_{+2}+2}\left(\hat{x}\right) \left(2 \epsilon  \left(2
   q_{+2}+5\right)+\hat{\omega }\right) \left(i \hat{\omega } \lambda
   _{+2}+\hat{\omega } \epsilon  \left(2 q_{+2}+3\right)-2 \epsilon  \left(\epsilon -i
   \lambda _{+2}\right)+\hat{\omega }^2\right)}{\epsilon  \left(q_{+2}+1\right)
   \left(q_{+2}+2\right)}\\
   &-8 H_{q_{+2}}\left(\hat{x}\right)
   \left(4 \hat{\omega } \epsilon  \left(q_{+2}+2\right)-4 \epsilon  \left(\epsilon -i
   \lambda _{+2}\right)+\hat{\omega }^2\right)\Bigg]\, ,
\end{align}

\noindent    
where in each case $\hat x=x \sqrt{\frac{2s n \omega }{L^2}}$.

\section{Boundary conditions from Hertz's reconstruction map}\label{app:Hertz}

A priori, it is not clear to which extent the Weyl scalars $\Psi_{0}$ and $\Psi_{4}$ encode all the information of a metric perturbation. Rather remarkably, though, once solutions for certain decoupled equations (in a specific sense) are known, there is an elegant procedure to reconstruct the whole perturbation. The ``master variables'' satisfying such equations are referred to as the \textit{Hertz potentials}. This was applied to perturbations of vacuum type-D spaces in \cite{PhysRevD.19.1641} and \cite{PhysRevD.11.2042}. The results in those references were proven in a more systematic and surprisingly simple form in \cite{PhysRevLett.41.203}. In the context of holography, this has proven to be very useful, particularly in the derivation of physical boundary conditions for perturbations in AdS space \cite{Dias:2013sdc} (see also \cite{Dias:2009ex,Cardoso:2013pza}). In this appendix we rederive our boundary conditions by explicit application of Hertz's reconstruction map.

In our type-D space a complex metric perturbation in a general polarisation state can be written as 
\begin{align}\label{metricpert}\notag
h_{\mu\nu}=& \left\{-k_{\mu}k_{\nu}\bar{\delta}\bar{\delta}-\bar{m}_{\mu}\bar{m}_{\nu}(D-\bar{\rho})(D+3\bar{\rho})+k_{(\mu}\bar{m}_{\nu)}\left[(D-\bar{\rho}+\rho)\bar{\delta}+\bar{\delta}(D+3\bar{\rho})\right]\right\}\bar{\varphi}^{IRG}\\ \notag
&+\{-l_{\mu}l_{\nu}\delta \delta-m_{\mu}m_{\nu}(\Delta-3\bar{\gamma}+\gamma+\bar{\mu})(\Delta -4\bar{\gamma}-3\bar{\mu})\\ 
&+l_{(\mu}m_{\nu)}\left[\delta(\Delta-4\bar{\gamma}-3\bar{\mu})+(\Delta -3\bar{\gamma}-\gamma+\bar{\mu}-\mu)\delta\right]\}\bar{\varphi}^{ORG}
\end{align} 
where ${\varphi}^{IRG}$ and ${\varphi}^{ORG}$ are the Hertz potentials of perturbations in traceless, ingoing ($h^{IRG}_{\mu\nu}k^{\mu}=0$) and outgoing ($h^{ORG}_{\mu\nu}l^{\mu}=0$) radiation gauge respectively, and satisfy the equations $\mathcal{O}_{0}^{\dagger}({\varphi}^{IRG})=0$ and $\mathcal{O}_{4}^{\dagger}({\varphi}^{ORG})=0$, where $\mathcal{O}_{0}$ and $\mathcal{O}_{4}$ are Teukolsky's operators and $\dagger$ denotes the operation of taking the adjoint, as defined in \cite{PhysRevLett.41.203}. Following the lines of  \cite{PhysRevD.11.2042}, we have taken $h_{\mu\nu}^{IRG}=2 \overline{\left[{S_{0}}^{\dagger}_{\mu\nu}\varphi^{IRG}\right]}$ and $h_{\mu\nu}^{ORG}=2 \overline{\left[{S_{4}}^{\dagger}_{\mu\nu}\varphi^{ORG}\right]}$. Here, $S_{0}$ and $S_{4}$ are defined by the identities $\mathcal{O}_{0}T_{0}(h)=S^{\mu\nu}_{0}\mathcal{E}_{\mu\nu}(h)$ and $\mathcal{O}_{4}T_{4}(h)=S^{\mu\nu}_{4}\mathcal{E}_{\mu\nu}(h)$ where $\mathcal{E}_{\mu\nu}$ is the linearised Einstein equation and $T_{0}$ and $T_{4}$ the operators that compute $\Psi_{0}$ and $\Psi_{4}$ out of $h_{\mu\nu}$, respectively (it is now clear, by the property $(AB)^{\dagger}=B^{\dagger}A^{\dagger}$ of composition of adjoints and the self-adjoint property $\mathcal{E}_{\mu\nu}^{\dagger}=\mathcal{E}_{\mu\nu}$, that a solution $\varphi$ of $\mathcal{O}^{\dagger}_{0}(\varphi)=0$ generates a solution $h_{\mu\nu}= {S_{0}}^{\dagger}_{\mu\nu}\varphi$ of $\mathcal{E}_{\mu\nu}(h)=0$, and similarly for $\mathcal{O}_{4}^{\dagger}$ and ${S_{4}}^{\dagger}_{\mu\nu}$). Solutions for $\varphi^{IRG}$ and $\varphi^{ORG}$ can be readily obtained by noticing the properties $\mathcal{O}_{0}^{\dagger}(\varphi)=\Psi_{2}^{-4/3}\mathcal{O}_{4}(\Psi_{2}^{4/3}\varphi)$ and $\mathcal{O}_{4}^{\dagger}(\varphi)=\Psi_{2}^{-4/3}\mathcal{O}_{0}(\Psi_{2}^{4/3}\varphi)$, and take the form
\begin{align}
\bar{\varphi}^{IRG}&=\Delta R^{\omega,q-4}_{(-2)}(r)\mathcal{H}_{q}(x)e^{-i\omega t}e^{iky}\\
\bar{\varphi}^{ORG}&=\bar{\Psi}_{2}^{-4/3}\frac{R^{\omega,q}_{(+2)}(r)}{\Delta}\mathcal{H}_{q-4}(x)e^{-i\omega t}e^{iky}
\end{align}
The radial functions $R^{\omega,q}_{(+2)}$ and $R^{\omega,q-4}_{(-2)}$ are solutions of \eqref{eq:R+eq} and \eqref{eq:R+eq'}, respectively, and we chose them to be related by the Teukolsky--Starobinsky identities \eqref{eq:TSI1} and \eqref{eq:TSI2}. Also, we recall that these radial functions are related to the $Y_{\pm2}$ variables according to \req{eq:radialvarR}. In addition, the angular functions $\mathcal{H}_{q}(x)$ are the solutions given in \eqref{eq:solang}. With this, it can be verified by direct application of $T_{0}$ and $T_{4}$ on \eqref{metricpert} that
\begin{align}
\Psi_{0}&=A_{(+2)}\frac{R^{\omega,q}_{(+2)}}{\Delta}\mathcal{H}_{q}(x)e^{-i\omega t}e^{iky} \\
\Psi_{4}&=A_{(-2)}\frac{\Delta R_{(-2)}^{\omega,q-4}}{(r+in)^{4}}\mathcal{H}_{q-4}(x)e^{-i\omega t}e^{iky}
\end{align}
where the constants $A_{(\pm2)}$ are not important for this discussion.

In order to determine the boundary conditions, we perform an asymptotic expansion of $R^{\omega,q}_{(+2)}$ and $R_{(-2)}^{\omega,q-4}$ near infinity, which follows that of the $Y_{\pm2}$ functions in \req{eq:Y2as} and is determined by the constants $a_{\pm2}$ and $b_{\pm2}$.
The boundary conditions are most conveniently identified by working in a gauge
\begin{equation}
\tilde{h}_{\mu\nu}=h_{\mu\nu}+2\nabla_{(\mu}\xi_{\nu)}
\end{equation} 
with $\tilde{h}_{r\mu}=0$. This can be achieved by expanding the gauge parameter as
\begin{equation}
\xi_{\mu}=e^{-i\omega t+i k y}r^2\sum_{i=0}^{\infty}r^{-i}(f_t^{(i)}(x),f_r^{(i)}(x)/r^2,f_x^{(i)}(x),f_y^{(i)}(x))\, ,
\end{equation}
which allows us to cancel as many $1/r^i$ terms in $\tilde{h}_{r\mu}$ as we want by choosing the functions $f_{\mu}^{(i)}(x)$ appropriately. Then, the resulting metric perturbation $\tilde{h}_{ab}$ typically contains terms that diverge as $r^2$, which should be removed according to the holographic boundary conditions in \eqref{eq:hab}. Some of these can be canceled with additional gauge transformations, but ultimately we find a constraint between the asymptotic expansions of $R^{\omega,q}_{(+2)}$ and $R_{(-2)}^{\omega,q-4}$ at $r\to \infty$, which establishes a relation between the constants $(a_{+2},b_{+2})$ and $(a_{-2},b_{-2})$. This, in turn, translates into a relation between the ratios of these quantities. $\lambda_{+2}$ and $\lambda_{-2}$ as defined in \eqref{rat}. 
On the other hand, the Teukolsky--Starobinsky identities \eqref{eq:TSI1} and \eqref{eq:TSI2} provide an additional relation involving $\lambda_{+2}$ and $\lambda_{-2}$ --- see \req{eq:lambda2rel2}. The solutions for $(\lambda_{+2},\lambda_{-2})$ of this pair of equations are precisely those given in \eqref{eq:lambda2sol} and \eqref{eq:lambdam2sol}.

\bibliographystyle{JHEP}
\bibliography{Gravities}

\newcommand{\noop}[1]{}

\providecommand{\href}[2]{#2}\begingroup\raggedright\begin{thebibliography}{10}

\bibitem{Maldacena}
J.~M. Maldacena, \emph{{The Large N limit of superconformal field theories and
  supergravity}}, \href{https://doi.org/10.1023/A:1026654312961}{\emph{Int. J.
  Theor. Phys.} {\bfseries 38} (1999) 1113--1133},
  [\href{https://arxiv.org/abs/hep-th/9711200}{{\ttfamily hep-th/9711200}}].

\bibitem{Witten}
E.~Witten, \emph{{Anti-de Sitter space and holography}}, {\emph{Adv. Theor.
  Math. Phys.} {\bfseries 2} (1998) 253--291},
  [\href{https://arxiv.org/abs/hep-th/9802150}{{\ttfamily hep-th/9802150}}].

\bibitem{Gubser}
S.~S. Gubser, I.~R. Klebanov and A.~M. Polyakov, \emph{{Gauge theory
  correlators from noncritical string theory}},
  \href{https://doi.org/10.1016/S0370-2693(98)00377-3}{\emph{Phys. Lett.}
  {\bfseries B428} (1998) 105--114},
  [\href{https://arxiv.org/abs/hep-th/9802109}{{\ttfamily hep-th/9802109}}].

\bibitem{Son:2007vk}
D.~T. Son and A.~O. Starinets, \emph{{Viscosity, Black Holes, and Quantum Field
  Theory}},
  \href{https://doi.org/10.1146/annurev.nucl.57.090506.123120}{\emph{Ann. Rev.
  Nucl. Part. Sci.} {\bfseries 57} (2007) 95--118},
  [\href{https://arxiv.org/abs/0704.0240}{{\ttfamily 0704.0240}}].

\bibitem{Gubser:2009md}
S.~S. Gubser and A.~Karch, \emph{{From gauge-string duality to strong
  interactions: A Pedestrian's Guide}},
  \href{https://doi.org/10.1146/annurev.nucl.010909.083602}{\emph{Ann. Rev.
  Nucl. Part. Sci.} {\bfseries 59} (2009) 145--168},
  [\href{https://arxiv.org/abs/0901.0935}{{\ttfamily 0901.0935}}].

\bibitem{Hartnoll:2009sz}
S.~A. Hartnoll, \emph{{Lectures on holographic methods for condensed matter
  physics}}, \href{https://doi.org/10.1088/0264-9381/26/22/224002}{\emph{Class.
  Quant. Grav.} {\bfseries 26} (2009) 224002},
  [\href{https://arxiv.org/abs/0903.3246}{{\ttfamily 0903.3246}}].

\bibitem{Herzog:2009xv}
C.~P. Herzog, \emph{{Lectures on Holographic Superfluidity and
  Superconductivity}},
  \href{https://doi.org/10.1088/1751-8113/42/34/343001}{\emph{J. Phys. A}
  {\bfseries 42} (2009) 343001},
  [\href{https://arxiv.org/abs/0904.1975}{{\ttfamily 0904.1975}}].

\bibitem{Berti:2009kk}
E.~Berti, V.~Cardoso and A.~O. Starinets, \emph{{Quasinormal modes of black
  holes and black branes}},
  \href{https://doi.org/10.1088/0264-9381/26/16/163001}{\emph{Class. Quant.
  Grav.} {\bfseries 26} (2009) 163001},
  [\href{https://arxiv.org/abs/0905.2975}{{\ttfamily 0905.2975}}].

\bibitem{Konoplya:2011qq}
R.~Konoplya and A.~Zhidenko, \emph{{Quasinormal modes of black holes: From
  astrophysics to string theory}},
  \href{https://doi.org/10.1103/RevModPhys.83.793}{\emph{Rev. Mod. Phys.}
  {\bfseries 83} (2011) 793--836},
  [\href{https://arxiv.org/abs/1102.4014}{{\ttfamily 1102.4014}}].

\bibitem{Birmingham:2001pj}
D.~Birmingham, I.~Sachs and S.~N. Solodukhin, \emph{{Conformal field theory
  interpretation of black hole quasinormal modes}},
  \href{https://doi.org/10.1103/PhysRevLett.88.151301}{\emph{Phys. Rev. Lett.}
  {\bfseries 88} (2002) 151301},
  [\href{https://arxiv.org/abs/hep-th/0112055}{{\ttfamily hep-th/0112055}}].

\bibitem{Son:2002sd}
D.~T. Son and A.~O. Starinets, \emph{{Minkowski space correlators in AdS / CFT
  correspondence: Recipe and applications}},
  \href{https://doi.org/10.1088/1126-6708/2002/09/042}{\emph{JHEP} {\bfseries
  09} (2002) 042}, [\href{https://arxiv.org/abs/hep-th/0205051}{{\ttfamily
  hep-th/0205051}}].

\bibitem{Starinets:2002br}
A.~O. Starinets, \emph{{Quasinormal modes of near extremal black branes}},
  \href{https://doi.org/10.1103/PhysRevD.66.124013}{\emph{Phys. Rev. D}
  {\bfseries 66} (2002) 124013},
  [\href{https://arxiv.org/abs/hep-th/0207133}{{\ttfamily hep-th/0207133}}].

\bibitem{Policastro:2002se}
G.~Policastro, D.~T. Son and A.~O. Starinets, \emph{{From AdS / CFT
  correspondence to hydrodynamics}},
  \href{https://doi.org/10.1088/1126-6708/2002/09/043}{\emph{JHEP} {\bfseries
  09} (2002) 043}, [\href{https://arxiv.org/abs/hep-th/0205052}{{\ttfamily
  hep-th/0205052}}].

\bibitem{Herzog:2002fn}
C.~P. Herzog, \emph{{The Hydrodynamics of M theory}},
  \href{https://doi.org/10.1088/1126-6708/2002/12/026}{\emph{JHEP} {\bfseries
  12} (2002) 026}, [\href{https://arxiv.org/abs/hep-th/0210126}{{\ttfamily
  hep-th/0210126}}].

\bibitem{Nunez:2003eq}
A.~Nunez and A.~O. Starinets, \emph{{AdS / CFT correspondence, quasinormal
  modes, and thermal correlators in N=4 SYM}},
  \href{https://doi.org/10.1103/PhysRevD.67.124013}{\emph{Phys. Rev. D}
  {\bfseries 67} (2003) 124013},
  [\href{https://arxiv.org/abs/hep-th/0302026}{{\ttfamily hep-th/0302026}}].

\bibitem{Kovtun:2005ev}
P.~K. Kovtun and A.~O. Starinets, \emph{{Quasinormal modes and holography}},
  \href{https://doi.org/10.1103/PhysRevD.72.086009}{\emph{Phys. Rev. D}
  {\bfseries 72} (2005) 086009},
  [\href{https://arxiv.org/abs/hep-th/0506184}{{\ttfamily hep-th/0506184}}].

\bibitem{Baier:2007ix}
R.~Baier, P.~Romatschke, D.~T. Son, A.~O. Starinets and M.~A. Stephanov,
  \emph{{Relativistic viscous hydrodynamics, conformal invariance, and
  holography}},
  \href{https://doi.org/10.1088/1126-6708/2008/04/100}{\emph{JHEP} {\bfseries
  04} (2008) 100}, [\href{https://arxiv.org/abs/0712.2451}{{\ttfamily
  0712.2451}}].

\bibitem{Aharony:2008ug}
O.~Aharony, O.~Bergman, D.~L. Jafferis and J.~Maldacena, \emph{{N=6
  superconformal Chern-Simons-matter theories, M2-branes and their gravity
  duals}}, \href{https://doi.org/10.1088/1126-6708/2008/10/091}{\emph{JHEP}
  {\bfseries 10} (2008) 091},
  [\href{https://arxiv.org/abs/0806.1218}{{\ttfamily 0806.1218}}].

\bibitem{Horowitz:1999jd}
G.~T. Horowitz and V.~E. Hubeny, \emph{{Quasinormal modes of AdS black holes
  and the approach to thermal equilibrium}},
  \href{https://doi.org/10.1103/PhysRevD.62.024027}{\emph{Phys. Rev. D}
  {\bfseries 62} (2000) 024027},
  [\href{https://arxiv.org/abs/hep-th/9909056}{{\ttfamily hep-th/9909056}}].

\bibitem{Cardoso:2001bb}
V.~Cardoso and J.~P. Lemos, \emph{{Quasinormal modes of Schwarzschild anti-de
  Sitter black holes: Electromagnetic and gravitational perturbations}},
  \href{https://doi.org/10.1103/PhysRevD.64.084017}{\emph{Phys. Rev. D}
  {\bfseries 64} (2001) 084017},
  [\href{https://arxiv.org/abs/gr-qc/0105103}{{\ttfamily gr-qc/0105103}}].

\bibitem{Cardoso:2003cj}
V.~Cardoso, R.~Konoplya and J.~P. Lemos, \emph{{Quasinormal frequencies of
  Schwarzschild black holes in anti-de Sitter space-times: A Complete study on
  the asymptotic behavior}},
  \href{https://doi.org/10.1103/PhysRevD.68.044024}{\emph{Phys. Rev. D}
  {\bfseries 68} (2003) 044024},
  [\href{https://arxiv.org/abs/gr-qc/0305037}{{\ttfamily gr-qc/0305037}}].

\bibitem{Musiri:2003rs}
S.~Musiri and G.~Siopsis, \emph{{Asymptotic form of quasinormal modes of large
  AdS black holes}},
  \href{https://doi.org/10.1016/j.physletb.2003.10.015}{\emph{Phys. Lett. B}
  {\bfseries 576} (2003) 309--313},
  [\href{https://arxiv.org/abs/hep-th/0308196}{{\ttfamily hep-th/0308196}}].

\bibitem{Cardoso:2001vs}
V.~Cardoso and J.~P. Lemos, \emph{{Quasinormal modes of toroidal, cylindrical
  and planar black holes in anti-de Sitter space-times}},
  \href{https://doi.org/10.1088/0264-9381/18/23/319}{\emph{Class. Quant. Grav.}
  {\bfseries 18} (2001) 5257--5267},
  [\href{https://arxiv.org/abs/gr-qc/0107098}{{\ttfamily gr-qc/0107098}}].

\bibitem{Miranda:2005qx}
A.~S. Miranda and V.~T. Zanchin, \emph{{Quasinormal modes of plane-symmetric
  anti-de Sitter black holes: A Complete analysis of the gravitational
  perturbations}},
  \href{https://doi.org/10.1103/PhysRevD.73.064034}{\emph{Phys. Rev. D}
  {\bfseries 73} (2006) 064034},
  [\href{https://arxiv.org/abs/gr-qc/0510066}{{\ttfamily gr-qc/0510066}}].

\bibitem{Miranda:2008vb}
A.~S. Miranda, J.~Morgan and V.~T. Zanchin, \emph{{Quasinormal modes of
  plane-symmetric black holes according to the AdS/CFT correspondence}},
  \href{https://doi.org/10.1088/1126-6708/2008/11/030}{\emph{JHEP} {\bfseries
  11} (2008) 030}, [\href{https://arxiv.org/abs/0809.0297}{{\ttfamily
  0809.0297}}].

\bibitem{Giammatteo:2005vu}
M.~Giammatteo and I.~G. Moss, \emph{{Gravitational quasinormal modes for Kerr
  anti-de Sitter black holes}},
  \href{https://doi.org/10.1088/0264-9381/22/9/021}{\emph{Class. Quant. Grav.}
  {\bfseries 22} (2005) 1803--1824},
  [\href{https://arxiv.org/abs/gr-qc/0502046}{{\ttfamily gr-qc/0502046}}].

\bibitem{Taub:1950ez}
A.~Taub, \emph{{Empty space-times admitting a three parameter group of
  motions}}, \href{https://doi.org/10.2307/1969567}{\emph{Annals Math.}
  {\bfseries 53} (1951) 472--490}.

\bibitem{Newman:1963yy}
E.~Newman, L.~Tamburino and T.~Unti, \emph{{Empty space generalization of the
  Schwarzschild metric}}, \href{https://doi.org/10.1063/1.1704018}{\emph{J.
  Math. Phys.} {\bfseries 4} (1963) 915}.

\bibitem{Hawking:1998ct}
S.~Hawking, C.~Hunter and D.~N. Page, \emph{{Nut charge, anti-de Sitter space
  and entropy}}, \href{https://doi.org/10.1103/PhysRevD.59.044033}{\emph{Phys.
  Rev. D} {\bfseries 59} (1999) 044033},
  [\href{https://arxiv.org/abs/hep-th/9809035}{{\ttfamily hep-th/9809035}}].

\bibitem{Chamblin:1998pz}
A.~Chamblin, R.~Emparan, C.~V. Johnson and R.~C. Myers, \emph{{Large N phases,
  gravitational instantons and the nuts and bolts of AdS holography}},
  \href{https://doi.org/10.1103/PhysRevD.59.064010}{\emph{Phys. Rev.}
  {\bfseries D59} (1999) 064010},
  [\href{https://arxiv.org/abs/hep-th/9808177}{{\ttfamily hep-th/9808177}}].

\bibitem{Imamura:2011wg}
Y.~Imamura and D.~Yokoyama, \emph{{N=2 supersymmetric theories on squashed
  three-sphere}}, \href{https://doi.org/10.1103/PhysRevD.85.025015}{\emph{Phys.
  Rev.} {\bfseries D85} (2012) 025015},
  [\href{https://arxiv.org/abs/1109.4734}{{\ttfamily 1109.4734}}].

\bibitem{Martelli:2012sz}
D.~Martelli, A.~Passias and J.~Sparks, \emph{{The supersymmetric NUTs and bolts
  of holography}},
  \href{https://doi.org/10.1016/j.nuclphysb.2013.04.026}{\emph{Nucl. Phys. B}
  {\bfseries 876} (2013) 810--870},
  [\href{https://arxiv.org/abs/1212.4618}{{\ttfamily 1212.4618}}].

\bibitem{Bobev:2016sap}
N.~Bobev, T.~Hertog and Y.~Vreys, \emph{{The NUTs and Bolts of Squashed
  Holography}}, \href{https://doi.org/10.1007/JHEP11(2016)140}{\emph{JHEP}
  {\bfseries 11} (2016) 140},
  [\href{https://arxiv.org/abs/1610.01497}{{\ttfamily 1610.01497}}].

\bibitem{Bobev:2017asb}
N.~Bobev, P.~Bueno and Y.~Vreys, \emph{{Comments on Squashed-sphere Partition
  Functions}}, \href{https://doi.org/10.1007/JHEP07(2017)093}{\emph{JHEP}
  {\bfseries 07} (2017) 093},
  [\href{https://arxiv.org/abs/1705.00292}{{\ttfamily 1705.00292}}].

\bibitem{Bueno:2018yzo}
P.~Bueno, P.~A. Cano, R.~A. Hennigar and R.~B. Mann, \emph{{Universality of
  Squashed-Sphere Partition Functions}},
  \href{https://doi.org/10.1103/PhysRevLett.122.071602}{\emph{Phys. Rev. Lett.}
  {\bfseries 122} (2019) 071602},
  [\href{https://arxiv.org/abs/1808.02052}{{\ttfamily 1808.02052}}].

\bibitem{Bueno:2020odt}
P.~Bueno, P.~A. Cano, R.~A. Hennigar, V.~A. Penas and A.~Ruip\'erez,
  \emph{{Partition functions on slightly squashed spheres and flux
  parameters}}, \href{https://doi.org/10.1007/JHEP04(2020)123}{\emph{JHEP}
  {\bfseries 04} (2020) 123},
  [\href{https://arxiv.org/abs/2001.10020}{{\ttfamily 2001.10020}}].

\bibitem{Misner:1963fr}
C.~W. Misner, \emph{{The Flatter regions of Newman, Unti and Tamburino's
  generalized Schwarzschild space}},
  \href{https://doi.org/10.1063/1.1704019}{\emph{J. Math. Phys.} {\bfseries 4}
  (1963) 924--938}.

\bibitem{Manko:2005nm}
V.~Manko and E.~Ruiz, \emph{{Physical interpretation of NUT solution}},
  \href{https://doi.org/10.1088/0264-9381/22/17/014}{\emph{Class. Quant. Grav.}
  {\bfseries 22} (2005) 3555--3560},
  [\href{https://arxiv.org/abs/gr-qc/0505001}{{\ttfamily gr-qc/0505001}}].

\bibitem{Astefanesei:2004ji}
D.~Astefanesei, R.~B. Mann and E.~Radu, \emph{{Breakdown of the entropy/area
  relationship for NUT-charged spacetimes}},
  \href{https://doi.org/10.1016/j.physletb.2005.05.057}{\emph{Phys. Lett. B}
  {\bfseries 620} (2005) 1--8},
  [\href{https://arxiv.org/abs/hep-th/0406050}{{\ttfamily hep-th/0406050}}].

\bibitem{Clement:2015cxa}
G.~Cl\'ement, D.~Gal'tsov and M.~Guenouche, \emph{{Rehabilitating space-times
  with NUTs}},
  \href{https://doi.org/10.1016/j.physletb.2015.09.074}{\emph{Phys. Lett. B}
  {\bfseries 750} (2015) 591--594},
  [\href{https://arxiv.org/abs/1508.07622}{{\ttfamily 1508.07622}}].

\bibitem{Clement:2015aka}
G.~Cl\'ement, D.~Gal'tsov and M.~Guenouche, \emph{{NUT wormholes}},
  \href{https://doi.org/10.1103/PhysRevD.93.024048}{\emph{Phys. Rev. D}
  {\bfseries 93} (2016) 024048},
  [\href{https://arxiv.org/abs/1509.07854}{{\ttfamily 1509.07854}}].

\bibitem{Kubiznak:2019yiu}
R.~A. Hennigar, D.~Kubiz\v{n}\'ak and R.~B. Mann, \emph{{Thermodynamics of
  Lorentzian Taub-NUT spacetimes}},
  \href{https://doi.org/10.1103/PhysRevD.100.064055}{\emph{Phys. Rev. D}
  {\bfseries 100} (2019) 064055},
  [\href{https://arxiv.org/abs/1903.08668}{{\ttfamily 1903.08668}}].

\bibitem{Bordo:2019tyh}
A.~B. Bordo, F.~Gray, R.~A. Hennigar and D.~Kubiz\v{n}\'ak, \emph{{Misner
  Gravitational Charges and Variable String Strengths}},
  \href{https://doi.org/10.1088/1361-6382/ab3d4d}{\emph{Class. Quant. Grav.}
  {\bfseries 36} (2019) 194001},
  [\href{https://arxiv.org/abs/1905.03785}{{\ttfamily 1905.03785}}].

\bibitem{Bordo:2019rhu}
A.~Ballon~Bordo, F.~Gray, R.~A. Hennigar and D.~Kubiz\v{n}\'ak, \emph{{The
  First Law for Rotating NUTs}},
  \href{https://doi.org/10.1016/j.physletb.2019.134972}{\emph{Phys. Lett. B}
  {\bfseries 798} (2019) 134972},
  [\href{https://arxiv.org/abs/1905.06350}{{\ttfamily 1905.06350}}].

\bibitem{Ciambelli:2020qny}
L.~Ciambelli, C.~Corral, J.~Figueroa, G.~Giribet and R.~Olea,
  \emph{{Topological Terms and the Misner String Entropy}},
  \href{https://doi.org/10.1103/PhysRevD.103.024052}{\emph{Phys. Rev. D}
  {\bfseries 103} (2021) 024052},
  [\href{https://arxiv.org/abs/2011.11044}{{\ttfamily 2011.11044}}].

\bibitem{Leigh:2011au}
R.~G. Leigh, A.~C. Petkou and P.~Petropoulos, \emph{{Holographic
  Three-Dimensional Fluids with Nontrivial Vorticity}},
  \href{https://doi.org/10.1103/PhysRevD.85.086010}{\emph{Phys. Rev. D}
  {\bfseries 85} (2012) 086010},
  [\href{https://arxiv.org/abs/1108.1393}{{\ttfamily 1108.1393}}].

\bibitem{Leigh:2012jv}
R.~G. Leigh, A.~C. Petkou and P.~Petropoulos, \emph{{Holographic Fluids with
  Vorticity and Analogue Gravity}},
  \href{https://doi.org/10.1007/JHEP11(2012)121}{\emph{JHEP} {\bfseries 11}
  (2012) 121}, [\href{https://arxiv.org/abs/1205.6140}{{\ttfamily 1205.6140}}].

\bibitem{Kalamakis:2020aaj}
G.~Kalamakis, R.~G. Leigh and A.~C. Petkou, \emph{{Aspects of Holography of
  Taub-NUT-AdS4}},  \href{https://arxiv.org/abs/2009.08022}{{\ttfamily
  2009.08022}}.

\bibitem{Astefanesei:2004kn}
D.~Astefanesei, R.~B. Mann and E.~Radu, \emph{{Nut charged space-times and
  closed timelike curves on the boundary}},
  \href{https://doi.org/10.1088/1126-6708/2005/01/049}{\emph{JHEP} {\bfseries
  01} (2005) 049}, [\href{https://arxiv.org/abs/hep-th/0407110}{{\ttfamily
  hep-th/0407110}}].

\bibitem{Brecher:2003rv}
D.~Brecher, P.~DeBoer, D.~Page and M.~Rozali, \emph{{Closed time - like curves
  and holography in compact plane waves}},
  \href{https://doi.org/10.1088/1126-6708/2003/10/031}{\emph{JHEP} {\bfseries
  10} (2003) 031}, [\href{https://arxiv.org/abs/hep-th/0306190}{{\ttfamily
  hep-th/0306190}}].

\bibitem{Godel:1949ga}
K.~Godel, \emph{{An Example of a new type of cosmological solutions of
  Einstein's field equations of graviation}},
  \href{https://doi.org/10.1103/RevModPhys.21.447}{\emph{Rev. Mod. Phys.}
  {\bfseries 21} (1949) 447--450}.

\bibitem{10.2307/2415999}
M.~M. Som and A.~K. Raychaudhuri, \emph{Cylindrically symmetric charged dust
  distributions in rigid rotation in general relativity}, {\emph{Proceedings of
  the Royal Society of London. Series A, Mathematical and Physical Sciences}
  {\bfseries 304} (1968) 81--86}.

\bibitem{Leahy:1982dj}
D.~Leahy, \emph{{SCALAR AND NEUTRINO FIELDS IN THE GODEL UNIVERSE. (TALK)}},
  \href{https://doi.org/10.1007/BF02650235}{\emph{Int. J. Theor. Phys.}
  {\bfseries 21} (1982) 703--753}.

\bibitem{Novello:1992hp}
M.~Novello, N.~Svaiter and M.~Guimaraes, \emph{{Synchronized frames for Godel's
  universe}}, \href{https://doi.org/10.1007/BF00758823}{\emph{Gen. Rel. Grav.}
  {\bfseries 25} (1993) 137--164}.

\bibitem{Radu:2001jq}
E.~Radu and D.~Astefanesei, \emph{{Quantum effects in a rotating space-time}},
  \href{https://doi.org/10.1142/S0218271802001962}{\emph{Int. J. Mod. Phys. D}
  {\bfseries 11} (2002) 715--732},
  [\href{https://arxiv.org/abs/gr-qc/0112029}{{\ttfamily gr-qc/0112029}}].

\bibitem{Bueno:2018uoy}
P.~Bueno, P.~A. Cano, R.~A. Hennigar and R.~B. Mann, \emph{{NUTs and bolts
  beyond Lovelock}}, \href{https://doi.org/10.1007/JHEP10(2018)095}{\emph{JHEP}
  {\bfseries 10} (2018) 095},
  [\href{https://arxiv.org/abs/1808.01671}{{\ttfamily 1808.01671}}].

\bibitem{Teukolsky:1973ha}
S.~A. Teukolsky, \emph{{Perturbations of a rotating black hole. 1. Fundamental
  equations for gravitational electromagnetic and neutrino field
  perturbations}}, \href{https://doi.org/10.1086/152444}{\emph{Astrophys. J.}
  {\bfseries 185} (1973) 635--647}.

\bibitem{Newman:1961qr}
E.~Newman and R.~Penrose, \emph{{An Approach to gravitational radiation by a
  method of spin coefficients}},
  \href{https://doi.org/10.1063/1.1724257}{\emph{J. Math. Phys.} {\bfseries 3}
  (1962) 566--578}.

\bibitem{Stephani:2003tm}
H.~Stephani, D.~Kramer, M.~A. MacCallum, C.~Hoenselaers and E.~Herlt,
  \emph{{Exact solutions of Einstein's field equations}}.
\newblock Cambridge Monographs on Mathematical Physics. Cambridge Univ. Press,
  Cambridge, 2003,
  \href{https://doi.org/10.1017/CBO9780511535185}{10.1017/CBO9780511535185}.

\bibitem{Young:1983dn}
R.~Young, \emph{{SEMICLASSICAL STABILITY OF ASYMPTOTICALLY LOCALLY FLAT
  SPACES}}, \href{https://doi.org/10.1103/PhysRevD.28.2420}{\emph{Phys. Rev. D}
  {\bfseries 28} (1983) 2420--2435}.

\bibitem{Warnick:2006ih}
C.~Warnick, \emph{{Semi-classical stability of AdS NUT instantons}},
  \href{https://doi.org/10.1088/0264-9381/23/11/008}{\emph{Class. Quant. Grav.}
  {\bfseries 23} (2006) 3801--3817},
  [\href{https://arxiv.org/abs/hep-th/0602127}{{\ttfamily hep-th/0602127}}].

\bibitem{Holzegel:2006gn}
G.~Holzegel, \emph{{A Note on the instability of Lorentzian Taub-NUT-space}},
  \href{https://doi.org/10.1088/0264-9381/23/11/017}{\emph{Class. Quant. Grav.}
  {\bfseries 23} (2006) 3951--3962},
  [\href{https://arxiv.org/abs/gr-qc/0602045}{{\ttfamily gr-qc/0602045}}].

\bibitem{Krtous:2018bvk}
P.~Krtou\v{s}, V.~P. Frolov and D.~Kubiz\v{n}\'ak, \emph{{Separation of Maxwell
  equations in Kerr\textendash{}NUT\textendash{}(A)dS spacetimes}},
  \href{https://doi.org/10.1016/j.nuclphysb.2018.06.019}{\emph{Nucl. Phys. B}
  {\bfseries 934} (2018) 7--38},
  [\href{https://arxiv.org/abs/1803.02485}{{\ttfamily 1803.02485}}].

\bibitem{Chandrasekhar:1985kt}
S.~Chandrasekhar, \emph{{The mathematical theory of black holes}}.
\newblock 1985.

\bibitem{Dias:2009ex}
O.~J.~C. Dias, H.~S. Reall and J.~E. Santos, \emph{{Kerr-CFT and gravitational
  perturbations}},
  \href{https://doi.org/10.1088/1126-6708/2009/08/101}{\emph{JHEP} {\bfseries
  08} (2009) 101}, [\href{https://arxiv.org/abs/0906.2380}{{\ttfamily
  0906.2380}}].

\bibitem{Dias:2013sdc}
O.~J.~C. Dias and J.~E. Santos, \emph{{Boundary Conditions for Kerr-AdS
  Perturbations}}, \href{https://doi.org/10.1007/JHEP10(2013)156}{\emph{JHEP}
  {\bfseries 10} (2013) 156},
  [\href{https://arxiv.org/abs/1302.1580}{{\ttfamily 1302.1580}}].

\bibitem{Cardoso:2013pza}
V.~Cardoso, O.~J.~C. Dias, G.~S. Hartnett, L.~Lehner and J.~E. Santos,
  \emph{{Holographic thermalization, quasinormal modes and superradiance in
  Kerr-AdS}}, \href{https://doi.org/10.1007/JHEP04(2014)183}{\emph{JHEP}
  {\bfseries 04} (2014) 183},
  [\href{https://arxiv.org/abs/1312.5323}{{\ttfamily 1312.5323}}].

\bibitem{Teukolsky:1974yv}
S.~A. Teukolsky and W.~H. Press, \emph{{Perturbations of a rotating black hole.
  III - Interaction of the hole with gravitational and electromagnet ic
  radiation}}, \href{https://doi.org/10.1086/153180}{\emph{Astrophys. J.}
  {\bfseries 193} (1974) 443--461}.

\bibitem{PhysRevD.19.1641}
L.~S. Kegeles and J.~M. Cohen, \emph{Constructive procedure for perturbations
  of spacetimes}, \href{https://doi.org/10.1103/PhysRevD.19.1641}{\emph{Phys.
  Rev. D} {\bfseries 19} (Mar, 1979) 1641--1664}.

\bibitem{PhysRevD.11.2042}
P.~L. Chrzanowski, \emph{Vector potential and metric perturbations of a
  rotating black hole},
  \href{https://doi.org/10.1103/PhysRevD.11.2042}{\emph{Phys. Rev. D}
  {\bfseries 11} (Apr, 1975) 2042--2062}.

\bibitem{PhysRevLett.41.203}
R.~M. Wald, \emph{Construction of solutions of gravitational, electromagnetic,
  or other perturbation equations from solutions of decoupled equations},
  \href{https://doi.org/10.1103/PhysRevLett.41.203}{\emph{Phys. Rev. Lett.}
  {\bfseries 41} (Jul, 1978) 203--206}.

\end{thebibliography}\endgroup
\label{biblio}
\end{document}